\documentclass[useAMS]{mn2e}

\usepackage{amssymb,amsmath,psfig,times}
\def\gsim{ \lower .75ex \hbox{$\sim$} \llap{\raise .27ex \hbox{$>$}} }
\def\lsim{ \lower .75ex\hbox{$\sim$} \llap{\raise .27ex \hbox{$<$}} }

\def\sc{Schwarzschild}
\def\u{\underline}

\title[Properties of bright Fermi blazars]
{General physical properties of bright Fermi blazars}  
\author[G. Ghisellini, et al.]
{G. Ghisellini$^1$\thanks{Email: gabriele.ghisellini@brera.inaf.it}, 
F. Tavecchio$^1$, L. Foschini$^1$, G. Ghirlanda$^1$, L. Maraschi$^2$, A. Celotti$^3$\\
$^1$ INAF -- Osservatorio Astronomico di Brera, Via Bianchi 46, I--23807 Merate, Italy\\
$^2$ INAF -- Osservatorio Astronomico di Brera, Via Brera 28, I--20100 Milano, Italy\\
$^3$ S.I.S.S.A., V. Beirut 2--4, I--34014 Trieste, Italy 
}

\begin{document}  

\maketitle

\begin{abstract}
We studied all blazars of known redshift detected by the {\it Fermi} satellite 
during its first three--months survey.
For the majority of them, pointed {\it Swift} observations
ensures a good multiwavelength coverage, enabling us to 
to reliably construct their spectral energy distributions (SED).
We model the SEDs using a one--zone leptonic model 
and study the distributions of the derived interesting physical parameters 
as a function of the observed $\gamma$--ray luminosity.
We confirm previous findings concerning the relation of the 
physical parameters with source luminosity which are 
at the origin of the blazar sequence.
The SEDs allow to estimate the luminosity of the accretion disk 
for the majority of broad emitting line blazars, while  
for the line--less BL Lac objects in the sample upper limits can be derived.
We find a positive correlation between the jet power and the
luminosity of the accretion disk in broad line blazars. 
In these objects we argue that the jet must be proton--dominated, and
that the total jet power is of the same order of (or slightly larger than) 
the disk luminosity.
We discuss two alternative scenarios to explain this result.
\end{abstract}
\begin{keywords}
BL Lacertae objects: general --- quasars: general ---
radiation mechanisms: non--thermal --- gamma-rays: theory --- X-rays: general
\end{keywords}

\section{Introduction}

The Large Area Telescope (LAT) onboard the {\it Fermi} Satellite, 
in the first months of operation succeeded to double
the number of know $\gamma$--ray emitting blazars 
(Abdo et al. 2009a, hereafter A09).
Excluding radio--galaxies, in the list of the LAT 3--months
survey we have 104 blazars, divided into 58 Flat Spectrum Radio Quasars
(FSRQs, including one Narrow Line Seyfert 1 galaxy), 
42 BL Lacs and 4 sources of uncertain classification.
Of these, 89 have a known redshift, and a large fraction of 
sources have been observed with the {\it Swift} satellite
within the 3 months survey, while others have been observed
by {\it Swift} at other epochs.

The combination of 
{\it Fermi} and {\it Swift} data are of crucial importance for the
modelling of these blazars, even if the 
LAT data are an average over the 3 months of the survey,
so that, strictly speaking, we cannot have a really simultaneous
optical to X--ray SED, even when the {\it Swift} observations
have been performed during the three months of the survey. 
Despite this, the optical/UV, X--ray and $\gamma$--ray
coverage can often define or strongly constrain
the position and the flux levels
of both peaks of the non--thermal emission of blazars.
Furthermore, as discussed in Ghisellini, Tavecchio \& Ghirlanda 
(2009, hereafter Paper 1),
the data from the UV Optical Telescope (UVOT) of {\it Swift} are often crucial
to disentangle the non--thermal beamed jet emission from thermal radiation
produced by the accretion disk. 
When the accretion disk is visible it is then possible to 
estimate both the mass of the black hole and the accretion luminosity
and to compare it with the power carried by the jet.
In Paper I we did this study for the 23 most luminous blazars,
exceeding, in the $\gamma$--ray range, an observed 
luminosity of $10^{48}$ erg s$^{-1}$.
In Tavecchio et al. (2009, hereafter T09) we considered the 
BL Lac objects detected by {\it Fermi} with
the aim to find, among those, the best candidates to be TeV emitters.
Here we extend these previous studies to the entire {\it Fermi} blazar sample
of the 3 months survey.
Besides finding the intrinsic physical properties characterising the
emitting region of these {\it Fermi} blazars, like magnetic 
field, particle density, size, beaming factor and so on, 
the main goal of our study is to investigate if there is a 
relation between the jet power and the disk luminosity,
as a function of the $\gamma$--ray luminosity.

This implies two steps. 
First we collect the data to construct
the SED of all sources, taking advantage
of possible {\it Swift} observations, that we analyse, and of archival
data taken from the NASA Extragalactic Database (NED, {\tt
http://nedwww.ipac.caltech.edu/}.
Secondly, we apply a model to fit these data, that returns the 
value of the physical parameters of the emitting region,
and, if possible, the value of the accretion disk luminosity
and of the black hole mass.
We do this for all the 85 sources (out of 89)
with a minimum coverage of the SED.
Among these blazars, there are several BL Lac objects for which
there is no sign of thermal emission produced by a disk, nor
of emission lines produced by the photo--ionising 
disk photons.
For them we derive an upper limit of the luminosity of a standard,
Shakura--Sunyaev (1973) disk.
We also derive the distributions of the needed physical 
parameters, and their dependence on the $\gamma$--ray luminosity
and on the presence/absence of a prominent accretion disk.
As expected, the jets of line--less BL Lacs have less power than
broad line FSRQs, have bluer spectra, and their emitting electrons 
suffer less radiative cooling.
This confirms earlier results (Ghisellini et al. 1998) explaining
the so called blazar sequence (Fossati et al. 1998).

The main result of our analysis concerns the relation
between the jet power and the accretion disk luminosity.
We find that they correlate in FSRQs, for which we can estimate 
both the black hole mass and the accretion luminosity.
We discuss two alternative scenarios to explain this behaviour.
For BL Lacs (with only an upper limit on the accretion 
luminosity), we suggest that the absence of any sign of
thermal emission, coupled to the presence of a relatively
important jet, strongly suggests a radiatively inefficient 
accretion regime.

In this paper we use a cosmology with $h=\Omega_\Lambda=0.7$ and $\Omega_{\rm M}=0.3$,
and use the notation $Q=Q_X 10^X$ in cgs units (except for the black hole masses,
measured in solar mass units).

\section{The sample}

Tab. \ref{sample} lists the 89 blazars with redshift 
of the A09 catalogue of {\it Fermi} detected blazars.
Besides their name, we report the redshift, 
the K--corrected $\gamma$--ray luminosity in the {\it Fermi}/LAT band, 
(see e.g. Ghisellini, Maraschi \& Tavecchio 2009), if there are observations
by the {\it Swift} satellite, if the  source was detected by EGRET and
the classification.
For the latter item we have maintained the same classification
as in A09, but we have indicated those sources classified as BL Lac objects 
but that do have broad emission lines.

The distinction between BL Lacs and FSRQs is traditionally 
based on the line equivalent width (EW) being smaller or larger than 5\AA.
With this definition, we classify as BL Lacs those sources having 
genuinely very weak or absent lines and also objects with
strong lines but whose non--thermal continuum is so enhanced to
reduce the line EW.
This second category of ``BL Lacs" should physically be associated 
to FSRQs. 
We have tentatively made this distinction on Tab. \ref{sample},
based on some information from the literature about the presence,
in these objects, of strong broad lines.
In the rest of the paper we will consider them as FSRQs.

The source PMN 0948--0022 belongs to still another class, being 
a Narrow Line Seyfert 1 (NLSy1; FWHM$\sim$1500 km s$^{-1}$). 
This source has been discussed in detail in Abdo et al. (2009b)
and Foschini et al. (2009):
its SED and general properties are indistinguishable from
FSRQs and we then assign this source to this group.

We then have 
57 FSRQs, 1 NLSy1  and 31 BL Lac objects.
Of the latter, 6 have strong broad lines or have been also classified as 
FSRQs (by Jauncey et al. 1989).
Four of the FSRQs (indicated with the superscript $n$ in Table \ref{sample})
do not have a sufficient data coverage to allow
a meaningful interpretation, and we will not discuss them.
The 23 blazars (21 FSRQs and 2 BL Lacs, but with broad emission lines)
with an average $\gamma$--ray luminosity exceeding $10^{48}$ erg s$^{-1}$ 
written in italics have been discussed in Paper 1.
The 25 underlined BL Lacs are discussed in T09.
One of these, (0814+425=OJ 425) is shown also here, since
it can be fitted both with a pure synchrotron self--Compton (SSC) 
and an ``External Compton" (EC) model (see next section).

In summary, in this paper we present the SED and the corresponding models
for 37 blazars: 32 FSRQs and 5 BL Lacs that are suspected to be FSRQs with
broad lines hidden by the beamed continuum.
However, when discussing the general properties of the {\it Fermi} blazars,
we will consider the entire blazar sample of Tab. \ref{sample}, with the 
only exception of the 4 FSRQs with a very poor data coverage. 
These are 85 sources.
To this aim we will use the results of Paper 1, concerning
the most $\gamma$--ray luminous blazars, and we will apply our  
model also to the BL Lacs shown in T09.
For the ease of the reader, we report in Table \ref{para} and Table \ref{powers} 
the physical parameters of all the 85 blazars.

\begin{table}
\centering
\begin{tabular}{lllllll}
\hline
\hline
Name  &Alias &$z$ &$\log L_\gamma$ &{\it S?} &E?  &Type\\
\hline
0017--0512$^{n}$ &CGRaBS  &0.227  &45.96 &Y &Y  &Q \\
\u{00311--1938} &\u{KUV}  &0.610  &46.40 &Y &   &B \\
{\it 0048--071}&{\it PKS} &1.975  &48.2  &  &   &Q \\
0116--219      &PKS        &1.165  &47.68 &  &   &Q \\
\u{0118--272}  &\u{PKS}    &0.559  &46.48 &  &UL &B \\
0133+47        &DA 55      &0.859  &47.41 &Y &UL &Q \\
0142--278      &PKS        &1.148  &47.63 &  &   &Q \\
{\it 0202--17} &{\it PKS}  &1.74   &48.2  &  &UL &Q \\
0208--512      &PKS        &1.003  &47.92 &Y &Y  &Q \\
{\it 0215+015} &{\it PKS}  &1.715  &48.16 &Y &UL &Q \\ 
0218+35        &B2         &0.944  &47.47 &Y &UL &Q \\ 
\u{0219+428}   &\u{3C66A}  &0.444  &47.16 &Y &Y  &B \\
{\it 0227--369}&{\it PKS}  &2.115  &48.6  &Y &   &Q \\
{\it 0235+164} &{\it AO}   &0.94   &48.4  &Y &Y  &B$^*$ \\
\u{0301--243}  &\u{PKS}    &0.260  &45.77 &Y &UL &B \\
0332--403      &PKS &1.445$^b$    &47.68 &Y &UL &B$^{**}$ \\ 
{\it 0347--211}&{\it PKS}  &2.944  &49.1  &Y &   &Q \\
{\it 0426--380}&{\it PKS}  &1.112  &48.06 &Y &UL &B$^*$ \\
\u{0447--439}  &\u{PKS}    &0.107  &46.03 &Y &   &B \\
{\it 0454--234}&{\it PKS}  &1.003  &48.16 &Y &Y  &Q \\
\u{0502+675}   &\u{1ES}    &0.314  &46.06 &Y &   &B \\
{\it 0528+134} &{\it PKS}  &2.04   &48.8  &Y &Y  &Q \\
0537--441      &PKS        &0.892  &47.99 &Y &Y  &B$^*$ \\   
0650+453       &B3         &0.933  &47.82 &Y &   &Q \\
0713+1935$^n$  &CLASS      &0.534  &46.84 &  &   &Q \\
0716+332       &TXS        &0.779  &47.12 &Y &   &Q \\
\u{0716+714}   &\u{TXS}    &0.26   &46.55 &Y &Y  &B \\
\u{0735+178}   &\u{PKS}    &0.424  &46.31 &Y &Y  &B \\
\u{0814+425}   &\u{OJ 425} &0.53   &46.87 &Y &UL &B \\
{\it 0820+560} &{\it S4}   &1.417  &48.01 &Y &   &Q \\
\u{0851+202}   &\u{OJ 287} &0.306  &46.18 &Y &UL &B \\ 
{\it 0917+449} &{\it TXS}  &2.1899 &48.4  &Y &Y  &Q \\
0948+0022      &PMN        &0.585  &46.95 &Y &   &NL$^a$\\
0954+556       &4C 55.17   &0.8955 &47.41 &Y &Y  &Q \\
\u{1011+496}   &\u{1ES}    &0.212  &45.83 &Y &   &B \\ 
1012+2439$^n$  &CRATES     &1.805  &47.99 &  &   &Q \\
{\it 1013+054} &{\it TXS}  &1.713  &48.2  &  &   &Q \\
1030+61        &S4         &1.401  &47.87 &Y &   &Q \\
\u{1050.7+4946}&\u{MS}     &0.140  &44.65 &Y &   &B \\
1055+018       &PKS        &0.89   &47.11 &Y &UL &Q \\
1057--79       &PKS        &0.569  &47.06 &Y &UL &B$^{**}$ \\
\u{10586+5628} &\u{RX}     &0.143  &45.14 &  &   &B \\   
\u{1101+384}   &\u{Mkn 421} &0.031 &44.52 &Y &Y  &B \\ 
\hline
\hline
\end{tabular}
\caption{The 89 {\it Fermi} blazars in the A09 catalogue 
with redshift, including two blazars that have no $z$ in the A09 lists
(0332--403 and 1553+11).
In the last 4 columns we indicate the logarithm of the average $\gamma$--ray
luminosity as observed by {\it Fermi} during the first 3 months of survey
(cgs units); if there are {\it Swift} observations; if the source was 
detected by EGRET (UL stands for an upper limit given
by Fichtel et al. 1994); the classification of the source (B=BL Lac; Q=FSRQs;
NL=Narrow line Seyfert galaxy; U=uncertain classification).
$a$: Narrow Line Seyfert 1, analysed in Abdo et al. (2009b) and Foschini et al. (2009).
$b$: redshift uncertain.
$n$: not studied in this paper due to lack of multiwavelength data; 
$*$: defined as BL Lacs for the EW of the lines, but broad 
emission lines are present.
$**$: classified as FSRQ in Jauncey et al. (1989).
}
\end{table}

\setcounter{table}{0}
\begin{table}
\centering
\begin{tabular}{lllllll}
\hline
\hline
Name  &Alias &$z$ &$\log L_\gamma$ &{\it S?} &E?  &Type\\
\hline   
1127--145      &PKS        &1.184  &47.70 &Y &Y  &Q \\
1144--379      &PKS        &1.049  &47.35 &Y &UL &Q \\
1156+295       &4C 29.45   &0.729  &47.15 &Y &Y  &Q \\
\u{1215+303}   &\u{B2}     &0.13   &45.57 &  &UL &B \\
\u{1219+285}   &\u{ON 231} &0.102  &45.25 &Y &Y  &B \\
1226+023       &3C 273     &0.158  &46.33 &Y &Y  &Q \\
1244--255      &PKS        &0.635  &46.86 &Y &UL &Q \\
1253--055      &3C 279     &0.536  &47.31 &Y &Y  &Q \\
1308+32        &B2         &0.996  &47.72 &Y &UL &Q \\
{\it 1329--049}&{\it PKS}  &2.15   &48.5  &  &   &Q \\
1333+5057$^n$  &CLASS      &1.362  &47.73 &  &   &Q \\
1352--104      &PKS        &0.332  &46.17 &  &UL &Q \\
{\it 1454--354}&{\it PKS}  &1.424  &48.5  &Y &Y  &Q \\
{\it 1502+106} &{\it PKS}  &1.839  &49.1  &Y &UL &Q \\
1508--055      &PKS        &1.185  &47.65 &Y &UL &Q \\
1510--089      &PKS        &0.360  &47.10 &Y &Y  &Q \\ 
\u{1514--241}  &\u{Ap Lib} &0.048  &44.25 &Y &Y  &B \\ 
{\it 1520+319} &{\it B2}   &1.487  &48.4  &Y &   &Q \\
{\it 1551+130} &{\it PKS}  &1.308  &48.04 &  &   &Q \\
\u{1553+11}    &\u{PG}     &0.36$^b$ &46.57 &Y &   &B \\ 
1622--253      &PKS        &0.786  &47.44 &  &Y  &Q \\
{\it 1633+382} &{\it 4C+38.41} &1.814  &48.6 &Y &Y &Q   \\
\u{1652+398}   &\u{Mkn 501}&0.0336 &43.95 &Y &Y  &B \\ 
\u{1717+177}   &\u{PKS}    &0.137  &45.50 &Y &UL &B \\
\u{1749+096}   &\u{OT 081} &0.322  &46.56 &Y &UL &B \\
1803+784       &S5         &0.680  &46.88 &Y &UL &B$^*$ \\
1846+322       &TXS        &0.798  &47.42 &Y &   &Q \\
1849+67        &S4         &0.657  &47.29 &Y &   &Q \\
1908--201      &PKS        &1.119  &47.99 &Y &Y  &Q \\
1920--211      &TXS        &0.874  &47.49 &Y &Y  &Q \\
\u{1959+650}   &\u{1ES}    &0.047  &44.30 &Y &   &B \\ 
\u{2005--489}  &\u{PKS}    &0.071  &44.51 &Y &Y  &B \\ 
{\it 2023--077}&{\it PKS}  &1.388  &48.6  &Y &Y  &Q \\
{\it 2052-47}  &{\it PKS}  &1.4910 &48.03 &  &Y  &Q \\
2141+175       &OX 169     &0.213  &45.93 &Y &UL &Q \\
2144+092       &PKS        &1.113  &47.85 &Y &UL &Q \\ 
\u{2155--304}  &\u{PKS}    &0.116  &45.76 &Y &Y  &B \\
2155+31        &B2         &1.486  &47.84 &Y &   &Q \\
\u{2200+420}   &\u{BL Lac} &0.069  &44.74 &Y &Y  &B \\
2201+171       &PKS        &1.076  &47.45 &Y &UL &Q \\
2204--54       &PKS        &1.215  &47.80 &Y &UL &Q \\
{\it 2227--088}&{\it PHL 5225} &1.5595 &48.2 &Y &UL &Q \\
2230+114       &CTA102     &1.037  &47.69 &Y &Y  &Q \\
{\it 2251+158} &{\it 3C 454.3} &0.859  &48.7 &Y &Y &Q  \\
{\it 2325+093} &{\it PKS}  &1.843  &48.5  &Y &   &Q  \\
2345--1555     &PMN        &0.621  &46.99 &Y &   &Q \\  
\hline
\hline 
\end{tabular}
\vskip 0.4 true cm
\caption{-- continue --
}
\label{sample}
\end{table}

\section{Swift observations and analysis}

For 33 of the 37 blazars studied in this paper there are {\it Swift} observations,
with several of them being observed during
the 3 months of the {\it Fermi} survey.
The data were analysed with the 
most recent software \texttt{SWIFT\_REL3.2} released as part of the 
\texttt{Heasoft v. 6.6.2}.
The calibration database is that updated to April 10, 2009. 
The XRT data were processed with the standard procedures 
({\texttt{XRTPIPELINE v.0.12.2}). 
We considered photon counting (PC) mode data with the standard 0--12 grade selection. 
Source events were extracted in a circular region of  aperture $\sim 47''$, 
and background was estimated in a same sized circular region far from the source. 
Ancillary response files were created through the \texttt{xrtmkarf} task. 
The channels with energies below 0.2 keV and above 10 keV were excluded 
from the fit and the spectra were rebinned in energy so to have at 
least 30 counts per bin. 
Each spectrum was analysed through XSPEC
with an absorbed power--law with a fixed 
Galactic column density from Kalberla et al. (2005).
The computed errors represent the 90\% confidence interval on the spectral parameters.
Tab. \ref{xrt} reports the log of the observations and the results of
the fitting the X--ray data with a simple power law model.

UVOT (Roming et al. 2005) source counts were extracted from 
a circular region $5''-$sized centred on the source position, 
while the background was extracted from 
a larger circular nearby source--free region.
Data were integrated with the \texttt{uvotimsum} task and then 
analysed by using the  \texttt{uvotsource} task.
The observed magnitudes have been dereddened according to the formulae 
by Cardelli et al. (1989) and converted into fluxes by using standard 
formulae and zero points from Poole et al. (2008).
Tab. \ref{uvot} list the observed magnitudes in the 6 filters of UVOT.


%
\begin{table*}
\centering
\begin{tabular}{llllllll}
\hline
\hline
source     &Obs. date   &$N^{\rm Gal}_{\rm H}$  &$\Gamma$    &$\chi2{\rm /dof}$ & $F_{\rm 0.2-10, unabs}$  &$F_{\rm 2-10, unabs}$ \\
           &dd/mm/yyyy  &$10^{20}$ cm$^{-2}$& & &10$^{-12}$  cgs &10$^{-12}$ cgs  \\
\hline
0133+47*   &18/11/2008         &11.4 &1.4$\pm$0.2   &19/13   &3.72$\pm $0.43  &2.57$\pm$0.33     \\
0208--512* &29/12/2008         &3.19 &1.9$\pm$0.2   &6/11    &2.6$\pm $0.3    &1.26$\pm$0.2      \\
0218+35*   &12/12/2008         &5.86 &2.7$_{-1.0}^{+1.1}$    &0.2/3          &2.0$\pm$0.4       & 0.29$\pm$0.05  \\
0332--403($^{a}$) &25/02/2009  &1.38 &1.4$\pm$0.2   &2/6   & 3.5$\pm$0.4      & 2.4$\pm$0.2  \\
0537--441  &08/10/2008         &3.94 &1.7$\pm$0.1   &23/20   &8.24$\pm $0.9   &4.63$\pm$0.47     \\
0650+453*  &14/02/2009         &8.64 &2.1$\pm$0.6  &5/5             &0.76$\pm$0.11 & 0.28$\pm$0.04  \\
0716+332   &27/01/2008         &5.93 &1.8$\pm$0.3   &3/3     &0.9$\pm$0.2     &0.4$\pm$0.2       \\
0948+0022**  &05/12/2008       &5.22 &1.8$\pm$0.2   & 5/8       & 4.4$\pm$0.2 &  2.2$\pm$0.1  \\
0954+556   &05/03/2009         &0.853&0.9$_{-1.8}^{+1.1}$  &0.1/2        & 2.2$\pm$0.5                & 1.9$\pm$0.5  \\
1030+61*   &03/07/2009         &0.63 &1.9$\pm$0.7   &0.3/4        & 0.62$\pm$0.10               &0.27$\pm$0.04  \\
1055+018*  &19/07/2008         &4.02 &1.8$\pm$0.5   &1/3     &3.5             &1.8               \\
1057--79   &20/01/2008         &8.76 &1.9$\pm$0.15  &6/10    &2.8$\pm$0.4     &1.3$\pm$0.3       \\
1127--145  &24/03/2007         &4.04 &1.31$\pm$0.05 &116/94  &9.8$\pm $0.5    &7.2$\pm$0.4       \\
1144--379* &21/11/2008         &7.5  &1.96$\pm$0.32 &1/5     &1.0$\pm $0.3    &0.46$\pm$1.2      \\
1156+295   &21/11/2008         &1.68 &1.52$\pm$0.14 &11/7    &2.7$\pm $0.4    &1.7$\pm$0.2       \\
1226+023   &10/05/2008         &1.79 &1.56$\pm$0.05 &88/68   &833$\pm$19       &521$\pm$26        \\
1244--255  &17/01/2007         &6.85 &1.8$\pm$0.5   &3/4     &2.1$\pm$0.5     &1.0$\pm$0.4       \\
1253--055($^{b}$) &08/08/2008  &2.12 &1.76$\pm$0.04             & 88/94       & 10.5$\pm$0.2  &  5.5$\pm$0.1    \\
1308+32    &20/08/2008         &1.27 &1.6$\pm$0.2  & 5/10       & 3.0$\pm$0.2      & 1.8$\pm$0.1   \\
1508--055($^{c}$) &20/02/2007  &6.09 &1.8$\pm$0.2   & 8/6        & 0.95$\pm$0.08    & 0.50$\pm$0.04  \\
1510--089  &10/01/2009         &7.8  &1.4$\pm$0.11  &194/316 &6.4$\pm $1.3    &4.4$\pm$0.4       \\
1803+784   &17/02/2007         &4.12 &1.5$\pm$0.1   &14/14   &2.8$\pm$0.4     &1.8$\pm$0.2       \\
1846+322($^{d}$)  &28/12/2008  &9.93 &1.7$\pm$0.2   & 10/7       & 0.78$\pm$0.05     &  0.45$\pm$0.03  \\
1849+67($^{e}$)   &16/11/2006  &4.66 &1.5$\pm$0.1   & 22/14       & 2.6$\pm$0.1 &   1.59$\pm$0.07       \\
1908--201($^{f}$) &08/03/2007  &9.24 &1.4$\pm$0.1 & 5/10  &  2.7$\pm$0.1  & 1.83$\pm$0.05 \\
1920--211  &07/08/2007         &5.69 &1.6$\pm$0.3     & 1/3  & 1.4$\pm$0.1  &  0.87$\pm$0.09  \\
2141+175($^{g}$) &19/04/2007   &7.35 &1.71$\pm$0.06  &   94/67     &  2.23$\pm$0.04  &   1.47$\pm$0.03 \\
2144+092($^{h}$) &28/04/2009   &4.56 &1.6$\pm$0.2 & 13/9       &   1.6$\pm$0.1   &  1.01$\pm$0.07   \\
2155+31($^{i}$)  &01/04/2009   &7.42 &1.0$\pm$0.6 & 0.6/1       &  1.0$\pm$0.1   &  0.8$\pm$0.1  \\
2201+171($^{l}$) &08/12/2009   &4.56 &1.80$_{-0.15}^{+0.08}$  & 13/17  & 1.22$\pm$0.05   &  0.62$\pm$0.03  \\
2204--54   &21/12/2008         &1.72 &1.6$\pm$0.1 &  9/17      & 2.2$\pm$0.1  & 1.37$\pm$0.07 \\
2230+114($^{m}$) &19/05/2005   &4.76 &1.50$\pm$0.05 &  74/66  &  6.5$\pm$0.1  &  4.2$\pm$0.1  \\
2345--155($^{n}$) &23/12/2008  &1.64 &1.6$\pm$0.4   & 0.3/1  & 0.39$\pm$0.05   & 0.23$\pm$0.03  \\
\hline
\hline
\end{tabular}
\vskip 0.4 true cm
\caption{Results of the X--ray analysis.
*: poorly determined spectrum, the C--Statistic was used to fit the  spectrum. 
Therefore, the $\chi2{\rm /dof}$ indicates instead the C-stat value and the PHA bins.
** Abdo et al. (2009b).
{\it a:} Average of 2 observations the same day. 
{\it b:} Average of 8 observations between 8 and 20 August 2008.
{\it c:} Average of 4 observations between 26/02/2007 and 13/09/2008.
{\it d:} Average of 2 observations on 28/12/2008 and 20/02/2009.
{\it e:} Average of 3 observations between 16/11/2006 and 04/08/2008.
{\it f:} Average of 4 observations between 8 and 17 March 2007.
{\it g:} Average of 9 observations between 19/04/2007 and 15/01/2009. 
{\it h:} Average of 2 observations the same day.
{\it i:} Average of 3 observations between 1 and 17 April 2009.
{\it l:} Average of 7 observations between 08/12/2006 and 18/04/2009.
{\it m:} Average of 5 observations between 19/05/2005 and 01/05/2007. 
{\it n:} Average of 2 observations on 23/12/2008 and 10/01/2009.
}
\label{xrt}
\end{table*}
\begin{table*}
\centering
\begin{tabular}{llllllll}
\hline
\hline
source        &OBS date    &$V$             &$B$             &$U$              &$W1$            & $M2$           &$W2$          \\
\hline
0133+47    &18/11/2008    &15.89$\pm$0.02 &16.46$\pm$0.01  &15.89$\pm$0.01 &16.30$\pm$0.02 &16.56$\pm$0.03 &16.74$\pm$0.02 \\
0208--512  &29/12/2008    &17.69$\pm$0.07 &18.06$\pm$0.05  &17.01$\pm$0.04 &16.79$\pm$0.03 &16.69$\pm$0.03 &17.03$\pm$0.03 \\
0218+35    &12/12/2008    &$>$19.4        &$>$20.3         &$>$19.9        &$>$20.3        &$>$20.1        &$>$21.0        \\
0332--403  &25/02/2009    &17.00$\pm$0.07 &17.59$\pm$0.05  &16.85$\pm$0.05 &17.23$\pm$0.06 &17.29$\pm$0.07 &18.20$\pm$0.07 \\
0537--441  &08/10/2008    &16.04$\pm$0.02 &16.48$\pm$0.02  &15.78$\pm$0.02 &16.01$\pm$0.02 &16.02$\pm$0.02 &16.26$\pm$0.02 \\
0650+453   &14/02/2009    &$>$19.8        &20.1$\pm$0.2    &19.3$\pm$0.2   &19.3$\pm$0.1   &19.4$\pm$0.1   &19.8$\pm$0.1   \\
0716+332   &27/01/2008    &--             &--              &--             &--             &--             &17.71$\pm$0.03 \\
0948+0022* &05/12/2008    &18.2$\pm$0.2   & 18.56$\pm$0.07 &17.79$\pm$0.06 &17.48$\pm$0.06 &17.50$\pm$0.06 &17.55$\pm$0.05 \\
0954+556   &05/03/2009    &17.6$\pm$0.2   & 17.97$\pm$0.06 &16.91$\pm$0.04 &16.80$\pm$0.04 &16.8$\pm$0.2   &16.94$\pm$0.08 \\
1030+61    &03/07/2009    &18.4$\pm$0.1   & 19.14$\pm$0.08 &18.36$\pm$0.07 &18.56$\pm$0.07 &18.5$\pm$0.1   &19.08$\pm$0.08 \\
1055+018   &19/07/2008    &--             &--              &--             &--             &16.97$\pm$0.04 &--             \\
1057--79   &20/01/2008    &--             &--              &--             &--             &17.31$\pm$0.03 &17.36$\pm$0.01 \\
1127--145  &24/03/2007    &16.46$\pm$0.02 & 16.70$\pm$0.01 &15.64$\pm$0.01 &15.51$\pm$0.01 &15.55$\pm$0.01 &15.79$\pm$0.01 \\
1144--379  &21/11/2008    &--             &--              &--             &--             &--             &20.0$\pm$0.1   \\
1156+295   &21/11/2008    &--             &--              &--             &16.85$\pm$0.01 &--             &--             \\
1226+023   &10/05/2008    &12.72$\pm$0.01 &12.92$\pm$0.01  &11.92$\pm$0.01 &11.48$\pm$0.01 &11.33$\pm$0.01 &11.34$\pm$0.01 \\
1244--255  &17/01/2007    &--             &--              &--             &--             &17.44$\pm$0.03 &--             \\
1253--055  &08/08/2008    &16.80$\pm$0.09 & 16.25$\pm$0.06 &16.41$\pm$0.05 &16.44$\pm$0.05 &16.35$\pm$0.06 &16.50$\pm$0.04 \\
1308+32    &20/08/2008    &--             &--              &--             &--             &16.71$\pm$0.03 &--             \\
1508--055  &20/02/2007    &17.4$\pm$0.2   &17.5$\pm$0.1    &16.56$\pm$0.06 &16.38$\pm$0.09 &16.8$\pm$0.2   &16.97$\pm$0.09 \\
1510--089  &10/01/2009    &16.76$\pm$0.08 &17.04$\pm$0.04  &16.18$\pm$0.02 &16.51$\pm$0.02 &16.42$\pm$0.04 &16.52$\pm$0.03 \\
1803+784   &17/02/2007    &16.51$\pm$0.03 &16.99$\pm$0.02  &16.32$\pm$0.02 &16.50$\pm$0.02 &16.55$\pm$0.02 &16.65$\pm$0.01 \\
1846+322   &28/12/2008    &18.9$\pm$0.1   &19.27$\pm$0.09  &18.50$\pm$0.07 &18.41$\pm$0.06 &18.45$\pm$0.06 &18.62$\pm$0.05 \\
1849+67    &16/11/2006    &17.6$\pm$0.1   &17.93$\pm$0.06  &17.28$\pm$0.06 &17.36$\pm$0.06 &16.13$\pm$0.03 &--             \\
1908--201  &08/03/2007    &16.8$\pm$0.1   &17.40$\pm$0.09  &16.80$\pm$0.08 &17.09$\pm$0.09 &17.34$\pm$0.08 &17.66$\pm$0.09 \\
1920--211  &07/08/2007    &--             &--              &--             &16.73$\pm$0.03 &--             &--             \\
2141+175   &19/04/2007    &16.16$\pm$0.02 &16.39$\pm$0.02  &15.21$\pm$0.02 &--             &15.09$\pm$0.03 &15.12$\pm$0.03 \\
2144+092   &28/04/2009    &18.2$\pm$0.2   &18.48$\pm$0.09  &17.67$\pm$0.07 &17.50$\pm$0.06 &17.68$\pm$0.07 &18.00$\pm$0.05 \\
2155+31    &01/04/2009    &--             &--              &--             &21.4$\pm$0.4   &--             &$>$ 20.9       \\
2201+171   &08/12/2009    &16.93$\pm$0.06 &17.71$\pm$0.05  &17.23$\pm$0.04 &17.62$\pm$0.05 &17.7$\pm$0.1   &17.86$\pm$0.05 \\
2204--54   &21/12/2008    &17.89$\pm$0.08 &18.18$\pm$0.04  &16.98$\pm$0.03 &16.83$\pm$0.03 &16.99$\pm$0.04 &17.40$\pm$0.04 \\
2230+114   &19/05/2005    &--             &--              &16.52$\pm$0.02 &--             &16.38$\pm$0.04 &16.70$\pm$0.03 \\
2345--155  &23/12/2008    &18.5$\pm$0.1   &18.61$\pm$0.06  &17.91$\pm$0.05 &17.57$\pm$0.07 &17.66$\pm$0.05 &17.78$\pm$0.04 \\
\hline
\hline
\end{tabular}
\vskip 0.4 true cm
\caption{UVOT Observed magnitudes.
* see also Abdo et al. (2009b) and Foschini et al. (2009).
}
\label{uvot}
\end{table*}

\section{The model}

We use the model described in detail in Ghisellini \& Tavecchio (2009,
hereafter GT09).
It is a relatively simple, one zone, homogeneous synchrotron 
and Inverse Compton model, aiming at accounting the several 
contributions to the radiation energy density produced externally 
to the jet, and their dependence upon the
distance of the emitting blob to the black hole.
Besides the synchrotron radiation produced internally to the jet,
we in fact consider radiation coming directly from the disk 
(i.e. Dermer \& Schlickeiser 1993),
the broad line region (BLR; e.g. Sikora, Begelman \& Rees 1994), a dusty torus
(see  B{\l}azejowski et al. 2000; Sikora et al. 2002), the host galaxy light
and the cosmic background radiation.

The emitting region is assumed spherical, of size $r_{\rm diss}$, moving 
with a bulk Lorentz factor $\Gamma$ and is located at a distance 
$R_{\rm diss}$ from the black hole of mass $M$.
The bolometric luminosity of the accretion disk is $L_{\rm d}$.
The jet accelerates in its inner parts with $\Gamma\propto R^{1/2}$ 
($R$ is the distance from the black hole), up to a value $\Gamma_{\rm max}$. 
In the acceleration region the jet is parabolic 
(following, e.g. Vlahakis \& K\"onigl 2004) and beyond this point the jet becomes
conical with a semi--aperture angle $\psi$ (assumed to be 0.1 for all sources).

The energy particle distribution $N(\gamma)$ [cm$^{-3}$]
is calculated solving the continuity 
equation where particle injection, radiative cooling and pair production
(via the $\gamma$--$\gamma \to e^\pm$ process) are taken into account.
The injection function $Q(\gamma)$ [cm$^{-3}$ s$^{-1}$]
is assumed to be a smoothly joining broken power--law,
with a slope $Q(\gamma)\propto \gamma^{-{s_1}}$ and
$\gamma^{-{s_2}}$ below and above a break energy $\gamma_{\rm b}$:
\begin{equation}
Q(\gamma)  \, = \, Q_0\, { (\gamma/\gamma_{\rm b})^{-s_1} \over 1+
(\gamma/\gamma_{\rm b})^{-s_1+s_2} }
\label{qgamma}
\end{equation}

The total power injected into the source in the form of relativistic
electrons is $P^\prime_{\rm i}=m_{\rm e}c^2 V\int Q(\gamma)\gamma d\gamma$,
where $V=(4\pi/3)r_{\rm diss}^3$ is the volume of the emitting region.

The injection process lasts for a light crossing time
$r_{\rm diss}/c$, and we calculate $N(\gamma)$ at this time.
This assumption comes from the fact that even if injection lasted longer,
adiabatic losses caused by the expansion of the source (which
is travelling while emitting) and the corresponding decrease 
of the magnetic field would make the observed flux to decrease. 
Therefore our calculated spectra correspond to the maximum of
a flaring episode.

The BLR is assumed for simplicity to be a thin spherical shell 
located at a distance $R_{\rm BLR}=10^{17}L_{\rm d, 45}^{1/2}$ cm.
A fraction $f_{\rm BLR}=0.1$ of the disk luminosity is re--emitted by broad lines.
Since $R_{\rm BLR}\propto L_{\rm d}^{1/2}$, the
radiation energy density of the broad line emission within the BLR is constant,
but is seen amplified by a factor $\sim\Gamma^2$ by the moving blob, as long
as $R_{\rm diss}<R_{\rm BLR}$.
A dusty torus, located at a distance 
$R_{\rm IR}=2.5\times 10^{18}L_{\rm d}^{1/2}$ cm,
reprocesses a fraction $f_{\rm IR}$ (of the order of 0.1--0.3) of $L_{\rm d}$
through dust emission in the far IR.
Above and below the accretion disk, in its inner parts, there is an X--ray emitting
corona of luminosity $L_{\rm X}$ (we almost always fixes it at a 
level of 30\% of $L_{\rm d}$).
Its spectrum is a power law of energy index $\alpha_X=1$ ending with a exponential cut
at $E_{\rm c}=$150 keV.
The specific energy density (i.e. as a function of frequency) of 
all these external components are calculated in the comoving frame, and
used to properly calculate the resulting External inverse Compton (EC) spectrum. 
The internally produced synchrotron emission is used to calculate the synchrotron
self Compton (SSC) flux.

\section{Some guidelines for the modelling}

In this section we follow (and somewhat repeat) the arguments presented 
in Paper 1, adding some considerations for line--less BL Lac objects.

A general comment concerns the flux at low (sub--mm to radio) frequencies.
The one--zone homogeneous model here adopted is aimed to explain the 
bulk of the emission, and necessarily requires a compact source,
self--absorbed (for synchrotron) at $\sim 10^{12}$ Hz.
The flux at radio frequencies must be produced further out in the jet.
Radio data, therefore, are not directly constraining the model.
Indirectly though, they can suggest a sort of continuity between the 
level of the radio emission and what the model predicts at higher
frequencies.

\subsection{Strong line objects}

Consider first sources whose inverse Compton flux is dominated by the EC process
with photons of the broad line region.
\begin{itemize}
\item 
When the UVOT data define an optical--UV bump, we interpret it as the 
direct emission from the accretion disc. This assumption allows us to determine 
both the black hole mass and the accretion rate. 
The maximum temperature (and hence the $\nu F_\nu$ peak of the disk luminosity)
occurs at $\sim$5 \sc\ radii and scales as
$T_{\rm max}\propto (L_{\rm d}/L_{\rm Edd})^{1/4}M^{-1/4}$.
The total optical--UV flux gives $L_{\rm d}$ 
[that of course scales as $(L_{\rm d}/L_{\rm Edd})\, M$].
Therefore we can derive both the black hole mass and the accretion rate.
For good UVOT data, the method is sensitive to variations of less than a factor 2
in both the black hole mass and the accretion rate 
(see the discussion and Fig. 2 in Ghisellini et al. 2009a).

\item
The ratio between the high to low energy emission humps ($L_{\rm C}/L_{\rm S}$)
is directly related to the ratio between the radiation to magnetic energy
density  $U^\prime_{\rm r}/U^\prime_{\rm B}$.
In this case the assumption 
$R_{\rm BLR}=10^{17} L_{\rm d,45}^{1/2}$ cm gives
\begin{equation}
{U^\prime_{\rm r} \over U^\prime_{\rm B}} \, =\, {L_{\rm C}\over L_{\rm S}}\, \to
U^\prime_{\rm B} \, =\, {L_{\rm S}\over L_{\rm C}} \, {\Gamma^2 \over 12\pi}\,
\to B=\Gamma\left( {2L_{\rm s}\over 3 L_{\rm C} } \right)^{1/2}
\label{b}
\end{equation}
where we have assumed that $U^\prime_{\rm r}\approx U^\prime_{\rm BLR}$.

\item
The peak of the high energy emission ($\nu_{\rm C}$)
is produced by the scattering of the line photons (mainly hydrogen Lyman--$\alpha$) with 
electrons at the break of the particle distribution ($\gamma_{\rm peak}$).
Its observed frequency is
$\nu_{\rm C}\sim 2\nu_{Ly\alpha} \Gamma\delta \gamma_{\rm peak}^2/(1+z)$.
A steep (energy spectral index $\alpha>1$) spectrum indicates a peak at 
energies below 100 MeV, and this constrains $\Gamma\delta \gamma_{\rm peak}^2$.

\item
Several sources whose Compton flux is dominated by the EC component
may have rather small values of $\gamma_{\rm peak}$.
Electrons with these energies may emit, by synchrotron, in the self--absorbed regime.
In these cases the peak of the synchrotron component is the self--absorption frequency.

\item
In powerful blazars the radiative cooling rate is almost
complete, namely even low energy electrons cool in a 
dynamical timescale $r_{\rm diss}/c$.
We call $\gamma_{\rm cool}$  
the random Lorentz factor of those electrons halving their energies
in a timescale $r_{\rm diss}/c$.
When the EC process dominates, $\gamma_{\rm cool}$ is small (a few).
Therefore the corresponding emitting particle distribution is weakly
dependent of the low energy spectral slope, $s_1$, of the injected
electron distribution.

\item
The strength of the SSC relative to the EC emission depends
on the ratio between the synchrotron over the external radiation energy densities,
as measured in the comoving frame, $U^\prime_{\rm s}/U^\prime_{\rm ext}$.
Within the BLR, $U^\prime_{\rm ext}$ depends only on $\Gamma^2$,
while $U^\prime_{\rm s}$ depends on the injected power, the 
size of the emission, and the magnetic field.
The larger the magnetic field, the larger the SSC component.
The shape of the EC and SSC emission is different:
besides the fact that the seed photon distributions are different,
we have that the flux at a given X--ray frequency
is made by electron of very different energies,
thus belonging to a different part of the electron distribution.
In this respect, the low frequency X--ray data of very hard X--ray spectra
are the most constraining,
since in these cases the (softer) SSC component must not exceed what observed.
This limits the magnetic field, the injected power (as measured in the comoving
frame) and the size.
Conversely, a relatively soft spectrum (but still rising, in $\nu F_\nu$) 
indicates a SSC origin, and this constrains the combination of $B$, $r_{\rm diss}$ 
and $P^\prime_{\rm i}$ even more.

\end{itemize}

\subsection{Line--less BL Lacs}

When the SED is dominated by the SSC process, as discussed by Tavecchio et al. (1998),
the number of observables (peak fluxes and frequencies of both the synchrotron and
SSC components, spectral slopes before and after the peaks, variability timescale) 
is sufficient to fix all the model parameters.
The sources whose high energy emission is completely
dominated by the SSC process are line--less BL Lac objects,
with less powerful and less luminous jets.
The lack of external photons and the weaker radiation losses
make $\gamma_{\rm cool}$ much larger in these sources than in FSRQs.
The low energy slope of the injected electrons, $s_1$, in these cases coincides 
with the low energy slope of the emitting distribution of electrons.

\subsection{Upper limits to the accretion luminosity and mass estimate}

For line--less BL Lac objects we find an upper limit to the 
accretion disk luminosity by requiring that the emission directly produced by the disk and 
by the associated emission lines are completely hidden by the non--thermal continuum.
By assuming a typical equivalent width of the emission lines observed in FSRQs 
($\sim 100$ \AA), and the one defining BL Lacs (EW$< 5$\AA), 
we require that the disk luminosity is a factor at least $\sim 20$ below  
the observed non--thermal luminosity.
Note that we {\it assume} that the disk is a standard Shakura \& Sunyaev (1973)
geometrically thin optically thick disk.
That this is probably {\it not} the case will be discussed later.

A crude estimate of the black hole mass is obtained in the following way.
Assume that the dissipation region, in units of \sc\ radii, is the same
in BL Lacs and FSRQs. 
Then an {\it upper limit} to the mass is derived assuming that light crossing times 
do not exceed the typical variability timescales observed.
A {\it lower limit} can be derived by considering the value
of the magnetic field in the emitting region.
Small black hole masses imply smaller
dimensions, thus smaller Poynting flux (for a given $B$--field).
To avoid implausibly small values of it, we then derive a lower limit to the 
black hole mass.
Together, this two conditions give masses in the range $10^8-10^9\, M_{\odot}$. 
As we discuss later (\S 6.1 and Tab. \ref{masses}), 
for BL Lacs in which independent mass estimates are available in literature, 
the masses derived in this way agree quite well.

\begin{table*} 
\centering
\begin{tabular}{lllllllllllllll}
\hline
\hline
Name   &$z$ &$R_{\rm diss}$ &$M$ &$R_{\rm BLR}$ &$P^\prime_{\rm i}$ &$L_{\rm d}$ &$B$ &$\Gamma$ &$\theta_{\rm v}$
    &$\gamma_{_0}$ &$\gamma_{\rm b}$ &$\gamma_{\rm max}$ &$s_1$  &$s_2$  \\
~[1]      &[2] &[3] &[4] &[5] &[6] &[7] &[8] &[9] &[10] &[11] &[12] &[13] &[14] &[15] \\
\hline   
00311--1938     &0.610 &72  (300)  &8e8  &$<$33 &1.1e--3 &$<$0.11 ($<$9e--4)   &1.7  &15   &3  &1   &1.6e4 &1e6   &0.5 &3   \\
{\it 0048--071} &1.975 &210 (700)  &1e9  &474   &0.025   &22.5 (0.15)          &2.4  &15.3 &3  &1   &400   &7e3   &1   &2.7 \\ 
0116--219       &1.165 &156 (650)  &8e8  &310   &7e-3    &9.6 (0.08)           &2.4  &14.7 &3  &1   &300   &3.5e3 &0.5 &2.5 \\
0118--272       &0.559 &75  (500)  &5e8  &$<$12 &1.2e--3 &$<$0.015 ($<$2.e--4) &0.75 &12.9 &3  &200 &1.5e4 &5e5   &1   &3.2 \\
0133+47         &0.859 &180 (600)  &1e9  &387   &0.013   &15 (0.1)             &11.1 &13   &3  &1   &20    &2.5e3 &1   &1.9 \\
0142--278       &1.148 &90  (500)  &6e8  &268   &0.02    &7.2 (0.08)           &3.91 &12.9 &3  &1   &100   &3.5e3 &1   &2.8 \\
{\it 0202--17}  &1.74  &300 (1000) &1e9  &671   &0.03    &45 (0.3)             &2.4  &15   &3  &1   &300   &5e3   &1   &3.1 \\ 
0208--512       &1.003 &126 (600)  &7e8  &383   &0.03    &14.7 (0.14)          &2.05 &10   &3  &1   &200   &8e3   &0   &2.5 \\
{\it 0215+015}  &1.715 &900 (1500) &2e9  &548   &0.04    &30 (0.1)             &1.1  &13   &3  &1   &2.5e3 &6e3   &--1 &3.5 \\
0218+35         &0.944 &54  (450)  &4e8  &77    &0.014   &0.6 (0.01)           &2.24 &12.2 &3  &1   &150   &3e3   &0   &3.2 \\
0219+428        &0.444 &60  (500)  &4.e8 &$<$35 &9e--3   &$<$0.12 ($<$2e--3)   &2.3  &12.9 &3  &1   &3.1e4 &1.5e5 &1.1 &4.3 \\
{\it 0227--369} &2.115 &420 (700)  &2e9  &547   &0.08    &30 (0.1)             &1.5  &14   &3  &1   &200   &5e3   &0   &3.1 \\
{\it 0235+164}  &0.94  &132 (440)  &1e9  &212   &0.042   &4.5 (0.03)           &2.9  &12.1 &3  &1   &400   &2.7e3 &--1 &2.1 \\
0301--243       &0.260 &144 (600)  &8e8  &$<$11 &4.8e--4 &$<$0.012 ($<$1e--4)  &0.42 &14   &3  &1   &9e3   &6e5   &1   &3.2 \\
0332--403       &1.445 &450 (300)  &5e9  &775   &0.03    &60 (0.08)            &4.25 &10   &3  &1   &90    &3e3   &1   &2.2 \\
{\it 0347--211} &2.944 &750 (500)  &5e9  &866   &0.12    &75 (0.1)             &1.5  &12.9 &3  &1   &500   &3e3   &--1 &3.0 \\
{\it 0426--380} &1.112 &156 (1300) &4e8  &600   &0.018   &36 (0.6)             &1.7  &13   &3  &1   &300   &6e3   &--1 &2.4 \\
0447--439       &0.107 &72 (400)   &6e8  &$<$25 &1.4e--4 &$<$0.063 ($<$7e--4)  &0.9  &15   &3  &1   &2e4   &4e5   &1   &3.5 \\
{\it 0454--234} &1.003 &338 (450)  &2.5e9 &433  &0.027   &18.8 (0.05)          &3    &12.2 &3  &1   &330   &4e3   &--1 &2.4 \\
0502+675        &0.314 &54 (300)   &6e8  &$<$23 &1.5e--3 &$<$0.054 ($<$6e--4)  &4    &15   &3  &1   &5e4   &1e6   &0   &2.8 \\
{\it 0528+134}  &2.04  &420 (1400) &1e9  &866   &0.13    &75 (0.5)             &2.6  &13   &3  &1   &150   &3e3   &--1 &2.8 \\
0537--441       &0.892 &216 (360)  &2e9  &346   &0.06    &12 (0.04)            &3.4  &11   &3.5&1   &90    &3e3   &0.5 &2.2 \\
0650+453        &0.933 &81 (900)   &3e8  &212   &0.018   &4.5 (0.1)            &1    &15   &3  &1   &90    &4.e3  &0   &2.6 \\
0716+332        &0.779 &81 (450)   &6e8  &212   &4.5e--3 &4.5 (0.05)           &4.1  &12.2 &3  &1   &150   &5e3   &0   &2.6 \\
0716+714        &0.26  &84 (700)   &4e8  &$<$42 &1.5e--3 &$<$0.18 ($<$3e--3)   &1.2  &15   &3  &1   &6e3   &6e5   &1.2 &3.2 \\
0735+178        &0.424 &142 (590)  &8e8  &$<$77 &3e--3   &$<$0.6  ($<$5e--3)   &0.66 &10   &3  &1   &1e3   &9.5e3 &1   &2   \\
0814+425        &0.53  &30 (500)   &2e8  &77    &2e-3    &0.6 (0.02)           &3.4  &12.9 &3  &1   &100   &4e3   &1   &2.1 \\
                &0.53  &54 (600)   &3e8  &$<$9.5&0.01    &$<$9e-3 ($<$2e--4)   &0.08 &14.1 &2.5&70  &800   &4e4   &2   &2.2 \\
{\it 0820+560}  &1.417 &261 (580)  &1.5e9 &581  &0.023   &34 (0.15)            &3.1  &13.9 &3  &1   &220   &3e3   &0   &3.4 \\
0851+202        &0.306 &90 (600)   &5e8  &$<$39 &5e--3   &$<$0.15 ($<$3e--3)   &1    &10   &3  &70  &2.3e3 &6e4   &1.2 &3.15\\
{\it 0917+449}  &2.19  &900 (500)  &6e9  &1341  &0.1     &180 (0.2)            &1.95 &12.9 &3  &1   &50    &4e3   &--1 &2.6 \\
0948+0022       &0.5846&72 (1600)  &1.5e8&300   &0.024   &9 (0.4)              &3.4  &10   &6  &1   &800   &1.6e3 &1   &2.2 \\
0954+556        &0.8955&315 (1050) &1e9  &173   &7.7e--3 &3 (0.02)             &0.7  &13   &2.5&1   &6e3   &9e3   &0.3 &2.1 \\
1011+496        &0.212 &36 (400)   &3e8  &$<$12 &7e--4   &$<$0.014 ($<$3e--4)  &3.5  &15   &3  &1   &6e4   &1e5   &0.2 &3.7 \\
{\it 1013+054}  &1.713 &252 (420)  &2e9  &300   &0.036   &9 (0.03)             &1.7  &11.8 &3  &1   &500   &3e3   &1   &2.4 \\
1030+61         &1.401 &405 (450)  &3e9  &424   &0.022   &18 (0.04)            &2.1  &12.2 &3  &1   &200   &7e3   &0   &3   \\
1050.7+4946     &0.140 &45 (500)   &3e8  &$<$6.7&3e--4   &$<$4.5e--3 ($<$1e--4)&0.08 &17   &3  &1   &2e4   &5e6   &0.7 &3.9 \\
1055+018        &0.89  &81 (450)   &6e8  &300   &9e--3   &9 (0.1)              &5.6  &12   &3  &1   &30    &4e3   &1   &2.3 \\
1057--79        &0.569 &99 (550)   &6e8  &300   &5.5e--3 &9 (0.1)              &4.6  &11   &3  &1   &200   &4e3   &--0.5&3.3 \\
10586+5628      &0.143 &60 (400)   &5e8  &$<$12 &1e--3   &$<$0.015 ($<$2e--4)  &0.55 &11.5 &5  &200 &1e3   &2.8e5 &1   &2.6 \\
1101+384        &0.031 &75 (500)   &5e8  &$<$0.9&6e--5   &$<$7.5e--5 ($<$1e--6)&0.25 &19   &1.8&100 &1.3e5 &5e5   &2   &3   \\
\hline
\hline 
\end{tabular}
\vskip 0.4 true cm
\caption{Input parameters used to model the SED.
Sources in italics have been discussed in Paper 1.
Note that $R_{\rm BLR}$ is a derived quantity, not an independent input parameter.
It is listed for an easy comparison with $R_{\rm diss}$.
Col. [1]: name;
Col. [2]: redshift;
Col. [3]: dissipation radius in units of $10^{15}$ cm and (in parenthesis) in units of $R_{\rm S}$;
Col. [4]: black hole mass in solar masses;
Col. [5]: size of the BLR in units of $10^{15}$ cm;
Col. [6]: power injected in the blob calculated in the comoving frame, in units of $10^{45}$ erg s$^{-1}$; 
Col. [7]: accretion disk luminosity in units of $10^{45}$ erg s$^{-1}$ and
        (in parenthesis) in units of $L_{\rm Edd}$;
Col. [8]: magnetic field in Gauss;
Col. [9]: bulk Lorentz factor at $R_{\rm diss}$;
Col. [10]: viewing angle in degrees;
Col. [11], [12] and [13]: minimum, break and maximum random Lorentz factors of the injected electrons;
Col. [14] and [15]: slopes of the injected electron distribution [$Q(\gamma)$] below and above $\gamma_{\rm b}$;
For all cases the X--ray corona luminosity $L_X=0.3 L_{\rm d}$.
Its spectral shape is assumed to be $\propto \nu^{-1} \exp(-h\nu/150~{\rm keV})$.
}
\end{table*}

\setcounter{table}{3}
\begin{table*} 
\centering
\begin{tabular}{lllllllllllllll}
\hline
\hline
Name   &$z$ &$R_{\rm diss}$ &$M$ &$R_{\rm BLR}$ &$P^\prime_{\rm i}$ &$L_{\rm d}$ &$B$ &$\Gamma$ &$\theta_{\rm v}$
    &$\gamma_{_0}$ &$\gamma_{\rm b}$ &$\gamma_{\rm max}$ &$s_1$  &$s_2$  \\
~[1]      &[2] &[3] &[4] &[5] &[6] &[7] &[8] &[9] &[10] &[11] &[12] &[13] &[14] &[15] \\
\hline   
1127--145       &1.184 &405 (450)  &3e9  &1061  &0.036   &112.5 (0.25)         &3.6  &12   &3  &1   &150   &4e3   &0.75&3.3 \\
1144--379       &1.049 &64.5 (430) &5e8  &173   &7e-3    &3 (0.04)             &3.7  &12   &3  &1   &300   &2e3   &1   &2.3 \\
1156+295        &0.729 &114  (380) &1e9  &245   &0.011   &6 (0.04)             &4    &11.3 &3  &1   &70    &5e3   &--1 &2.8 \\
1215+303        &0.13  &45 (500)   &3e8  &$<$12 &4e--4   &$<$0.014 ($<$3e--4)  &0.3  &15   &3  &100 &6e3   &5e5   &1   &3.3 \\
1219+285        &0.102 &75 (500)   &5e8  &$<$12 &2.3e--4 &$<$0.015 ($<$2e--4)  &0.45 &12.9 &3  &100 &2.5e4 &6e5   &1.7 &3.3 \\
1226+023        &0.158 &120 (500)  &8e8  &693   &0.015   &48 (0.4)             &11.6 &12.9 &3  &1   &40    &2e4   &1   &3.4 \\
1244--255       &0.635 &71 (340)   &7e8  &205   &5e--3   &4.2 (0.04)           &4.8  &10   &3  &1   &100   &4e3   &1   &2.3 \\
1253--055       &0.536 &74 (310)   &8e8  &173   &0.012   &3 (0.025)            &4.5  &10.2 &3  &1   &200   &3e3   &0.5 &2.6 \\
1308+32         &0.996 &115 (550)  &7e8  &435   &0.015   &18.9 (0.18)          &4.8  &13   &3  &1   &300   &3e3   &1   &2.5 \\
{\it 1329--049} &2.15  &450 (1000) &1.5e9 &822  &0.07    &67.5 (0.3)           &1.4  &15   &3  &1   &300   &5e3   &1   &3.3 \\
1352--104       &0.332 &23.4 (390) &2e8  &77    &2.5e--3 &0.6 (0.02)           &5.5  &11.4 &3.5 &1  &50    &4e3   &0   &2.7 \\
{\it 1454--354} &1.424 &150 (250)  &2e9  &671   &0.25    &45 (0.15)            &2    &20.  &3  &1   &1e3   &4e3   &--1 &2.0 \\
{\it 1502+106}  &1.839 &450 (500)  &3e9  &764   &0.16    &58.5 (0.13)          &2.8  &12.9 &3  &1   &600   &4e3   &--1 &2.1 \\
1508--055       &1.185 &360 (600)  &2e9  &775   &9e--3   &60 (0.2)             &2.7  &13   &3  &1   &400   &5e3   &1   &2.9 \\
1510--089       &0.360 &126 (600)  &7e8  &205   &6e--3   &4.2 (0.04)           &3.7  &14.1 &3  &1   &150   &4e3   &1   &3   \\
1514--231       &0.048 &105 (700)  &5e8  &$<$4.7 &2e--3  &$<$2e--3 ($<$3e--5)  &0.4  &15   &4.3 &200 &3e3  &2e4   &2.5 &4.6 \\
{\it 1520+319}  &1.487 &1500 (2000) &2.5e9 &237 &0.04    &5.6 (0.015)          &0.06 &15   &3  &1   &2e3   &3e4   &0.8 &2.6 \\
{\it 1551+130}  &1.308 &330 (1100) &1e9  &755   &0.02    &57 (0.38)            &2    &13   &3  &1   &200   &6e3   &--1 &2.4 \\
1553+11         &0.36  &96 (800)   &4e8  &$<$35 &7.5e--3 &$<$0.12 ($<$2e-3)    &6    &15   &3  &1   &4e4   &6e5   &0.3 &3   \\
1622--253       &0.786 &360 (300)  &4e9  &300   &0.01    &9 (0.015)            &1.5  &10   &3  &1   &300   &6e3   &--1 &2.6 \\
{\it 1633+382}  &1.814 &750 (500)  &5e9  &866   &0.07    &75 (0.1)             &1.5  &12.9 &3  &1   &230   &6e3   &0   &2.9 \\
1652+398        &0.0336 &63 (300)  &7e8  &$<$1.4 &1e-4   &$<$2e--4 ($<$2e--6)  &0.11 &16   &3  &200 &2e5   &2e6   &2   &3   \\
1717+177        &0.137 &45 (500)   &3e8  &$<$13 &1.8e--3 &$<$0.018 ($<$4e--4)  &0.1  &12.9 &3  &100 &1e3   &1e6   &1.8 &2.8 \\
1749+096        &0.322 &172 (820)  &7e8  &$<$79 &3.5e--3 &$<$0.63 ($<$6e--3)   &1    &15   &3  &1   &2e3   &1e5   &0.9 &2.8 \\
1803--784       &0.680 &66 (440)   &5e8  &137   &6.5e--3 &1.88 (0.025)         &9.8  &12   &3  &1   &20    &3.5e3 &1   &2.2 \\
1846+322        &0.798 &150 (1000) &5e8  &312   &6.e--3  &9.75 (0.13)          &2.5  &14   &3  &1   &200   &3e3   &1   &2.5 \\
1849+67         &0.657 &72 (400)   &6e8  &212   &5.5e--3 &4.5 (0.05)           &4.6  &13   &3  &1   &250   &3e3   &1   &2.3 \\
1908--201       &1.119 &195 (650)  &1e9  &548   &0.012   &30 (0.2)             &6.9  &14.7 &2.4 &1  &300   &2.7e3 &1   &2.5 \\
1920--211       &0.874 &150 (500)  &1e9  &387   &8e--3   &15 (0.1)             &5.2  &12.9 &3  &1   &100   &5e3   &0   &2.4 \\
1959+650        &0.047 &60 (1000)  &2e8  &$<$7.7 &7e--5  &$<$6e--3 ($<$2e--4)  &1.1  &18   &3  &1   &2e5   &6e5   &1.2 &3   \\
2005--489       &0.071 &83 (550)   &5e8  &$<$12 &7e--5   &$<$0.015 ($<$2e--4)  &0.9  &13.5 &3  &100 &6e4   &1e5   &1.5 &4.3 \\
{\it 2023--077} &1.388 &378 (420)  &3e9  &474   &0.07    &22.5 (0.05)          &1.8  &11.8 &3  &1   &350   &4e3   &0   &2.6 \\ 
{\it 2052--47}  &1.491 &210 (700)  &1e9  &612   &0.045   &37.5 (0.25)          &2.6  &13   &3  &1   &100   &7e3   &--1 &3.0 \\
2141+175        &0.213 &60 (500)   &4e8  &268   &1.3e--3 &7.2 (0.12)           &4.3  &10   &4  &1   &200   &1.5e4 &0   &3.2 \\
2144+092        &1.113 &195 (650)  &1e9  &387   &0.02    &15 (0.1)             &2.4  &14.7 &3  &1   &200   &5e3   &0.5 &3.2 \\
2155--304       &0.116 &72 (300)   &8e8  &$<$29 &1e--3   &$<$0.084 ($<$7e--4)  &3.5  &16   &3  &1   &3e4   &2e5   &0.5 &3.9 \\
                &0.116 &120 (500)  &8e8  &$<$29 &3e--4   &$<$0.084 ($<$7e--4)  &0.7  &16   &3  &1   &1.8e4 &5e5   &1   &3.3 \\
2155+31         &1.486 &96 (800)   &4e8  &110   &0.03    &1.2 (0.02)           &1.2  &16   &3  &1   &140   &5e3   &0.5 &3.2 \\
2200+420        &0.069 &75 (500)   &5e8  &$<$17 &2.5e--3 &$<$0.03 ($<$4e--4)   &1    &10   &3  &80  &150   &1e5   &1   &3.1 \\
2201+171        &1.076 &180 (300)  &2e9  &346   &0.016   &12 (0.04)            &5.9  &10   &3  &1   &300   &3e3   &1.2 &2.2 \\
2204--54        &1.215 &195 (650)  &1e9  &520   &0.023   &27 (0.18)            &3.1  &14   &3  &1   &150   &3e3   &0.5 &3.1 \\
{\it 2227--088} &1.5595 &211 (470) &1.5e9 &497  &0.06    &24.8 (0.11)          &3.3  &12   &3  &1   &200   &5e3   &0.5 &3.2 \\
2230+114        &1.037 &195 (650)  &1e9  &670   &0.025   &45 (0.3)             &4.1  &14   &3  &1   &110   &3e3   &0.5 &3.1 \\
{\it 2251+158}  &0.859 &240 (800)  &1e9  &548   &0.14    &30 (0.2)             &4.1  &13   &3  &1   &250   &4e3   &1   &2.7 \\
{\it 2325+093}  &1.843 &420 (1400) &1e9  &671   &0.08    &45 (0.3)             &1.6  &16   &3  &1   &190   &5e3   &0   &3.5 \\
2345--155       &0.621 &132 (1100) &4e8  &190   &3.5e--3 &3.6 (0.06)           &1.8  &13   &3  &1   &100   &4e3   &--1 &2.8 \\
\hline
$\langle$FSRQ$\rangle$   &1   &189 (630)  &1e9 &387     &0.02  &15 (0.1)            &2.6 &13 &3 &1  &300   &3e3   &1   &2.7 \\
$\langle$BL Lac$\rangle$ &0.1 &75.6 (630) &4e8 &$<$24.5 &8e--4 &$<$0.06 ($<$1e--3)  &0.8 &15 &3 &1  &1.5e4 &8.e5  &1   &3.3 \\
\hline
\hline 
\end{tabular}
\vskip 0.4 true cm
\caption{-- continue --}
\label{para}
\end{table*}

\begin{figure}
\vskip -0.4cm
\hskip -1cm
\psfig{figure=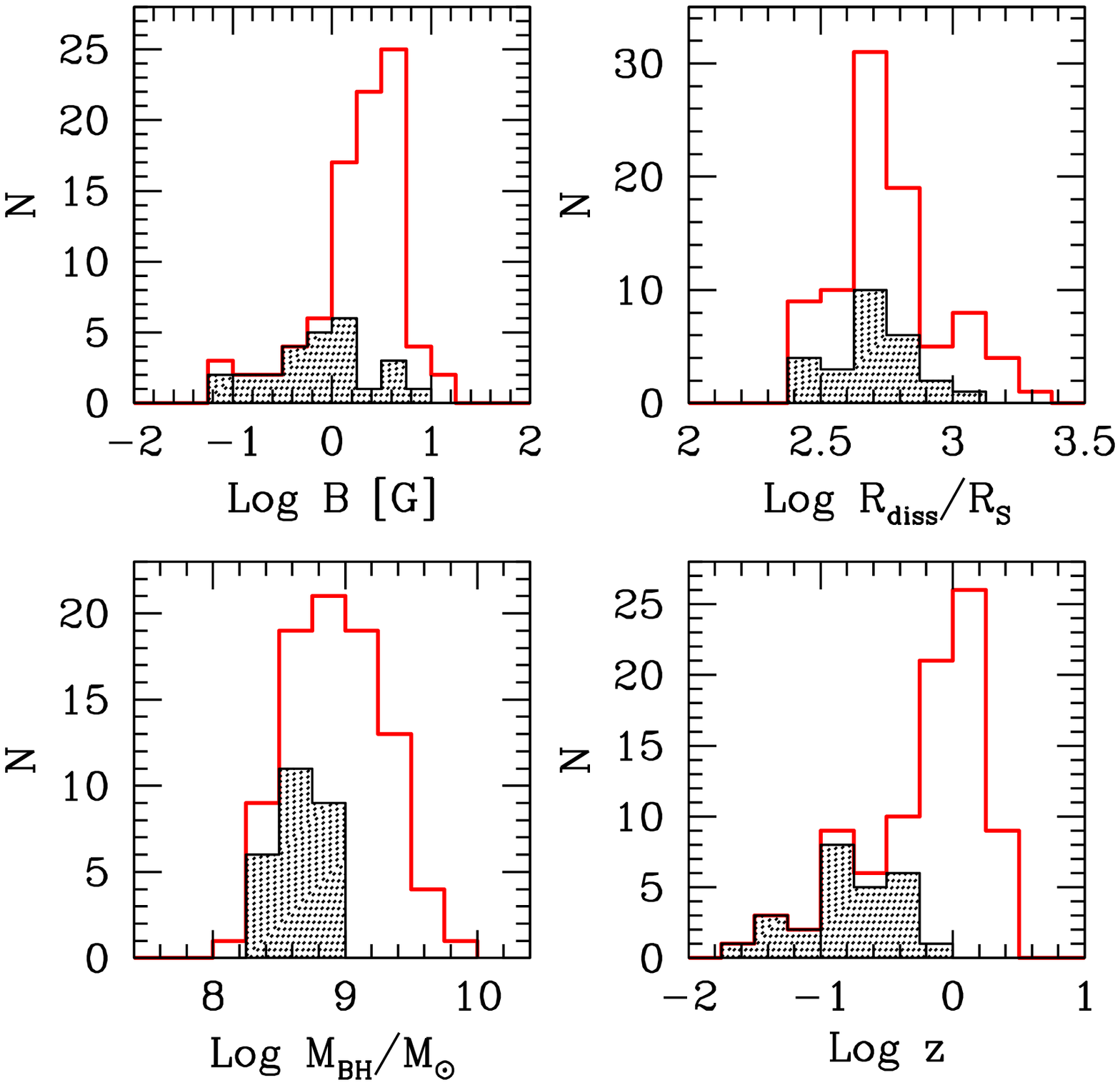,width=10cm,height=10cm}
\vskip -1.5 cm
\hskip -1cm
\psfig{figure=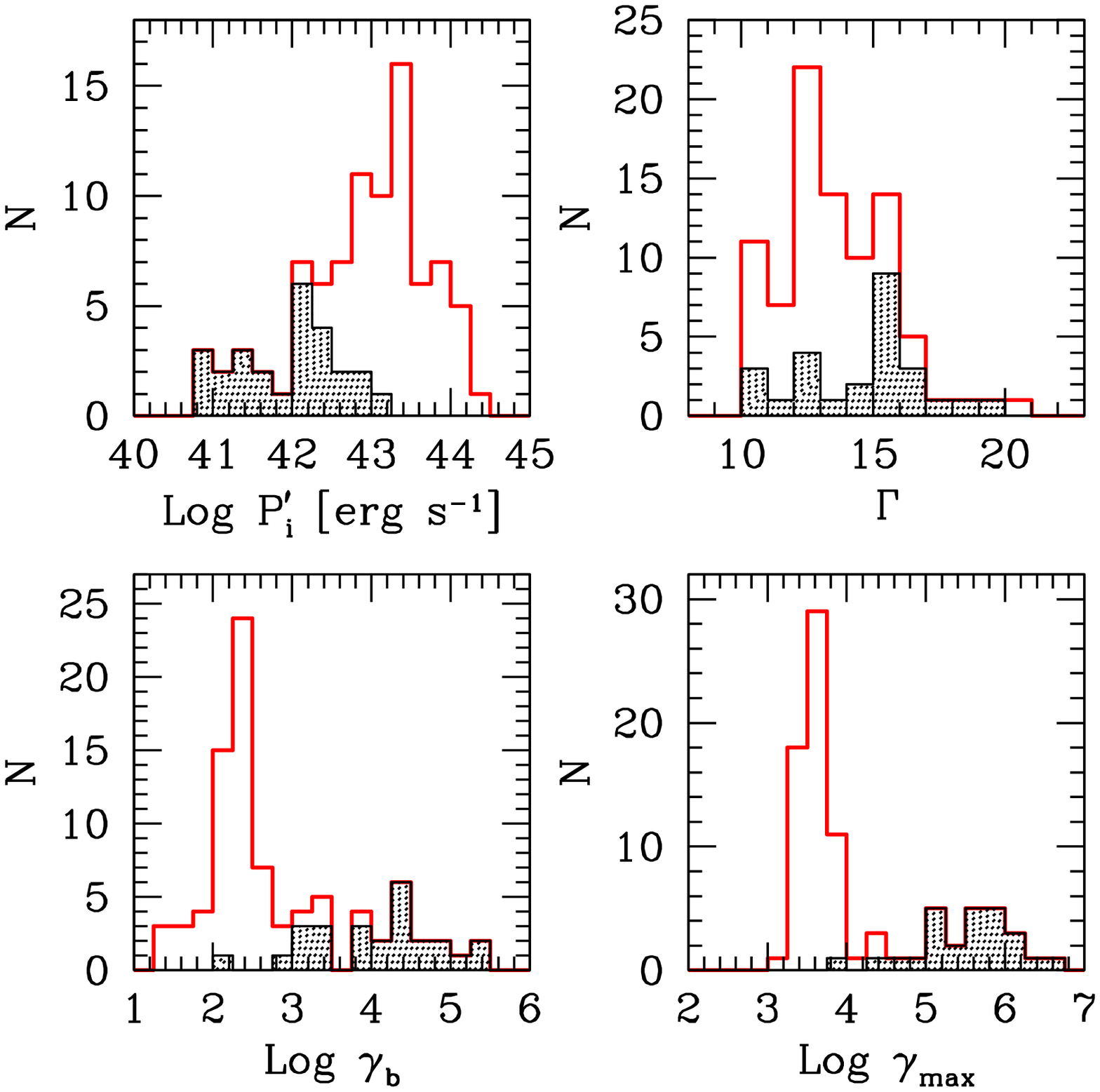,width=10cm,height=10cm}
\vskip -1  cm
\caption{Distribution of the magnetic field,
location of the dissipation region in units of the \sc\ radius,
black hole mass, redshift, power $P^\prime_{\rm i}$
injected in relativistic electrons (as measured in the comoving frame),
bulk Lorentz factor, break energy ($\gamma_{\rm b}$)  
and maximum energy ($\gamma_{\rm max}$) 
of the injected distribution of electrons for all the 85 blazars.
Shaded areas correspond to BL Lacs with only an upper limit
on their disk accretion luminosity.
}
\label{isto1}  
\end{figure}

\begin{figure}
\vskip -0.4cm
\hskip -1cm
\psfig{figure=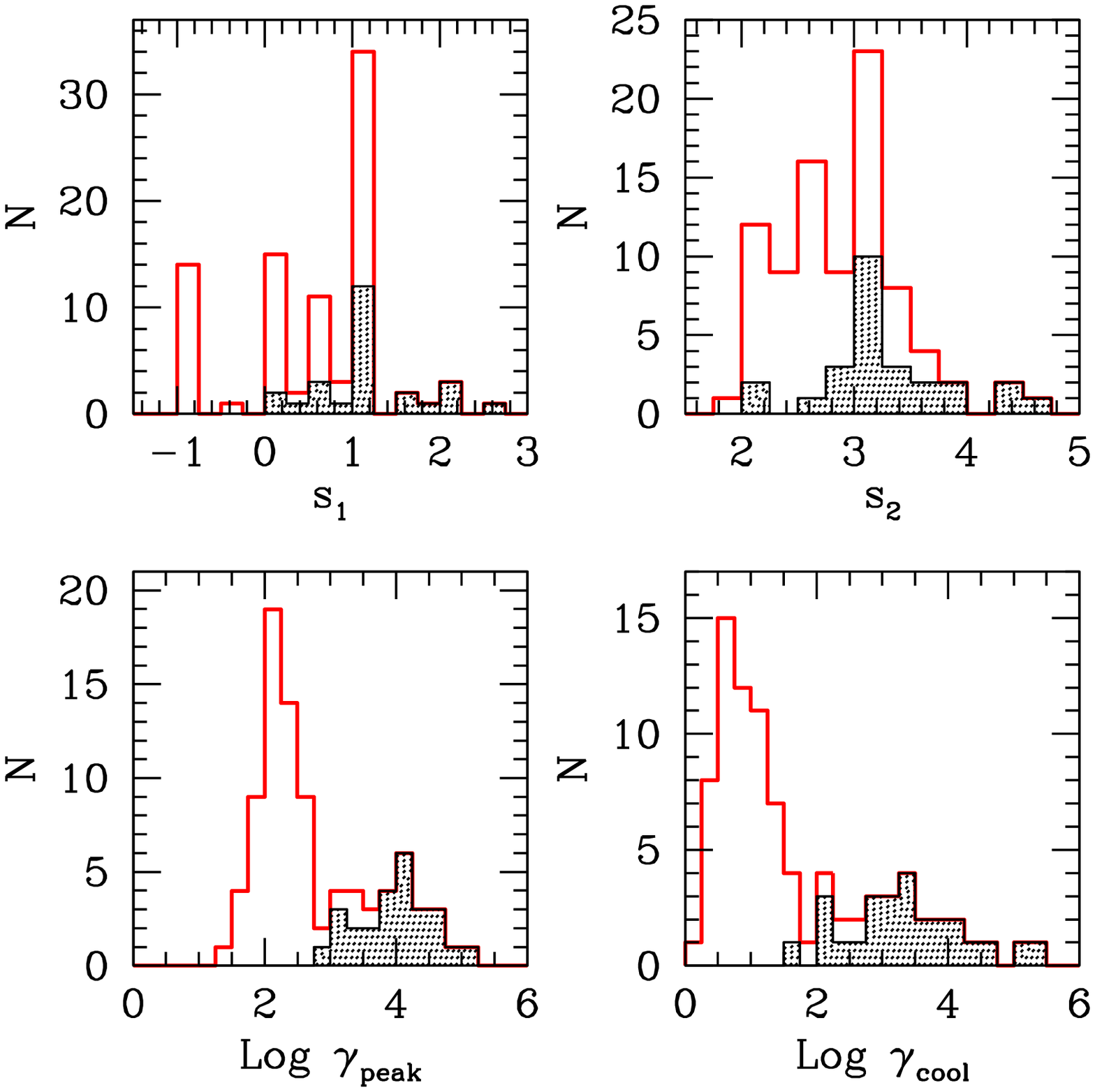,width=10cm,height=10cm}
\vskip -1.5 cm
\hskip -1cm
\psfig{figure=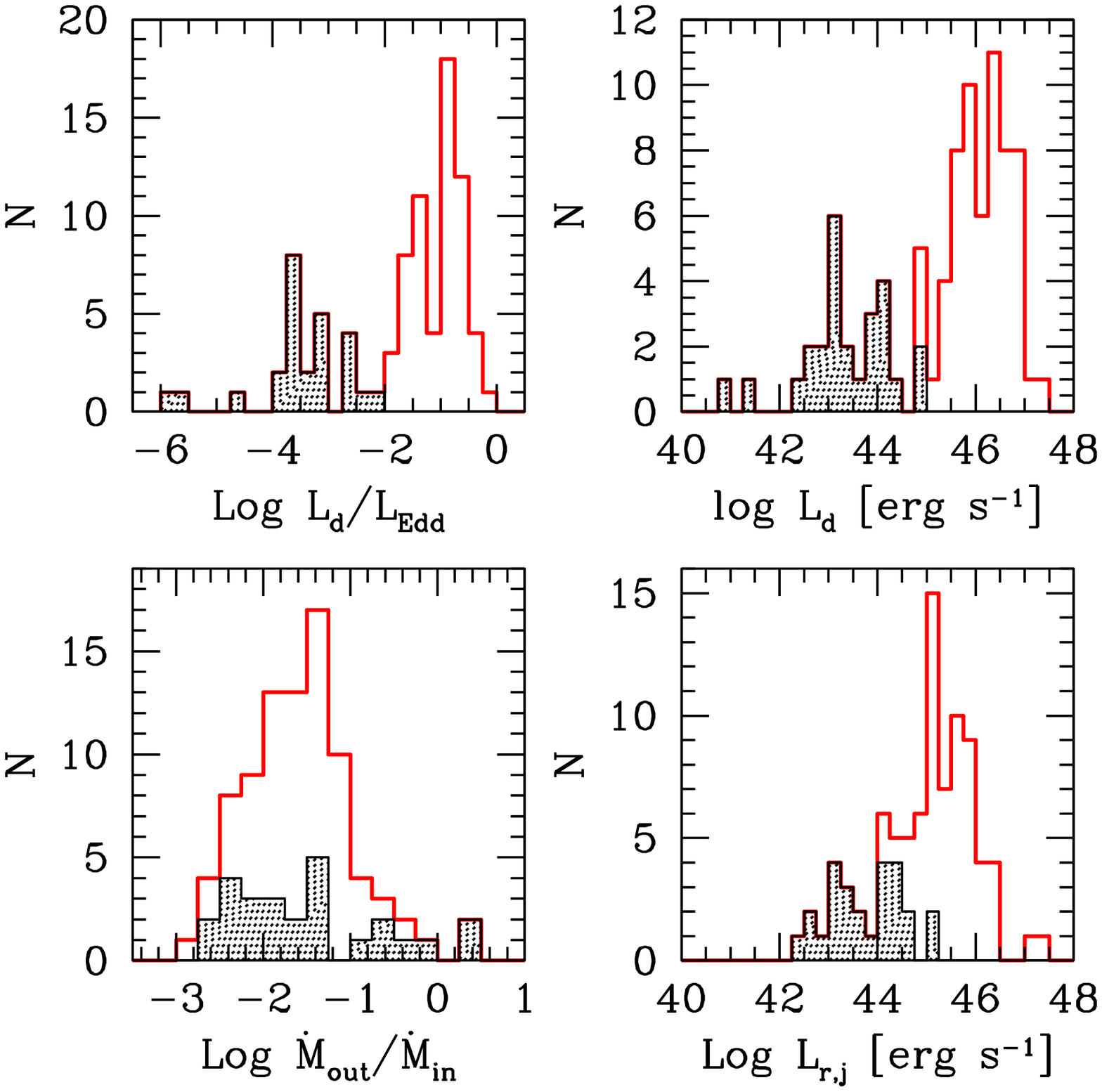,width=10cm,height=10cm}
\vskip -1.  cm
\caption{
Distributions of the slopes of the injected
electron distribution ($s_1$ and $s_2$ are the slopes
before and after $\gamma_{\rm b}$, respectively),
the value of the random Lorentz factor $\gamma_{\rm peak}$
of the electrons radiating at the peaks of the SED,
the value of the random Lorentz factor $\gamma_{\rm cool}$
of electrons cooling in one light crossing time $r_{\rm diss}/c$,
accretion disk luminosities 
(also in units of the Eddington one), the ratio between the
outflowing and accretion mass rate and the power spent
by the jet to produce the radiation we see.
Shaded areas correspond to BL Lacs with only an upper limit
on their disk accretion luminosity.
}
\label{isto2}  
\end{figure}

\begin{table} 
\centering
\begin{tabular}{lllll}
\hline
\hline
Name   &$\log P_{\rm r}$ &$\log P_{\rm B}$ &$\log P_{\rm e}$ &$\log P_{\rm p}$ \\
\hline   
00311--1938     &44.37 &44.10 &43.46 &43.93 \\
{\it 0048--071} &45.75 &45.35 &44.69 &47.10 \\
0116--219       &45.16 &45.06 &44.09 &46.26 \\
0118--272       &44.20 &43.30 &43.77 &43.91 \\
0133+47         &45.22 &46.41 &44.43 &47.01 \\
0142--278       &45.39 &44.89 &44.98 &47.26 \\
{\it 0202--17}  &45.80 &45.65 &44.80 &47.31 \\
0208--512       &45.44 &44.39 &44.63 &46.44 \\
{\it 0215+015}  &45.81 &45.78 &44.86 &46.09    \\
0218+35         &45.21 &43.96 &44.76 &46.51 \\ 
0219+428        &45.16 &44.08 &44.12 &45.44 \\
{\it 0227--369} &46.18 &45.49 &44.97 &47.34   \\
{\it 0235+164}  &45.78 &44.97 &44.61 &46.60   \\
0301--243       &43.66 &43.42 &43.56 &43.93  \\
0332--403       &45.41 &46.21 &44.51 &47.05 \\ 
{\it 0347--211} &46.30 &45.91 &44.55 &46.92    \\
{\it 0426--380} &45.48 &44.67 &44.30 &46.36   \\
0447--439       &45.41 &46.21 &44.51 &47.05 \\
{\it 0454--234} &45.60 &45.80 &44.16 &46.40   \\
0502+675        &44.53 &44.59 &42.75 &43.29 \\
{\it 0528+134}  &47.39 &45.86 &45.87 &48.31   \\
0537--441       &45.80 &45.39 &44.97 &47.19 \\   
0650+453        &45.55 &43.73 &44.90 &46.88 \\
0716+332        &44.77 &44.84 &44.05 &45.92  \\
0716+714        &44.36 &43.93 &44.06 &45.21  \\
0735+178        &44.19 &43.51 &44.12 &45.12  \\
0814+425 (EC)   &44.40 &43.81 &43.97 &45.94  \\
0814+425 (SSC)  &44.23 &41.35 &44.91 &45.64  \\
{\it 0820+560}  &45.60 &45.75 &44.57 &46.85  \\
0851+202        &44.45 &43.48 &44.28 &44.96  \\ 
{\it 0917+449}  &46.20 &46.29 &45.00 &47.57   \\
0948+0022       &45.30 &44.35 &44.71 &46.68 \\
0954+556        &45.04 &44.48 &44.52 &45.07  \\
1011+496        &44.19 &44.12 &42.80 &43.21 \\ 
{\it 1013+054}  &45.68 &45.02 &44.66 &46.99   \\
1030+61         &45.50 &45.65 &44.33 &46.62  \\
1050.7+4946     &42.53 &41.14 &43.35 &42.81  \\
1055+018        &44.93 &45.04 &44.57 &46.90  \\
1057--79        &44.75 &44.97 &44.06 &45.85  \\
10586+5628      &43.64 &42.76 &43.72 &44.02  \\   
1101+384        &42.60 &42.67 &43.03 &43.50 \\ 
\hline
\hline 
\end{tabular}
\vskip 0.4 true cm
\caption{
Jet power in the form of radiation, Poynting flux,
bulk motion of electrons and protons (assuming one proton
per emitting electron).
Sources in italics have been analysed in Paper I
and their jet powers are here reported for
completeness.
}
\label{powers}
\end{table}

\setcounter{table}{4}
\begin{table} 
\centering
\begin{tabular}{lllll}
\hline
\hline
Name   &$\log P_{\rm r}$ &$\log P_{\rm B}$ &$\log P_{\rm e}$ &$\log P_{\rm p}$ \\
\hline   
1127--145       &45.66 &46.06 &44.80 &47.28  \\
1144--379       &44.92 &44.49 &44.34 &46.41   \\
1156+295        &45.04 &45.12 &44.45 &46.38   \\
1215+303        &43.20 &43.46 &43.46 &42.18  \\
1219+285        &43.15 &42.92 &43.24 &43.61   \\
1226+023        &45.05 &46.09 &44.90 &47.48  \\
1244--255       &44.56 &44.64 &44.12 &46.24   \\
1253--055       &45.00 &44.71 &44.40 &46.29   \\
1308+32         &45.34 &45.28 &44.57 &46.83  \\
{\it 1329--049} &46.18 &45.53 &45.07 &47.65   \\
1352--104       &44.18 &43.94 &44.02 &45.71  \\
{\it 1454--354} &47.01 &45.13 &45.20 &47.47   \\
{\it 1502+106}  &46.43 &46.12 &44.63 &46.91   \\
1508--055       &45.16 &45.78 &44.07 &46.54  \\
1510--089       &44.99 &45.26 &44.41 &46.77 \\ 
1514--241       &43.08 &43.17 &43.98 &44.59  \\ 
{\it 1520+319}  &45.91 &43.84 &45.22 &46.50   \\
{\it 1551+130}  &45.53 &45.45 &44.21 &46.48   \\
1553+11         &45.23 &45.44 &43.16 &44.29  \\  
1622--253       &44.95 &45.04 &44.17 &45.65 \\
{\it 1633+382}  &46.06 &45.91 &44.60 &47.05    \\
1652+398        &42.45 &41.66 &43.05 &43.20  \\ 
1717+177        &43.14 &41.23 &44.09 &44.76  \\
1749+096        &44.73 &44.40 &44.43 &45.37  \\
1803+784        &44.73 &45.35 &44.41 &46.80  \\
1846+322        &45.02 &44.99 &44.25 &46.59 \\
1849+67         &44.90 &44.83 &44.23 &46.38 \\
1908--201       &45.36 &46.18 &44.37 &46.82  \\
1920--211       &45.07 &45.58 &44.19 &46.30 \\
1959+650        &43.32 &43.72 &42.51 &43.17  \\ 
2005--489       &42.96 &43.66 &42.65 &43.01  \\ 
{\it 2023--077} &45.98 &45.41 &44.65 &46.87    \\
{\it 2052--47}  &45.84 &45.27 &45.01 &47.24   \\
2141+175        &44.00 &44.39 &43.51 &45.18  \\
2144+092        &45.60 &45.25 &44.69 &46.98 \\ 
2155--304 (high)&44.40 &44.78 &43.06 &43.85  \\
2155--304 (low) &43.76 &43.83 &43.40 &43.89  \\
2155+31         &45.19 &43.89 &44.98 &46.36 \\
2200+420        &43.35 &43.32 &43.94 &44.90 \\
2201+171        &45.12 &45.62 &44.36 &46.74 \\
2204--54        &45.60 &45.44 &44.80 &47.09 \\
{\it 2227--088} &45.89 &45.40 &45.07 &47.29    \\
2230+114        &45.62 &45.66 &44.89 &47.23 \\
{\it 2251+158}  &46.33 &45.78 &45.43 &47.87    \\
{\it 2325+093}  &46.30 &45.65 &45.06 &47.51  \\
2345--1555      &44.72 &44.56 &43.97 &45.92  \\  
\hline
\hline 
\end{tabular}
\vskip 0.4 true cm
\caption{-- continue --
}
\end{table}

\section{Results}

In Fig. \ref{isto1} and Fig. \ref{isto2} we show the distributions 
of the parameters derived from model fitting the SED of the entire 
sample of {\it Fermi} blazars (85 sources).
Shaded areas in these histograms correspond to the 25 BL Lacs for which
only an upper limit to the accretion luminosity could be derived
(the plotted areas correspond to 26 objects because
they include two states of PKS 2155--304).
They can be considered as ``genuine" BL Lacs, namely blazars
whose emission lines are intrinsically weak or absent, and not
hidden by the beamed continuum.

The SED of the 37 blazar studied in this paper, together with the 
best fitting model, are shown in Figs. \ref{f1}--\ref{f10} of the Appendix.
In Tab. \ref{xrt} and Tab. \ref{uvot} we list the result of our
XRT and UVOT analysis for the 33 blazars (out of the 37 blazars
studied in this paper) with {\it Swift} data.
All optical--UV fluxes shown in Fig. \ref{f1} -- \ref{f10} have been
de--reddened according to the Galactic $A_{\rm V}$ given in the
NED database. 

Tab. \ref{para} reports the parameters used to compute the 
theoretical SEDs and Tab. \ref{powers} lists the power 
carried by the jet in the form of radiation, electrons, magnetic field
and protons (assuming one proton per emitting electron).
For the ease of the reader, in these tables we report also the values
found and presented in Paper 1.

\subsection{Redshifts and black hole masses}

The redshift distribution of {\it Fermi} FSRQs extends to larger
values than for BL Lacs (Fig. \ref{isto1}). 
This has been already pointed out in A09 and is the consequence of BL Lacs
being less $\gamma$--ray luminous than FSRQs.

The derived black hole masses are in the range (1--60)$\times 10^8M_\odot$
(Fig. \ref{isto1}).
The object for which we could estimate the least massive black hole is
PMN 0948+0022, that is a NLSy1 discussed in Abdo et al. (2009b),
where a mass of 1.5$\times 10^8 M_\odot$ was found.

We have searched in the literature other estimates 
of the black hole masses of our blazars, and report them
in Tab. \ref{masses}.
They have been mainly derived from the FWHM of the emission lines,
through the assumption of virial velocity of the broad line clouds.
There is a rough agreement between our and the other estimates,
but note that for specific objects the reported estimates 
vary by a factor 3--10.
It is comforting that the black hole masses assumed here for line--less  
BL Lacs are consistent 
with the existing estimates present in the literature.


\subsection{Injected power, location of the dissipation region and bulk Lorentz factors}

The injected power in relativistic electrons, as measured in 
the comoving frame, is 
in the range $P^\prime_{\rm i}=10^{43}$--$10^{44}$ erg s$^{-1}$ for FSRQs and a 
$10^{41}$--$10^{43}$ erg s$^{-1}$ for BL Lacs.
Note that we should not compare this power with
the power the jet carries in the form of 
bulk motion of particles and fields, since $P^\prime_{\rm i}$
is measured in the comoving frame.
To have comparable quantities, we should multiply 
$P^\prime_{\rm i}$ by $\Gamma^2$.

Among FSRQs, 4 sources have $R_{\rm diss}>R_{\rm BLR}$
(0215+015 and 1520+319, discussed in Paper 1, plus 0954+556 and 1622--253).
This means a reduced radiation energy density that in turn implies 
a weaker radiative cooling and a larger $\gamma_{\rm peak}$.
All other FSRQs dissipate within the BLR, at a distance
of the order of 300--1000 $R_{\rm S}$ (Fig. \ref{isto1}).
For BL Lacs we have the same range of $R_{\rm diss}$.

The distribution of the bulk Lorentz factor is rather narrow,
being contained within the 10--15 range, with few BL Lacs
having $\Gamma$ between 15 and 20.
The average $\Gamma$ for BL Lacs is somewhat larger than for FSRQs.

\subsection{Magnetic field} 

On average, we need a slightly larger magnetic field $B$
in FSRQs (1--10 G) than in BL Lacs (0.1--1 G).
Note that in our list we lack TeV BL Lacs not detected
by {\it Fermi} (for them, see T09), that have more extreme
properties (and smaller magnetic fields) than the sources
considered here.
Note also that there is one FSRQ with a small magnetic field:
this is 1520+319, whose $R_{\rm diss}$ is at very large
distances.

\subsection{Particle distribution}

The average properties of the injected particle distribution
can be seen in Fig. \ref{isto1} and Fig. \ref{isto2}.
Note that the {\it injected} particle distribution $Q(\gamma$)
[cm$^{-3}$ s$^{-1}$]
is different from the {\it emitting} particle distribution
$N(\gamma)$ [cm$^{-3}$], that 
is the solution of the continuity equation evaluated
at the light crossing time $r_{\rm diss}/c$.
It can be seen that the diversity of the blazar spectra 
requires a rather broad range of $s_2$, the injected slope for the high
energy electrons (between 2 and 4.5), steeper for BL Lac objects.
Consider also that when radiative cooling is important (almost always
in FSRQs, see the distribution of $\gamma_{\rm cool}$ 
in Fig. \ref{isto2}), the high energy emitting particle distribution
will be characterised by a power law slope $s_2+1$, i.e. steeper still.
The fact that BL Lacs are characterised by a flatter $\gamma$--ray spectrum
in the {\it Fermi} band (A09, Ghisellini, Maraschi \& Tavecchio 2009)
{\it is not due} to a flatter slope of the $N(\gamma)$ distribution above
$\gamma_{\rm b}$, but to
their SED peaking at energies close to or higher than a few GeV.

Also the distribution of $s_1$ is broad, 
with FSRQs having harder slopes.
But in this case the radiative cooling is stronger, 
and $N(\gamma)\propto \gamma^{-2}$ from $\gamma_{\rm cool}$
to $\gamma_{\rm b}$, making $s_1$ less important.
For BL Lacs, instead, the cooling is much 
weaker, and in several cases $N(\gamma)\propto \gamma^{-s{_1}}$ 
at intermediate energies, mostly contributing to the X--ray band.
This requires a softer $s_1$.
There are however extreme cases (TeV emitting BL Lacs, 
with an hard spectrum) present in T09, but not here (they
are not detected by {\it Fermi}) that require a very hard
$s_1$ or even a cutoff in the electron distribution (i.e.
no electrons injected below some critical value).

The distributions of $\gamma_{\rm b}$, $\gamma_{\rm max}$ 
(Fig. \ref{isto1}) are clearly different for FSRQs and BL Lacs,
with BL Lacs requiring much larger values.
This, together with the weaker cooling for BL Lacs
(implying a larger $\gamma_{\rm cool}$)
determines the distribution of $\gamma_{\rm peak}$, the
value of Lorentz factors of those electrons producing most 
of the radiation we see.
This confirms earlier results concerning the interpretation
of the blazar sequence (i.e. Ghisellini et al. 1998; 
Celotti \& Ghisellini 2008).

The relation of $\gamma_{\rm peak}$ with the energy densities
and with $\gamma_{\rm cool}$ is shown in Fig. \ref{gpeak}.
The top panel shows $\gamma_{\rm peak}$ as a function of the 
sum of the magnetic and radiation energy density as measured
in the comoving frame. The grey symbols are the values
for the blazars studied in Celotti \& Ghisellini (2008).

\subsection{Disk luminosities}

The accretion disk luminosities of FSRQs (Fig \ref{isto2})
derived by the model fits
are in the range $10^{45}$--$10^{47}$ erg s$^{-1}$,
and the upper limits for BL Lacs indicate, always,
values below $10^{45}$ erg s$^{-1}$.
In Eddington units, FSRQs always have values above $10^{-2}$,
and BL Lacs always values below.
Bearing in mind that for some FSRQs our estimates are poor
(when the beamed non--thermal flux hide the thermal emission
or when {\it Swift} data are missing) this result
is intriguing.
It is in perfect agreement with the ``blazar's divide"
between broad line and line--less objects proposed
by Ghisellini, Maraschi \& Tavecchio (2009).
It also agrees with the scenario proposed by Cavaliere 
\& D'Elia (2002) and B\"ottcher \& Dermer (2002).
We will further discuss this point later.

\subsection{Accretion and outflow mass rates}

Fig. \ref{isto2} shows the ratio of the accretion 
($\dot M_{\rm in}$) and outflow ($\dot M_{\rm out}$)
mass rates.
For FSRQs we set $\dot M_{\rm out}\equiv P_{\rm j}/(\Gamma c^2)$:
therefore $\dot M_{\rm out}$ suffers from the same uncertainties 
of $P_{\rm j}$, derived assuming one proton per emitting electron
and that {\it all} electrons emit.
For completeness we show this distribution also for BL Lacs,
but the values are in this case completely uncertain, for two reasons.
First, since only an upper limit on $L_{\rm d}$ is derived, 
we have a corresponding upper limit for $\dot M_{\rm in}$.
On the other hand, the disks in BL Lacs may well be 
radiatively inefficient: if so, they will have larger accretion rates
than the ones corresponding to a standard disk.
We will estimate $\dot M_{\rm in}$ for BL Lacs later, 
using the assumption of $P_{\rm j}=\dot M_{\rm in} c^2$.

For FSRQs, instead, the distribution is more meaningful
(bearing in mind the limitations mentioned above)
and indicates that the mass outflowing rate, on average,
is 1--10\% of the mass accretion rate.
The distribution is rather narrow, and this may indicate that the
mass outflow rate of the jet (derived assuming one proton per
emitting electron) is linked with the mass accretion rate.
In other words, the matter of the jet may come directly
from the accreting one, with other process, like entrainment,
less important.

\begin{table}
\centering
\begin{tabular}{llllllll}
\hline
\hline
Name      &      &              &Ref   &    &Ref   &  &Ref  \\
\hline 
0502+675  &8.78  &8.80  &F03b                              \\
0851+202  &8.7   &8.79  &W04                               \\
1011+496  &8.48  &8.71  &F03b &8.28  &W08    &8.32  &W02   \\  
          &      &8.25  &W02  &8.25  &W02  \\
1101+384  &8.7   &8.29  &W08  &8.42  &W08    &8.97  &W08   \\
          &      &9.13  &W02  &8.88  &W02  \\
1215+303  &8.48  &8.83  &F03b &7.53  &W02    &7.62  &W02   \\ 
1514--241 &8.7   &8.74  &F03a &8.09  &B03    &8.74  &F03b    \\
          &      &8.40  &Wo05 &9.10  &W02    &8.86  &W02    \\
1652+398  &8.84  &9.21  &W08  &8.78  &W08    &8.62  &Wo05  \\
          &      &8.94  &F03  &9.21  &B03        \\
1749+096  &8.84  &8.66  &F03b &7.21  &W02    &7.36  &W02   \\ 
1959+650  &8.3   &8.08  &W08  &8.28  &W08    &7.96  &Wo05 \\
          &      &8.56  &F03a &8.30  &W02    &8.22  &W02  \\
          &      &8.53  &F03b  \\
2005--489 &8.7   &8.14  &W08  &8.89  &F03b  &8.66   &W02 \\
          &      &8.51  &W02  \\
2200+420  &8.7   &8.08  &W08  &8.28  &W08    &8.35  &W04 \\ 
          &      &8.77  &F03b &8.56  &W02    &8.43  &W02  \\
\hline

0537--441 &9.3   &8.74     &W04  &7.7   &P05  &8.33  &L06  \\  
0820+560  &9.18  &9.49     &C09                             \\
0917+449  &9.78  &9.88     &C09                             \\
0948+002 &8.17  &8.26      &C09                             \\
0954+566  &9     &9.20     &C09  &7.7   &P05  &7.87  &L06  \\
1013+054  &9.3   &9.78     &C09                             \\
1030+61   &9.48  &9.39     &C09                             \\
1055+018  &8.78  &9.25     &C09                             \\ 
1144--379 &8.7   &7.6      &P05                             \\
1156+295  &9     &9.11     &C09  &8.56  &X05  &8.54  &X05  \\
          &      &8.63     &P05  &8.54  &L06                \\
1226+023  &8.9   &9.10     &F03b &9.21  &W04  &8.6   &P05 \\
          &      &8.92     &L06                                \\
1253--055 &8.9   &8.48     &W04  &7.9   &P05  &8.28  &L06  \\
1308+32   &8.84  &8.94     &C09  &9.24  &W04    \\
1502+106  &9.48  &9.50     &C09  &8.74  &L06     \\
1510--089 &8.84  &8.62     &W04  &8.31  &X05             \\
          &      &8.22     &X05  &8.20  &L06   \\
1551+130  &9     &9.19     &C09                 \\
1633+382  &9.7   &9.74     &C09  &8.67  &L06 \\
1803+784  &8.7   &8.57     &W04  &7.92  &L06 \\
2141+175  &8.6   &8.98     &F03b &8.14  &X05  &8.05     &X05 \\
          &      &7.95     &L06 \\
2227--088 &9.18  &9.40     &C09 \\
2230+114  &9     &8.5      &P05 \\
2251+158  &9     &9.10     &W04  &8.5  &P05   &8.83     &L06  \\ 
\hline
\hline 
\end{tabular}
\vskip 0.4 true cm
\caption{
Estimates of the mass of the black hole  
for the blazars in our sample.
Values are given for the logarithm of the black hole mass
measured in solar masses.
We list here only those blazars for which we have found 
another, independent, mass estimate.  
The first value (Col. 2) is the estimate found in this paper.
In the top part of the table we list BL Lacs for which
we found only an upper limit on the disk luminosity.
In this case our estimates of the black hole mass is very uncertain.
References: 
B03: Barth et al. (2003);
C09: cheng et al. (2009);
F03a: Falomo et al. (2003a);
F03b: Falomo et al. (2003a);
L06: Liu et al. (2006);
P05: Pian et al. (2005);
W02: Wu et al. (2002);
W04: Wang et al. (2004);
W08: Wagner (2008);
Wo05: Woo et al. (2005).
X05: Wie et al. (2005);
}
\label{masses}
\end{table}

\subsection{Emission mechanisms}

For all but four FSRQs (0215+015, 0954+556, 1520+319 and 1622--253)
the dissipation region is within the BLR, that
provides most of the seed photons scattered at high frequencies.
The main emission processes are then synchrotron and thermal emission
from the accretion disk for the low frequency parts, and External Compton 
for the hard X--rays and the $\gamma$--ray part of the spectrum.
The X--ray corona and the SSC flux marginally contribute to the soft X--ray part of
the SED.
When $R_{\rm diss}<R_{\rm BLR}$ the overall non--thermal
emission is rather insensitive to the presence/absence of the IR torus,
since the bulk of the seed photons are provided by the broad lines.
Instead, for the 4 FSRQs with $R_{\rm diss}>R_{\rm BLR}$, the external
Compton process with the IR radiation of the torus is crucial
to explain their SED.

For BL Lacs (see the SEDs in T09) the main mechanism is SSC,
but in few cases this process is unable to account for a very
large separation between the synchrotron and the inverse Compton
peaks of the SED, without invoking extremely large bulk Lorentz
factors (larger than 100).
These are the cases where the SED is better modelled invoking an
extra source of seed photons for the IC process, besides the ones
produced by synchrotron in the same zone.
One possibility is offered by the spine/layer scenario
(Ghisellini, Tavecchio \& Chiaberge 2005), in which a slower
layer surrounds the fast spine of the jet.
The radiative interplay between the two structures enhances the IC 
flux and can account for the observed SED in these cases.


Another problem, with the standard one--zone SSC scenario, concerns
the ultrafast (i.e. minutes) variability sometimes seen at
high energies (Aharonian et al. 2007, Albert et al. 2007). 
This cannot be accounted for by the simple models, and require
other emitting zones or extra population of electrons
(see e.g. Ghisellini \& Tavecchio 2008; Ghisellini et al. 2009b; 
Giannios, Uzdensky \& Begelman 2009).

For all our sources the importance of the $\gamma$--$\gamma\to e^\pm$ process
(which is included in our model) is very modest, and does not 
influence the observed spectrum, nor the derived jet power,
discussed below.

\begin{figure}
\vskip -0.5cm
\hskip -1cm
\psfig{figure=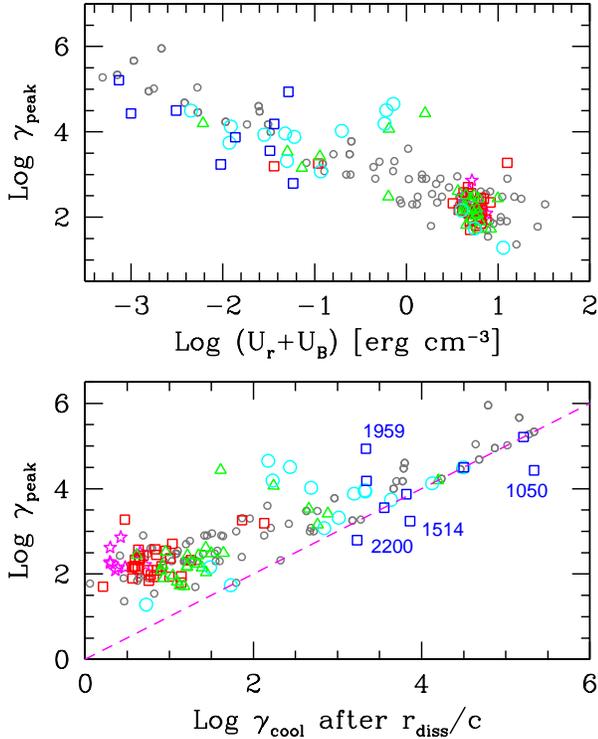,width=11cm,height=11cm}
\vskip -0.5 cm
\caption{Top: $\gamma_{\rm peak}$ vs $U_B+U^\prime_{\rm r}$.
Bottom:  $\gamma_{\rm peak}$ vs $\gamma_{\rm cool}$
calculated after one light crossing time.
The dashed line indicates equality.
See text for details.
Different symbols refer to different $\gamma$--ray luminosity bins,
as in Fig \ref{pjld}.
For comparison, we show (little grey circles) the blazars 
studied in Celotti \& Ghisellini (2008).
}
\label{gpeak}  
\end{figure}

\section{Jet power}

Table \ref{powers} lists the power carried by the jet in the form of
radiation ($P_{\rm r}$), magnetic field ($P_{\rm B}$), electrons
($P_{\rm e}$) and cold protons ($P_{\rm p}$, assuming one proton
per emitting electron). 
All the powers are calculated as
\begin{equation}
P_i  \, =\, \pi r_{\rm diss}^2 \Gamma^2\beta c \, U^\prime_i
\end{equation}
where $U^\prime_i$ is the energy density of the $i$ component,
as measured in the comoving frame.
We comment below on each contribution:

\begin{itemize}

\item
The power carried in the form of the produced radiation,
$P_{\rm r} =\pi r_{\rm diss}^2 \Gamma^2\beta c \, U^\prime_{\rm rad}$,
can be re--written as [using $U^\prime_{\rm rad}=L^\prime/(4\pi r_{\rm diss}^2 c)$]:
\begin{equation}
P_{\rm r}  \, =\,  L^\prime {\Gamma^2 \over 4} \, =\, L {\Gamma^2 \over 4 \delta^4}
\, \sim \, L {1 \over 4 \delta^2}
\end{equation} 
where $L$ is the total observed non--thermal luminosity
($L^\prime$ is in the comoving frame) and $U^\prime_{\rm rad}$ is the 
radiation energy density produced by the jet (i.e.
excluding the external components).
The last equality assumes $\theta_{\rm v}\sim 1/\Gamma$.
This is a almost model--independent quantity,
since it depends only on the adopted $\delta$, that can be estimated 
also by other means, namely superluminal motions.

\item
When calculating
$P_{\rm e}$ (the jet power in bulk motion of emitting electrons)
we include their average energy, i.e.
$U^\prime_{\rm e}= n_{\rm e} \langle\gamma\rangle m_{\rm e} c^2$.
Usually, when estimating this quantity, we have the problem of determining 
$\gamma_{\rm min} m_{\rm e} c^2$, the minimum energy of the electron
distribution [where, for steep distribution functions $N(\gamma)$, most
of the electrons are].
This problem is much alleviated here, since 
the assumed form of the particle injection function $Q(\gamma)$ is
rather flat at low energies.
The amount of the electrons at low energies then mainly depends on
cooling, making $N(\gamma)\propto \gamma^{-2}$ down to $\gamma_{\rm cool}$,
and flatter below.
Thus the total amount of electrons contributing to $P_{\rm e}$ 
depends on cooling, not on a pre--assigned shape of the particle distribution
(including a pre--assigned $\gamma_{\rm min}$).
In this sense the $P_{\rm e}$ derived here is less arbitrary.
Furthermore, in the case of luminous FSRQs, the X--ray flux can be reliably
associated to the EC mechanism (this occurs when the slope is very hard,
because the SSC component tends to be rather softer, see the discussion of this
point in Celotti \& Ghisellini 2008).
In this case the low energy X--ray data are crucial to fix $\gamma_{\rm min}$:
a too high value makes the modelled X--ray flux to underestimate the observed one.
Since in this sources the radiative cooling is severe, this agrees with the fact that 
$\gamma_{\rm min}$  must be small, of the order of $\gamma_{\rm cool}$ or less.
As a final comment, consider that the estimate of $P_{\rm e}$ includes
only those electrons contributing to the emission.
Since it is unlikely that all the electrons present in the emitting region
are accelerated, $P_{\rm e}$ is a lower limit.
On the other hand, when $\langle \gamma\rangle$ is greater than a few, 
the contribution of the accelerated electrons to $P_{\rm e}$ may
dominate over the contribution of the cool (not accelerated) ones.

\item
For $P_{\rm p}$ (the jet power in bulk motion of cold protons)
we have assumed that there is one proton per emitting electron,
i.e. electron--positron pairs are negligible.
This is a crucial assumption. 
Partly, it is justified within the context of our model
because we take into account the pair production process, 
and we find that pairs are always negligible.
If a substantial amount of pairs comes from the inner regions of the jet
we must explain why they have survived annihilation, important 
in the inner, more compact and denser regions (Ghisellini et al. 1992).
If they have survived because they were hot (thus they had a smaller
annihilation cross section) then they should have produced a large
amount of radiation (that we do not observe). 
In doing so, they should have cooled rapidly, and then annihilate.
These considerations (see also similar comments in Celotti \& Ghisellini 2008)
lead us to accept the assumption of one proton per electron
as the most reliable. 

For BL Lacs the presence or not of electron--positron pairs is less of
a problem, because the mean energy of the emitting electrons
is large, approaching the rest mass--energy of a proton.
In this cases $P_{\rm e}\sim P_{\rm p}$.

\item
$P_{\rm B}$ is derived using the magnetic field found from the model fitting.
There can be the (somewhat contrived) possibility that the size of 
the emitting region is smaller than the one considered here, i.e.
inside the jet there could be smaller volumes where the magnetic field
lines reconnect, and in this case the total Poynting flux of the jet
can be larger than what we estimate here. 
We consider that this is unlikely.

\end{itemize}

Fig. \ref{istokin} shows the distributions of the jet powers
and the bottom panel, for comparison, shows the distribution of 
the accretion disk luminosities.
Grey shaded ares correspond to BL Lacs for which we could
estimate only an upper limit to their disk luminosity.
Fig. \ref{epsilon} shows the fraction of the total jet power (i.e. 
$P_{\rm p}+P_{\rm e}+P_{\rm B}$) transformed in radiation ($\epsilon_{\rm r}$),
carried by electrons ($\epsilon_{\rm e}$) and Poynting flux ($\epsilon_{\rm B}$).

For FSRQs the power carried in radiation ($P_{\rm r}$) is larger than $P_{\rm e}$.
This is a consequence of fast cooling: electrons convert their energy into radiation
in a time shorter than $r_{\rm diss}/c$ and the radiation component can in this
time accumulate more energy than what remains in the electrons (even if they are
continuously injected during this time).
The distribution of $P_{\rm B}$ is at slightly smaller values than the 
distribution of $P_{\rm r}$, indicating that the Poynting flux cannot
be at the origin of the radiation we see.
As described in Celotti \& Ghisellini (2008), this is a direct consequence
of the large values of the so called Compton dominance (i.e. the ratio of 
the Compton to the synchrotron luminosity), since
this limits the value of the magnetic field.

\begin{figure}
\vskip -0.2cm
\hskip -0.5 cm
\psfig{figure=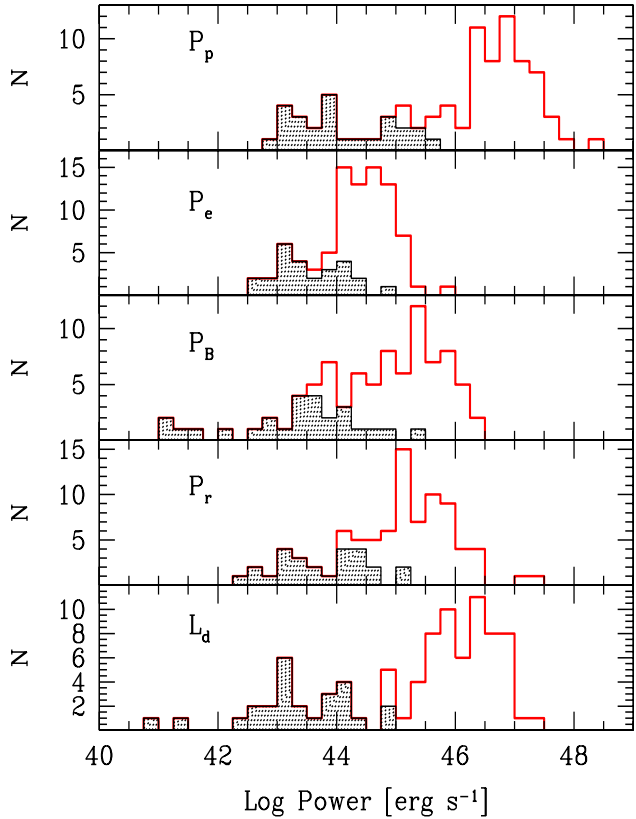,width=9.7cm,height=11.5cm}
\vskip -0.2 cm
\caption{From top to bottom:
distributions of the jet power in the form of 
bulk kinetic power of cold protons ($P_{\rm p}$, assuming one proton per
emitting electron); emitting electrons ($P_{\rm e}$, including
their average random energy $\langle\gamma\rangle m_{\rm e}c^2$);
Poynting flux ($P_{\rm B}$), radiation ($P_{\rm r}$).
The bottom panel shows the distribution of the luminosities $L_{\rm d}$ 
of the accretion disk. 
The shaded area in all panels 
corresponds to sources with only upper limits on $L_{\rm d}$.
}
\label{istokin}  
\end{figure}

To justify the power that the jet carries in radiation (we insist: it is the
least controversial quantity) we are forced to consider the power carried
by the jet in the form of protons.
Following the consideration made above, the simplest and most reasonable 
assumption is to assume that there is one proton per electrons.
If so, $P_{\rm p}$ for FSRQs is a factor $\sim$10--100 larger than $P_{\rm r}$,
meaning an efficiency of 1--10\% for the jet to convert its bulk kinetic
motion into radiation (see also the top panel of Fig. \ref{epsilon}).
This is reasonable: most of the jet power in FSRQs goes to form and energise the 
large radio structures, and not into radiation.
On the other hand, we do not have yet a firm handle on how much power the 
radio--lobes require (this estimate, among other things, depends on 
the proton energy density, still a very poorly known 
quantity).
Another inference comes from blazars emitting X--rays at large ($\sim$10--100 kpc) 
distances as observed by Chandra. 
For them the leading emission model (e.g. Tavecchio et al. 2004, 2007; but 
see. e.g., Kataoka et al. 2008) requires that the jet is still relativistic
at those scales (with $\Gamma$--factors similar to the ones derived in the 
inner regions) and this in turn suggests that the jet has not lost much of its 
power in producing radiation.

Consider now BL Lacs: we still have that $P_{\rm r}\sim P_{\rm e}~ \gsim~ P_{\rm B}$,
but now also $P_{\rm p}$ is of the same order.
This means that we are using virtually all the available jet power
to produce the radiation we see.
This is also illustrated in Fig. \ref{epsilon}.
Thus for BL Lacs we either assume that not all electrons are accelerated, 
allowing for an extra reservoir of power in bulk motion of the protons,
or, more intriguingly, we conclude {\it that the jet noticeably decelerates}.
The latter option is in agreement with recent findings on the absence 
of fast superluminal motion in TeV BL Lacs (Piner \& Edward 2004; Piner, Pant \& Edwards 2008), 
with the absence of 
strong extended radio structures, and with the result of the {\it Chandra} 
observations of extended X--ray jets at large scales, showing sub-luminal speed
at large scales (e.g. Worrall et al. 2001; Kataoka \& Stawarz 2005).
Moreover, this issue of jet deceleration of BL Lac jets has been debated
recently on the theoretical point of view (Georganopoulos \& Kazanas 2003;
Ghisellini, Tavecchio \& Chiaberge 2005).

We conclude that the jet of FSRQs are powerful, matter dominated
and transforming a few per cent of their kinetic power into radiation.
The jet of BL Lacs are less powerful, with the different forms of power
in rough equipartition, transforming a larger fraction of their
kinetic power into radiation, and probably decelerating.
Despite these different characteristics, there is no discontinuity
between FSRQs and BL Lacs.
All the different properties can be explained with the difference
in jet power accompanied by a different environment, in turn
caused by a different regime of accretion.
This important point is discussed below.

\begin{figure}
\vskip -0.7cm
\hskip -0.5 cm
\psfig{figure=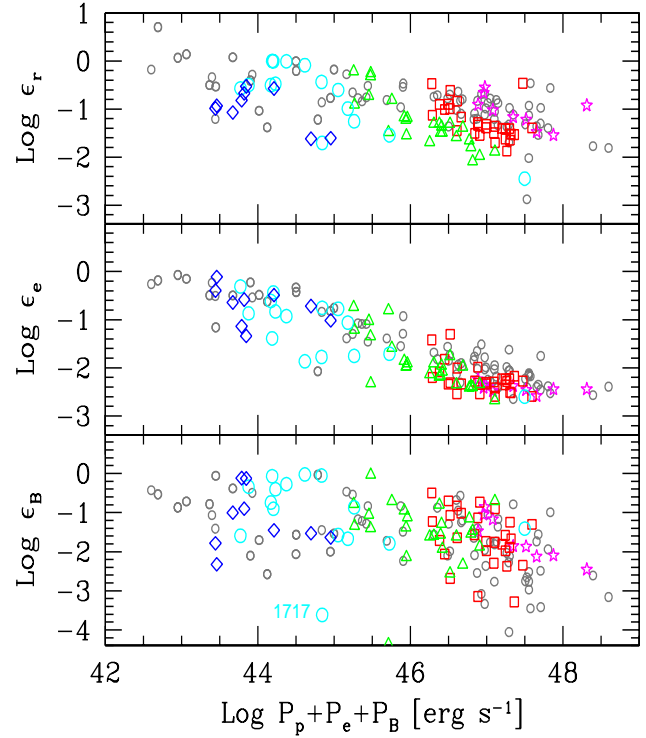,width=9.7cm,height=11.5cm}
\vskip -0.5 cm
\caption{
The fraction of $L_{\rm jet}$ radiated 
($\epsilon_{\rm r}$, top panel), in relativistic leptons 
($\epsilon_{\rm e}$, mid panel) and in magnetic fields 
($\epsilon_B$, bottom panel) as functions of 
$P_{\rm jet}=P{\rm p}+P_{\rm e}+P_{\rm B}$. 
Different symbols refer to different $\gamma$--ray luminosity bins,
as in Fig \ref{pjld}.
For comparison, we show (little grey circles) the blazars 
studied in Celotti \& Ghisellini (2008).
}
\label{epsilon}  
\end{figure}

\begin{figure}
\vskip -0.5cm
\psfig{figure=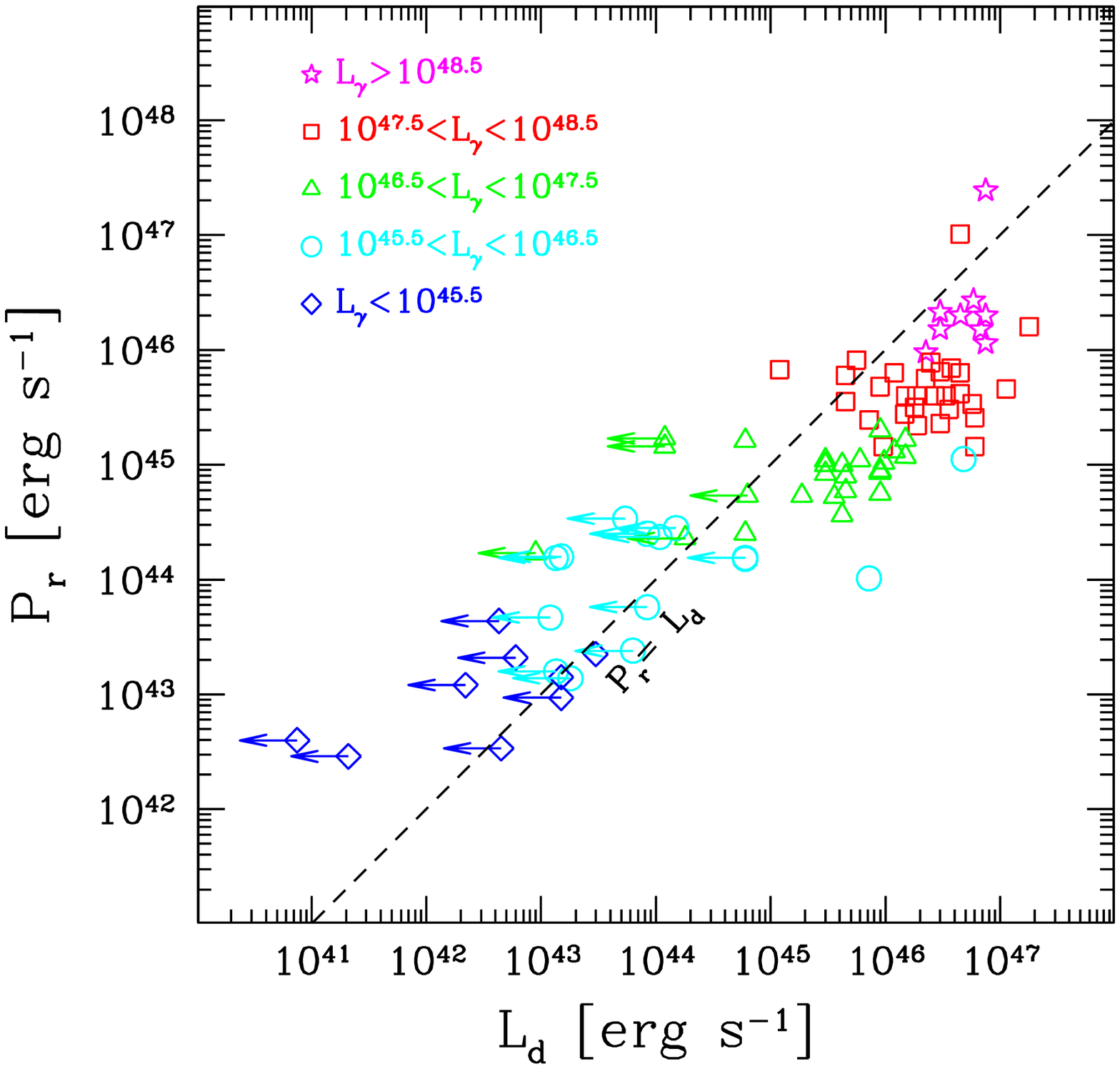,width=9cm,height=9cm}
\vskip -1 cm
\psfig{figure=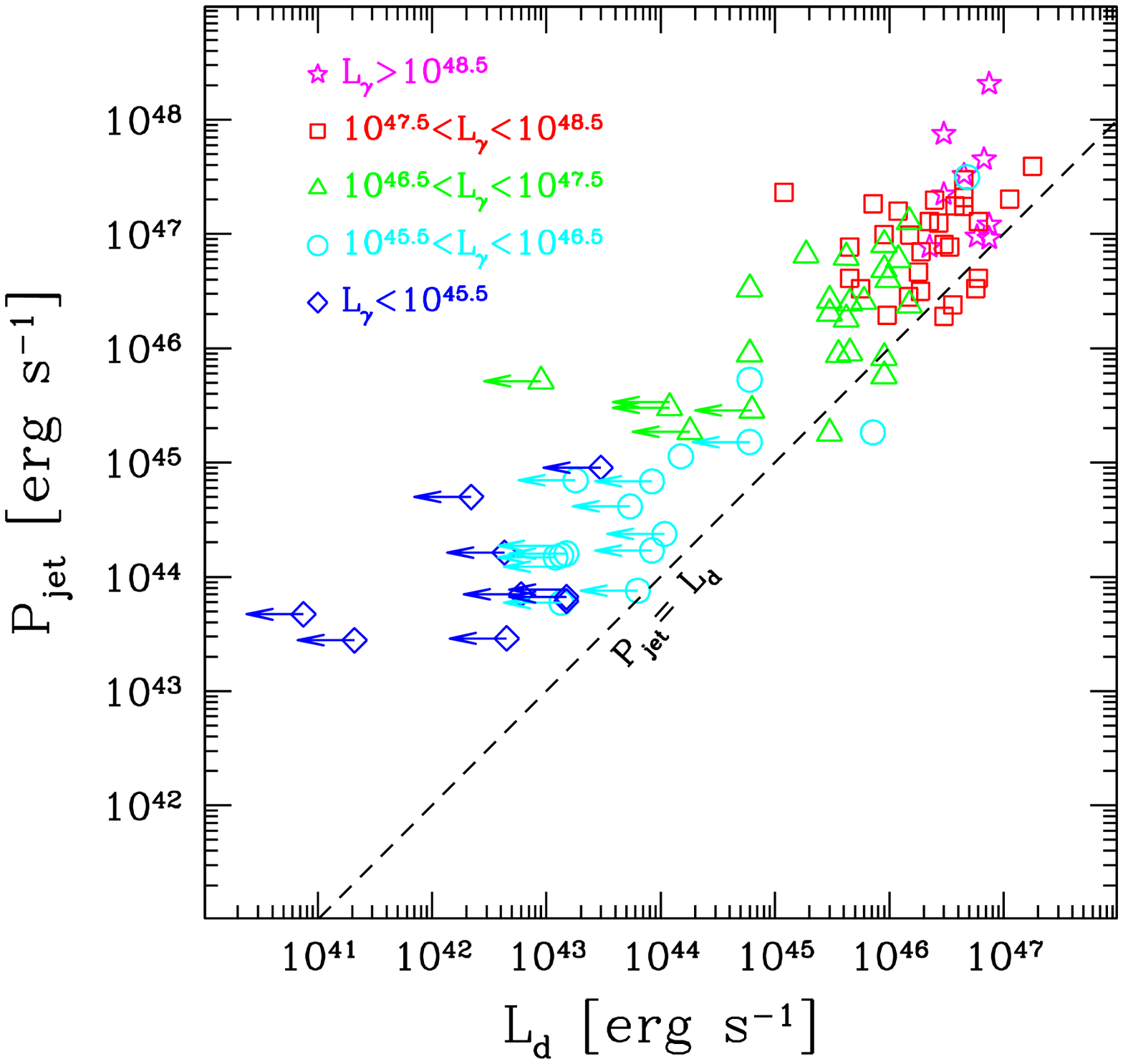,width=9cm,height=9cm}
\vskip -0.5 cm
\caption{Top: 
The radiation power produced by the jet as a function
of the accretion disk luminosity.
Bottom panel: the jet power (i.e. 
$P_{\rm jet}=P_{\rm p}+P_{\rm e}+P_{\rm B}$)
as a function of the accretion disk luminosity.      
The different symbols (as labelled) correspond
to blazars of different $\gamma$--ray 
observed luminosities.
}
\label{pjld}  
\end{figure}

\subsection{Jet power vs accretion luminosity}

The availability of the {\it Swift}/UVOT data for many of our blazars 
made possible to estimate the accretion disk luminosity for several of them.
We can then discuss one of the crucial problem in
jet physics: the disk/jet connection.

Fig. \ref{pjld} shows what we think is the main result of our work (see also Maraschi 
\& Tavecchio 2003, Sambruna et al. 2006 for earlier results).
The top panel shows $P_{\rm r}$ as a function of the accretion disk 
luminosity $L_{\rm d}$, while the bottom panel shows $P_{\rm jet}$ vs $L_{\rm d}$.
The different symbols correspond to different bins of the observed $\gamma$--ray luminosity
$L_\gamma$, as labelled.

Consider first the top panel:
\begin{itemize}
\item
Blazars with different $L_\gamma$ form a sequence in 
the $P_{\rm r}$--$L_{\rm d}$ plane.
That $L_\gamma$ correlates with $P_{\rm r}$ is not a surprise, since we already
knew (from EGRET) that the $\gamma$--ray luminosity is dominating the bolometric output.
What is interesting is that the most luminous $\gamma$--ray blazars have also a more 
powerful accretion disk.

\item
For our analysis we have considered the average value of
the $\gamma$--ray luminosity during the 3 months survey.
Then it should not be extreme, given 
the large amplitude and rapid variability
shown by blazars, especially at high energies.
In other words, the shown $P_{\rm r}$ is more indicative of an
``average" state, not of an extremely high state, even if, in a flux
limited sample of variable sources, like the {\it Fermi} one, 
sources in high states are always over--represented.
Variability of $P_{\rm r}$ is however an issue, and we can 
consider that the single blazar can vary at least by 
a factor 10--30 around the shown $P_{\rm r}$.
This contributes to the somewhat large scatter around the 
$P_{\rm r}$--$L_{\rm d}$ relation.

\item
Considering only FSRQs, we have that a least square fit yields
$\log P_{\rm r, 45} = 0.73 \log L_{\rm d, 45} - 0.36 $,
with a probability for the correlation to be random of $P=5\times 10^{-8}$.
The same least square fit yields
$\log L_{\rm d, 45} = 0.65 \log P_{\rm r, 45} + 0.82 $,
indicating that a slope around unity is consistent with the data.
Since both $P_{\rm r}$ and $L_{\rm d}$ depends on redshift,
we have also applied a partial correlation analysis, 
as explained in Padovani (1992).
Using Eq. 1 of that paper, we have verified that 
$P_{\rm r}$ and  $L_{\rm d}$, once the redshift dependence is
excluded, still correlate, although 
the probability to be random increases to $P=2\times 10^{-4}$.

\item 
Line--less BL Lacs are shown with their corresponding
upper limits on $L_{\rm d}$.
These are nevertheless important, showing that they must deviate
from the general trend defined by FSRQs (see also Maraschi \& Tavecchio 2003).

\end{itemize}

Consider now the bottom panel, showing $P_{\rm jet}$ vs $L_{\rm d}$.
As mentioned, the jet power of FSRQs is dominated by the bulk motion of cold protons,
while in BL Lacs it is more equally distributed among electrons, protons and
magnetic field. 
Also in this plane the more $\gamma$--ray luminous blazars have
the most powerful jet.

Considering only FSRQs, we have that a least square fit yields
$\log P_{\rm jet, 45} = 0.62 \log L_{\rm d, 45} + 1.07  $,
with a probability for the correlation to be random of $P=6\times 10^{-7}$.
The same least square fit yields
$\log L_{\rm d, 45} = 0.56 \log P_{\rm jet, 45} + 0.14 $,
indicating that a slope around unity is consistent with the data.
Excluding the dependence of redshift by applying a partial correlation analysis, 
the probability to be random
increases to $P=3.4\times 10^{-6}$.

Also in this plane the line--less BL Lacs (with upper limits for $L_{\rm d}$)
deviate from the trend defined by FSRQs, implying that their jet is much more
powerful than the luminosity emitted by their accretion disks.

We now assume, as an ansatz, that the jet power is always
of the order of $\dot M_{\rm in} c^2$, for FSRQ as well as for BL Lacs (see 
Ghisellini \& Tavecchio 2008 for a discussion).
This allows to estimate the accretion rate for BL Lacs independently
from their (invisible) disk luminosity.
Furthermore we can form the ratio $\dot M_{\rm in}/\dot M_{\rm Edd}$
given by
\begin{equation}
{\dot M_{\rm in} \over \dot M_{\rm Edd} } \, \equiv\, 
{\dot M_{\rm in} c^2 \over 1.3\times 10^{38} (M/M_\odot)}
\end{equation}
For FSRQs we have used $\dot M_{\rm in}$ and $M$
derived from our modelling ($\dot M_{\rm in}$ is 
given by $\dot M_{\rm in} = L_{\rm d} / (\eta c^2)$, with
$\eta=0.08$ for all sources).
For BL Lacs we simply set $\dot M_{\rm in} =P_{\rm jet}/c^2$.
The resulting distributions are shown in Fig. \ref{mdot}.
Since for BL Lacs the mass we have used is very uncertain, 
although in agreement with other independent estimates,
we show, beside the values of $\dot M_{\rm in}/\dot M_{\rm Edd}$
obtained using the masses listed in Tab. \ref{para} (thick solid line),
also the distribution obtained by adopting the same mass of $10^8$ 
(shaded cyan) and $10^9 M_\odot$ (shaded gray) for all BL Lacs.

Fig. \ref{mdot} shows that there is a ``divide" between
BL Lacs and FSRQs occurring at 
$\dot M_{\rm in}/\dot M_{\rm Edd} \sim 0.1$,
equivalent to $L_{\rm d}/L_{\rm Edd}\sim 0.01$,
in striking agreement with the value proposed
by Ghisellini, Maraschi \& Tavecchio (2009) and
very similar to the value proposed by Ghisellini \& Celotti (2001)
for the division between FR 1 and FR 2 radio--galaxies, based
on completely different arguments (see also Xu, Cao \& Wu 2009).
This division can be readily interpreted as the change
in the accretion regime of the disk, becoming radiatively
inefficient when $\dot M_{\rm in}$ is less than $\sim 10\%$ 
of the Eddington value (and $L_{\rm d}$ is less than 
$\sim1\%$ of $L_{\rm Edd}$). Notably, there are hints that 
similar results holds for {\it radio-quiet} sources (e.g. Ho 2009;
see also Ho 2008 for review).

\begin{figure}
\vskip -0.5cm
\hskip -0.5 cm
\psfig{figure=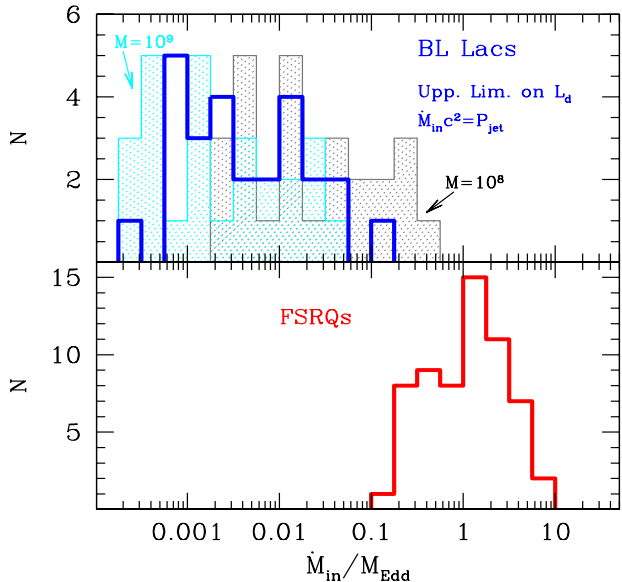,width=9.5cm,height=9cm}
\vskip -0.5 cm
\caption{
Distributions of the accretion rates in Eddington units for BL Lacs,
calculated assuming $P_{\rm j}=\dot M_{\rm in}c^2$ (top panel)
and for FSRQs (bottom).
The assumed black hole mass is listed in Tab. \ref{para}. For BL Lacs,
given the uncertain estimates of the mass, we also report two other 
distributions, assuming a mass of $M=10^8\, M_{\odot}$ and $M=10^9\, M_{\odot}$ 
(grey and cyan histograms, respectively) for all the sources.}
\label{mdot}  
\end{figure}


\section{Summary of results}

We have studied the entire sample of blazars detected during the first
3--months survey of {\it Fermi} and have a known redshift and
a reasonable data coverage of their SED. 
By studying the resulting 85 objects we have found the following main results:
\begin{itemize}

\item
The simultaneous or quasi--simultaneous {\it Swift} observations for 
a large fraction of our sources allowed to have an unprecedented view
on the optical to $\gamma$--ray SED of blazars.
In addition, the optical--UV data were very important to 
separate the thermal emission produced by the accretion disk from
the beamed non--thermal continuum.
In this way, for FSRQs, we could estimate the black hole mass
and the accretion rate. This in turn allowed to study the connection 
between the power of the jet and the luminosity emitted by the 
accretion disk. We found that they correlate.

\item 
The estimated black hole masses are in the 
range between $10^8$ and several times $10^9$ solar masses
for FSRQs.
For BL Lacs the poorly constrained masses are in the range
$10^8-10^9 M_\odot$.
These values are consistent with those found in the literature 
for the same objects, but existing estimates vary.

\item
The luminosity emitted by the accretion disks of FSRQs is
between 1 and 60 per cent of the Eddington one.
Upper limits to the disk emission of BL Lacs
indicate $L_{\rm d}/L_{\rm Edd}<10^{-2}$.

\item
The ``divide" between FSRQs and BL Lacs, in terms of
the accreting mass rate (Ghisellini, Maraschi \& Tavecchio 2009), is fully
confirmed. It occurs when the accretion mass rate becomes smaller
than 10 per cent of the Eddington one, or, equivalently,
when the disk luminosity becomes smaller than 1 per cent Eddington.

\item
The $\gamma$--ray luminosity is a good tracer both of the accretion disk luminosity 
(for FSRQs) and of the jet power (for all blazars).

\item 
As for the jet emission processes, the EC component almost always dominates
the emission beyond the X--ray band in FSRQs, with the SSC contributing
to soft and mid--energy X--rays in some cases.
In BL Lacs, most of the sources can be fitted by a pure SSC model,
but some of them require an extra component when the separation, in energy,
of the synchrotron and Compton peaks is too large.
This can be provided by a spine/layer structure of the jet, that avoids
the need of extremely large $\Gamma$--factors.

\item
The seed photons for the EC mechanism can be provided by a fairly standard
BLR, as assumed here and in the ``canonical" scenario for powerful blazars.
The majority of FSRQs dissipate within the BLR, while 4 of them 
are better fitted assuming a dissipation region between the BLR and an IR
emitting torus, at distances greater than the BLR.

\item 
The jet dissipation region is located between a few hundred and
a thousand \sc\ radii for all sources.

\item
Bulk Lorentz factors are in the range 10--15.

\item
The magnetic field in the emitting region of FSRQs is between 1 and 10 G,
and 10 times less for BL Lacs, on average.

\item
Jets in FSRQs must be matter dominated, while in BL Lacs there
can be equipartition between the power in bulk motion of
the emitting electrons, cold protons and magnetic field.

\item 
In FSRQs it is likely that the 
electron-positron pair component is negligible.
If so, the jet power in these sources is dominated (by a factor 10--30)
by the cold proton component, and it is a factor $\sim$3--5 
larger than the luminosity emitted by the accretion disk.
The outflowing mass rate is around a few per cent of the accreting mass rate.

\item
We confirm that the SEDs of blazar form a sequence, explained
in terms of different radiative cooling suffered by the electrons,
with higher energy electrons present in jets of lower power.

\end{itemize}

\section{Discussion}

The results listed in the previous section confirm
earlier findings, largely based on blazars with an EGRET detection, 
and/or detections in the TeV band.
Because of the factor $\sim20$ better sensitivity of {\it Fermi}/LAT
with respect to EGRET, we are now starting to explore sources
that are not in ``extraordinary" bright states in $\gamma$--ray band.
Sources in our sample should be closer to the average state of blazars,
even if, given the still limited sensitivity and the large amplitude
variability (even by factor 30--100) our blazars are most likely
emitting above their average.

These {\it Fermi} blazars confirm that the jet of blazars form a
sequence whose main parameter is their emitted luminosity.
This may seem strange, given the strong dependence of
the observed luminosity to the Doppler beaming, then on the 
viewing angle.
On the other hand, the results of our model fitting show that 
our blazars are all viewed at small angles, with no misaligned jet 
entering the sample.
Misaligned sources, therefore, are fainter than the current
{\it Fermi} blazars and should appear in deeper catalogues.

\begin{figure}
\vskip -0.5cm
\psfig{figure=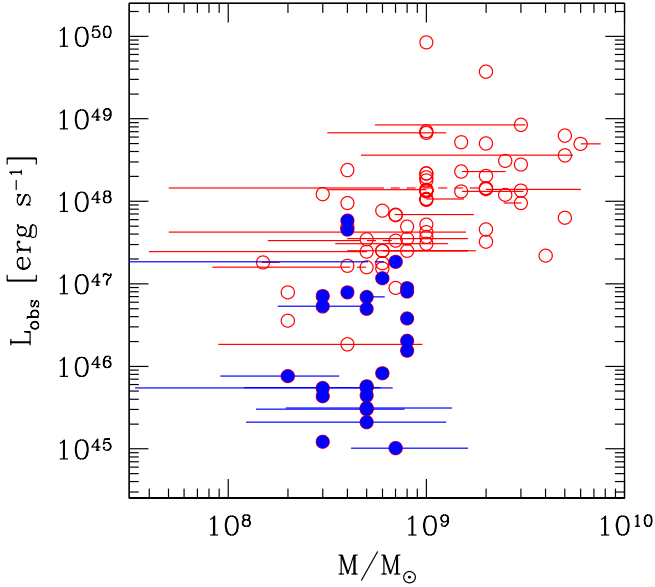,width=9.cm,height=9cm}
\vskip -0.5 cm
\caption{The observed bolometric luminosity
produced by the jet (calculated from the
fitting model) as a function of the black hole mass.
Empty circles are FSRQs with estimated black hole masses and accretion 
luminosities, filled circles are BL Lacs with only an upper limit on their 
disk luminosities, and whose black hole mass is uncertain.
We also show the range of the estimates of the black hole mass found in the literature.
}
\label{mlr}  
\end{figure}

\subsection{Black hole mass and the blazar sequence}

There is, in our opinion (Ghisellini \& Tavecchio 2008),
another important parameter, besides the jet power,
controlling the look of the emitted SED and its bolometric
luminosity: the mass of the black hole.
Blazars with small black hole masses and accreting at a rate
greater than a critical value should be ``red" (i.e. they should have
relatively small peak frequencies and large Compton dominance)
even if their observed bolometric luminosity is relatively small,
contrary to what the simplest version of the blazar sequence would predict.
Fig. \ref{mlr} shows the observed bolometric luminosity
(as derived by the model) as a function of the 
black hole mass estimated in this paper.
Empty circles are FSRQs with estimated black hole masses and accretion 
luminosities, filled circles are BL Lacs with only an upper limit on their 
disk luminosities, and whose black hole mass is uncertain.
We also show the range of black hole masses existing in the literature and
reported in Tab. \ref{masses}.
Within FSRQs there is indeed a tendency for
larger luminosities to correspond to larger black hole masses.
Viceversa, below $L_{\rm obs}=10^{47}$ erg s$^{-1}$ all black hole masses are
smaller than $10^9 M_\odot$.
Fig. \ref{mlr} shows that there can be ``red" blazars with $L_{\rm obs}$
similar to (bluer) BL Lacs, but this happens when their black hole mass 
is relatively small.

Having the distribution of all relevant physical parameters,
we can do the exercise to construct the ``average" SED
of FSRQs and BL Lacs, respectively, of our sample.
This is illustrated by Fig. \ref{average}, for which we have 
used for the average FSRQs and BL Lacs the parameters 
listed at the end of Tab. \ref{para}.
Note that in our sample there are no ``extreme" TeV BL Lacs,
since, as discussed in T09, these sources have so large 
Compton peak frequencies to make difficult a detection by {\it Fermi}.
So Fig. \ref{average} corresponds to the average BL Lac
detected by {\it Fermi}, and not to the average BL Lac in general.

\begin{figure}
\vskip -0.4cm
\hskip -1cm
\psfig{figure=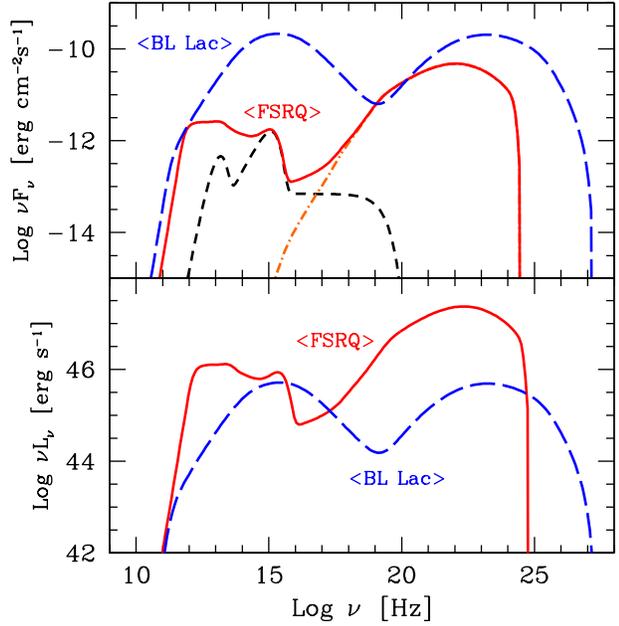,width=10cm,height=10cm}
\vskip -0.8 cm
\caption{The average SED of FSRQs and of BL Lacs without 
any sign of disc emission.
These SED have been constructed taking the (logarithmic)
average of the parameters of the sources belonging to the two subclasses
(see Table \ref{para}).
The top and the bottom panels show the fluxes and luminosities,
respectively.
The shown frequencies are calculated in the rest frame
of the source for the luminosity plot, and are the observed
ones for the flux plot. 
The assumed redshift are $z=1$ for $\langle$FSRQ$\rangle$ 
and $z=0.1$ for  $\langle$BL Lac$\rangle$.
}
\label{average}  
\end{figure}

\begin{figure}
\vskip -0.4cm
\hskip -0.7cm
\psfig{figure=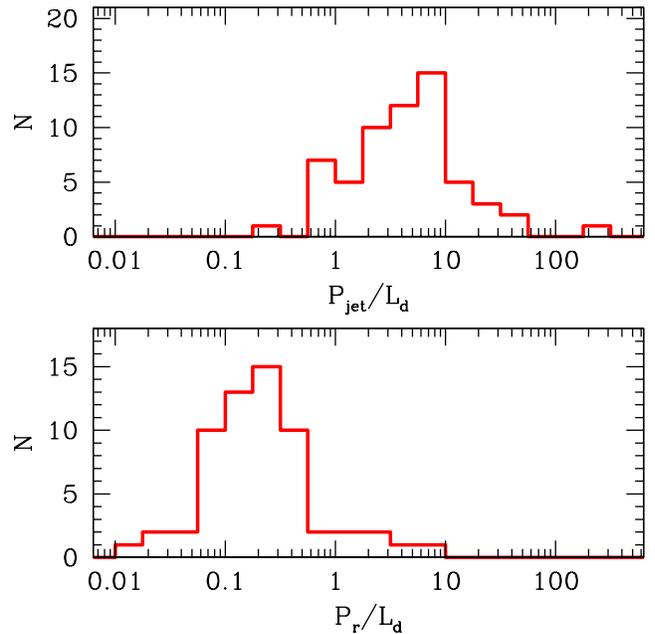,width=10cm,height=10cm}
\vskip -0.8 cm
\caption{
Top panel: the distribution of the ratio between 
the total jet power $P_{\rm jet}$
and the accretion disk luminosity $L_{\rm d}$.
Bottom panel: The distribution of the ratio between $P_{\rm r}$
(the power carried by the jet in the form of produced radiation)
and the accretion disk luminosity $L_{\rm d}$.
Both distributions are for FSRQs only.
}
\label{isto5}  
\end{figure}

\subsection{Jet power and accretion}

The jet powers derived here are large, reaching, at the high end
of their distributions, values greater than 10 times the disk
luminosity (see the upper panel of Fig. \ref{isto5}).
For these extreme objects $P_{\rm jet}\gsim \dot M_{\rm in} c^2$.
This is admittedly a model--dependent statement.
It assumes a one--zone leptonic model and that there is  one proton
per emitting electron. 
A more robust statement is that the amount of power
the jet spends to produce and carry the non--thermal radiation 
is also very large, being in some case equal to the disk luminosity
produced by accretion (see the bottom panel of Fig. \ref{isto5}), 
and more often a factor $\sim$3--10 smaller.
Consider that the jet power cannot be simply be represented
by $P_{\rm r}$: if this were the case, the entire jet power is used to produce the radiation
we see, so that the jet would decelerate significantly.
Instead, the jet {\it must} continue to be relativistic up to large distances,
as required by the existence of strong radio lobes and the X--ray radiation
seen by the {\it Chandra} satellite at distances (from the black hole) of hundreds of kpc.
Therefore a reasonable {\it lower limit} on $P_{\rm jet}$ should be a factor
3--10 greater than $P_{\rm r}$.

What is then the source of the power of the jet?
Is it only the gravitational energy of the accreting matter or do we 
necessarily need also the rotational energy of a spinning black hole?
We here discuss two possible alternatives, that can both explain 
our results, but are drastically different for the ultimate energy
source for the jet.

\vskip 0.3 cm
\noindent
{\bf Jets powered by accretion only ---} 
Jolley et al. (2009), building up on previous ideas
put forward in Jolley \& Kuncic (2008), propose that,
in jetted sources, a sizeable fraction of the accretion power
goes to power the jet.
As a result, the remaining power for the disk
luminosity is less than usually estimated
by setting $L_{\rm d}=\eta \dot M_{\rm in} c^2$, with $\eta\sim0.08$--0.1.
This implies that the mass accretion rate needed to sustain a given $L_{\rm d}$ 
is {\it larger} than what we have estimated.
Also the total accretion power is larger, and it is sufficient to
explain the large jet power we have derived.
Assume that the total power extractable from the accretion process is
$\eta_{\rm tot} \dot M_{\rm in} c^2$, and that a fraction $\eta_{\rm d}$
($\eta_{\rm j}$) of $\dot M_{\rm in} c^2$ is used to produce the disk
luminosity (the jet power).
We have:
\begin{eqnarray}
\eta_{\rm tot} \dot M_{\rm in} c^2\, &=& \, L_{\rm d} + P_{\rm j}  \nonumber \\
\eta_{\rm tot} \dot M_{\rm in} c^2\, &=& \,  \eta_{\rm d}\dot M_{\rm in} c^2+ 
\eta_{\rm j}\dot M_{\rm in} c^2 \, \, \to \nonumber \\
\eta_{\rm tot} \, &=&\, \eta_{\rm d}+ \eta_{\rm j} 
\end{eqnarray}
Our results imply i) $\eta_{\rm j} > \eta_{\rm d}$ and ii)
that $\eta_{\rm j}/\eta_{\rm d}\sim$ const in different blazars to account for
the observed $P_{\rm j}$--$L_{\rm d}$ correlation.

\vskip 0.3 cm
\noindent
{\bf Jets powered by the black hole spin ---} 
The rotational energy of a maximally spinning black hole
is 29\% of the hole rest mass energy (i.e. up to $5\times 10^{62} M_9$ erg),
amply sufficient to power a strong jet for its entire lifetime.
In principle, in this case one can have $P_{\rm j}>L_{\rm d}$,
given a sufficiently efficient way to extract the energy
of the spinning black hole.
In this case we can ``decouple" $P_{\rm j}$ and $L_{\rm d}$,
since they have a different energy source.

%

On the other hand, for FSRQs, we do see a relation between the $P_{\rm j}$
and $L_{\rm d}$, and at first glance this seems to suggest that it is the 
accretion, not the spin, to power the jet.
We can envisage a possible solution to this apparently contradictory
issue, by linking the extraction 
of the hole rotational energy to the accretion process.
The main idea is the following:
the energy density $\rho_0 v_{\rm \psi}^2$ of the accreting material 
close to the black hole horizon can sustain a maximum
magnetic energy density $B_0^2/(8\pi)$ of the same order
(see also Ghisellini \& Celotti 2002).
Here $v_{\rm \psi}$ is the circular (keplerian) velocity of
the matter.
The magnetic field sustained by the accreting matter
can then tap the rotational energy of the hole.
The mechanism able to do this task is the
Blandford--Znajek (1977) process (hereafter BZ), whose efficiency has been
debated in recent years 
(e.g. Moderski \& Sikora 1996; Ghosh \& Abramowicz 1997; 
Livio, Ogilvie \& Pringle 1999; McKinney 2005; Garofalo 2009;
Krolik \& Hawley 2002).
Here, for simplicity, without entering in the technical discussion
on the efficiency of this mechanism, we assume that the
jet power is of the form:
\begin{equation}
P_{\rm j} \, \sim \, k a^2\pi  R_0^2 {B_0^2\over 8\pi} c
\label{pjet}
\end{equation}
where $a\le 1$ is the dimension--less angular momentum of the hole,
$R_0$ is some fiducial distance of the order of the black hole
horizon, and $B_0$ is the magnetic field at that radius.
Eq. \ref{pjet} is nothing else than a Poynting flux.
The factor $k$ includes our uncertainties about the
efficiency of the BZ process.

Assume that at a distance $R_0$ from the black hole, the disk has
a height $H_0$ above the equatorial plane.
If $v_{\rm r}$ is the radial in--fall velocity, we have
\begin{equation}
\dot M_{\rm in} \, =\, 4\pi R_0 H_0 \rho_0   v_{\rm R}  \, \to \, \rho_0\, =\, 
{\dot M_{\rm in} \over  4\pi R_0 H_0 v_{\rm R} }
\end{equation}
If the energy density of the magnetic field is a fraction $\epsilon_{\rm B}$
of the kinetic energy density of the matter orbiting around the black hole, we have
\begin{equation}
{B_0^2\over 8\pi} \, =\, {1\over 2}\,\epsilon_{\rm B}\, \rho_0 v^2_\phi 
\end{equation}
If a magnetic field with the same magnitude is threading the spinning hole,
then we have
\begin{equation}
P_{\rm j} \, \sim \, {\dot M_{\rm in} c^2\epsilon_{\rm B} k a^2 \over  8 (H_0/R_0) }   
{\beta^2_\phi \over \beta_{\rm R} } \, 
=\,   L_{\rm d}\, {\epsilon_{\rm B} k a^2 \over  8 \eta (H_0/R_0) }   
{\beta^2_\phi \over \beta_{\rm R} }
\end{equation}
Close to the gravitational or \sc\ radius, we may set $H_0/R_0\lsim 1$.
The ratio $\beta^2_\phi/\beta_{\rm R}$ can be slightly larger than
unity, depending on viscosity.
For $\epsilon_{\rm B}\sim 1$, the jet power is
maximum, and for $k$ not much less than unity can be 
of the same order of the disk luminosity.
The (rather strong) requirement that our data are posing on the BZ
mechanism is therefore on its efficiency, that must be large.

\section*{Acknowledgments}
This work was partly financially supported by a 2007 COFIN--MIUR and
an ASI I/088/06) grant.
This research made use of the NASA/IPAC Extragalactic Database (NED) 
which is operated by the Jet Propulsion Laboratory, Caltech,
under contract with the NASA.
We acknowledge the use of public data from the {\it Swift} data archive. 
The research made use of data obtained from the High Energy Science Archive 
Research Center (HEASARC), provided by NASA's GSFC.


\section{appendix: Spectral Energy Distributions}

\begin{figure}
\vskip -0.2 cm
\psfig{figure=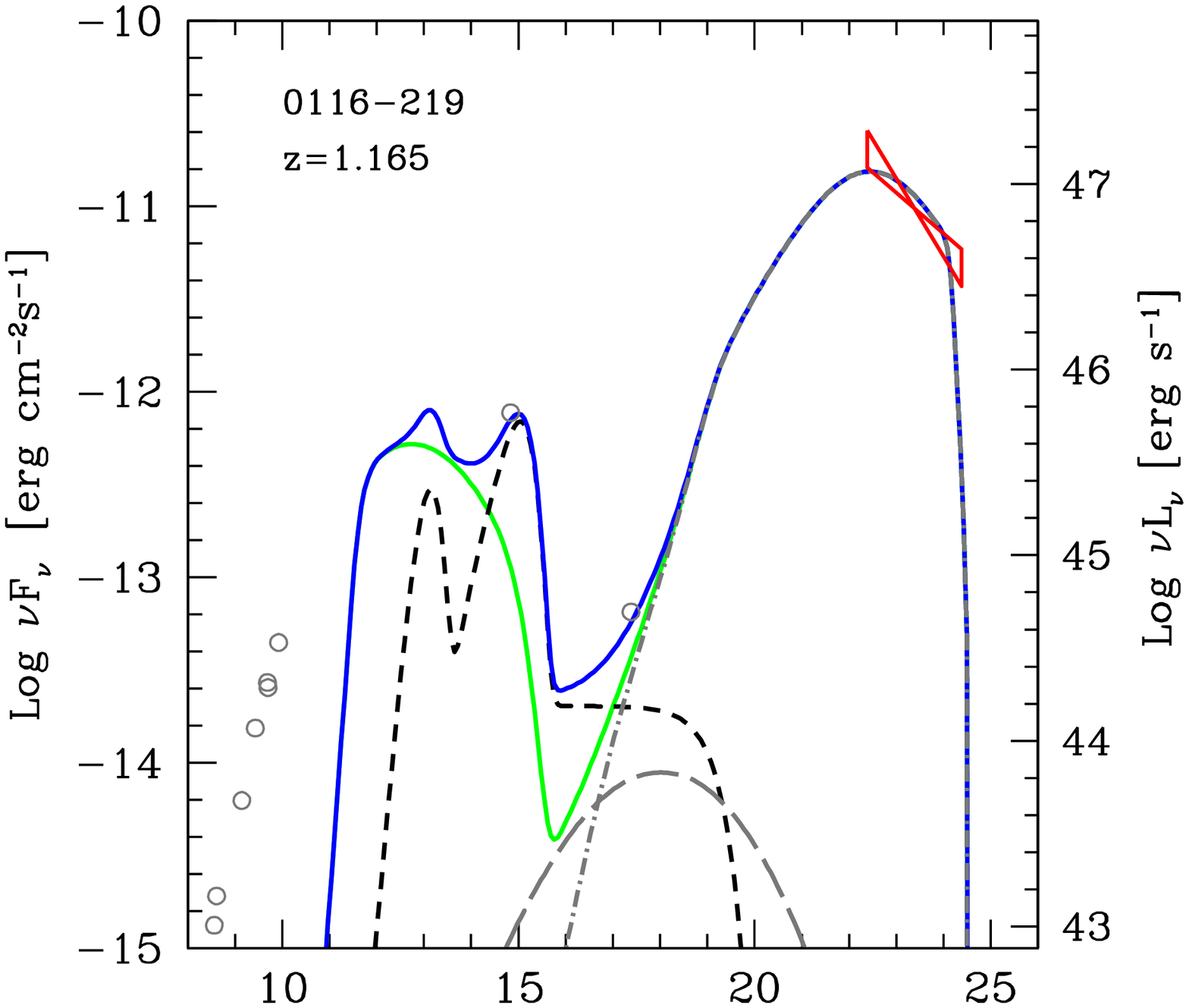,width=9cm,height=7cm}
\vskip -1.4 cm
\psfig{figure=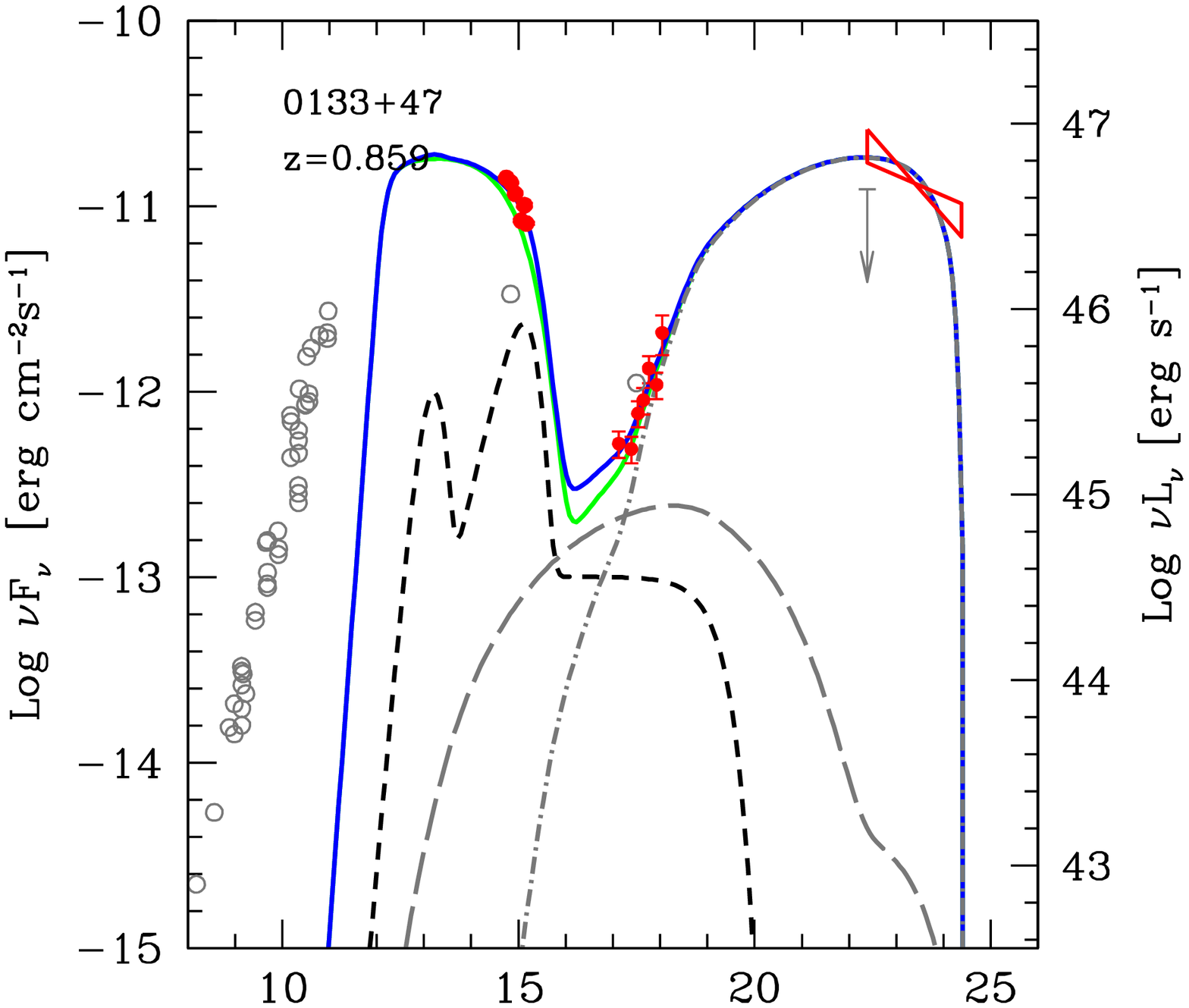,width=9cm,height=7cm}
\vskip -1.4cm
\psfig{figure=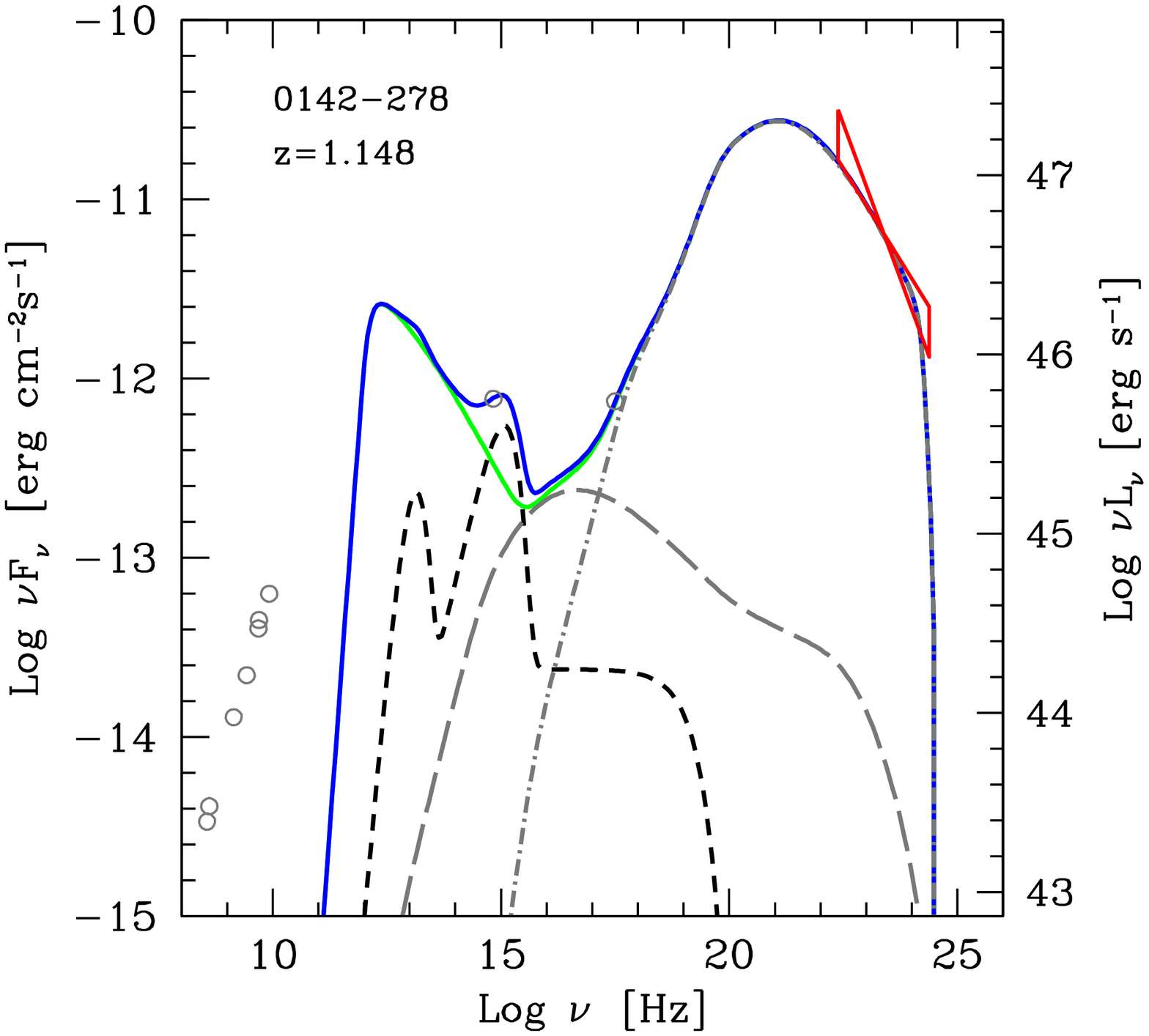,width=9cm,height=7cm}
\vskip -0.6 cm
\caption{SED of PKS 0166--219, 0133+47 (DA 55) and 
PKS 0142--278, together with the fitting models,
with parameters listed in Tab. \ref{para}.
{\it Fermi} and {\it Swift} data are indicated
by dark grey symbols (red in the electronic version),
while archival data (from NED) are in light grey.
The short--dashed line is the emission from the IR torus, the accretion disk
and its X--ray corona; the long--dashed line is the SSC contribution and
the dot--dashed line is the EC emission.
The solid light grey line (green in the electronic version) is the
non thermal flux produced by the jet, the solid dark grey line (blue
in the electronic version) is the sum of the non--thermal and thermal 
components.
}
\label{f1}
\end{figure}

\begin{figure}
\vskip -0.2cm
\psfig{figure=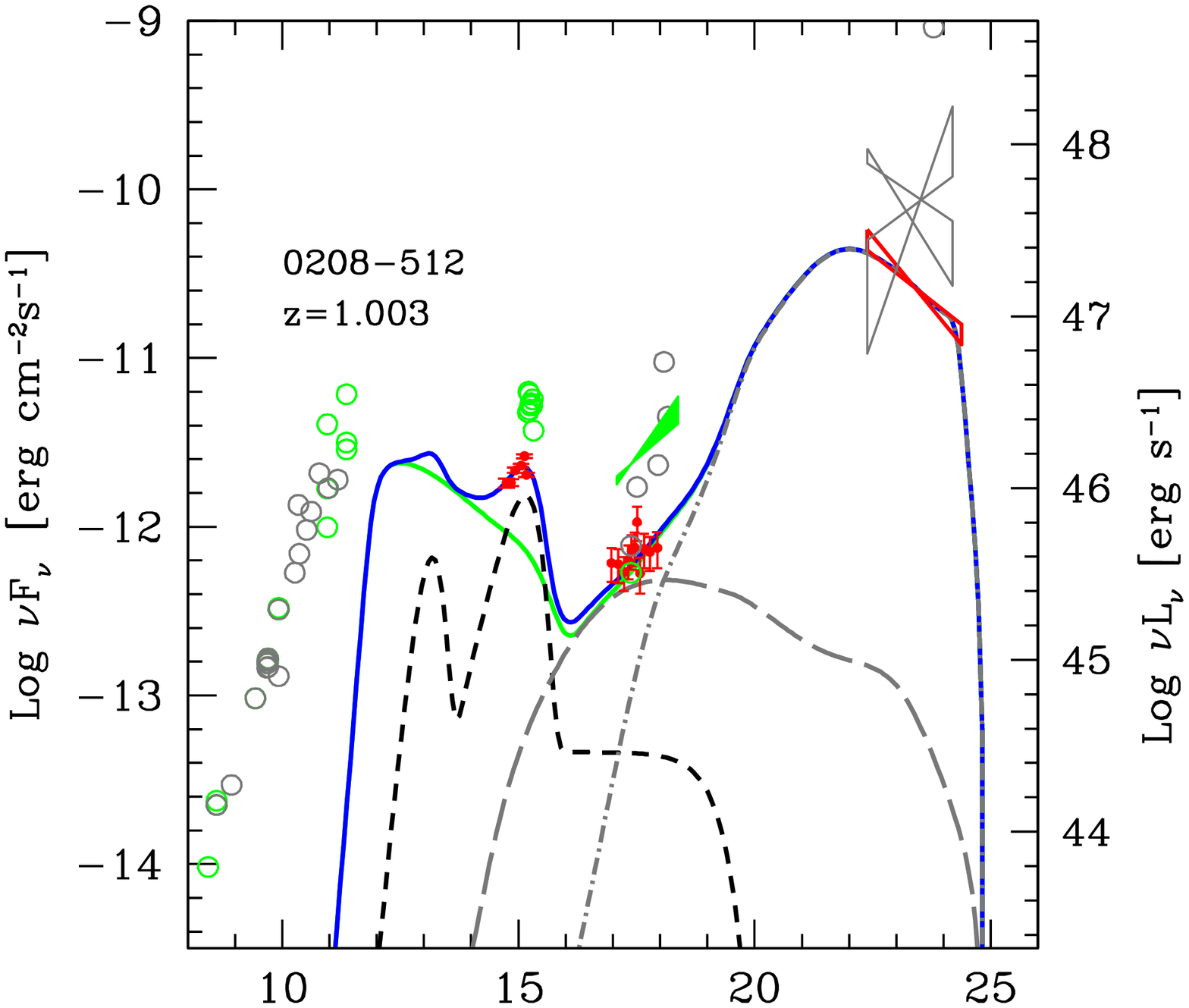,width=9cm,height=7cm}
\vskip -1.4cm
\psfig{figure=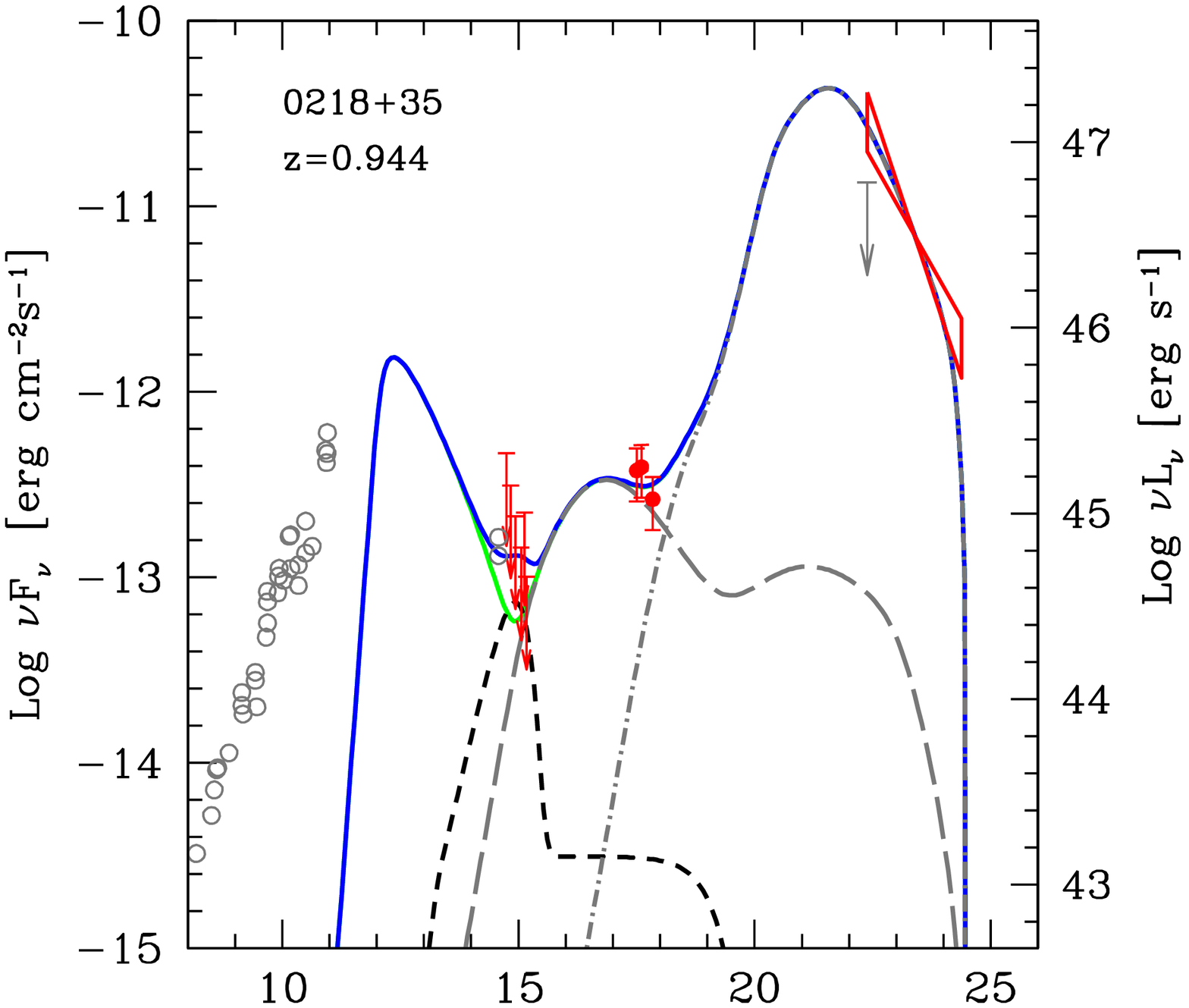,width=9cm,height=7cm}
\vskip -1.4cm
\psfig{figure=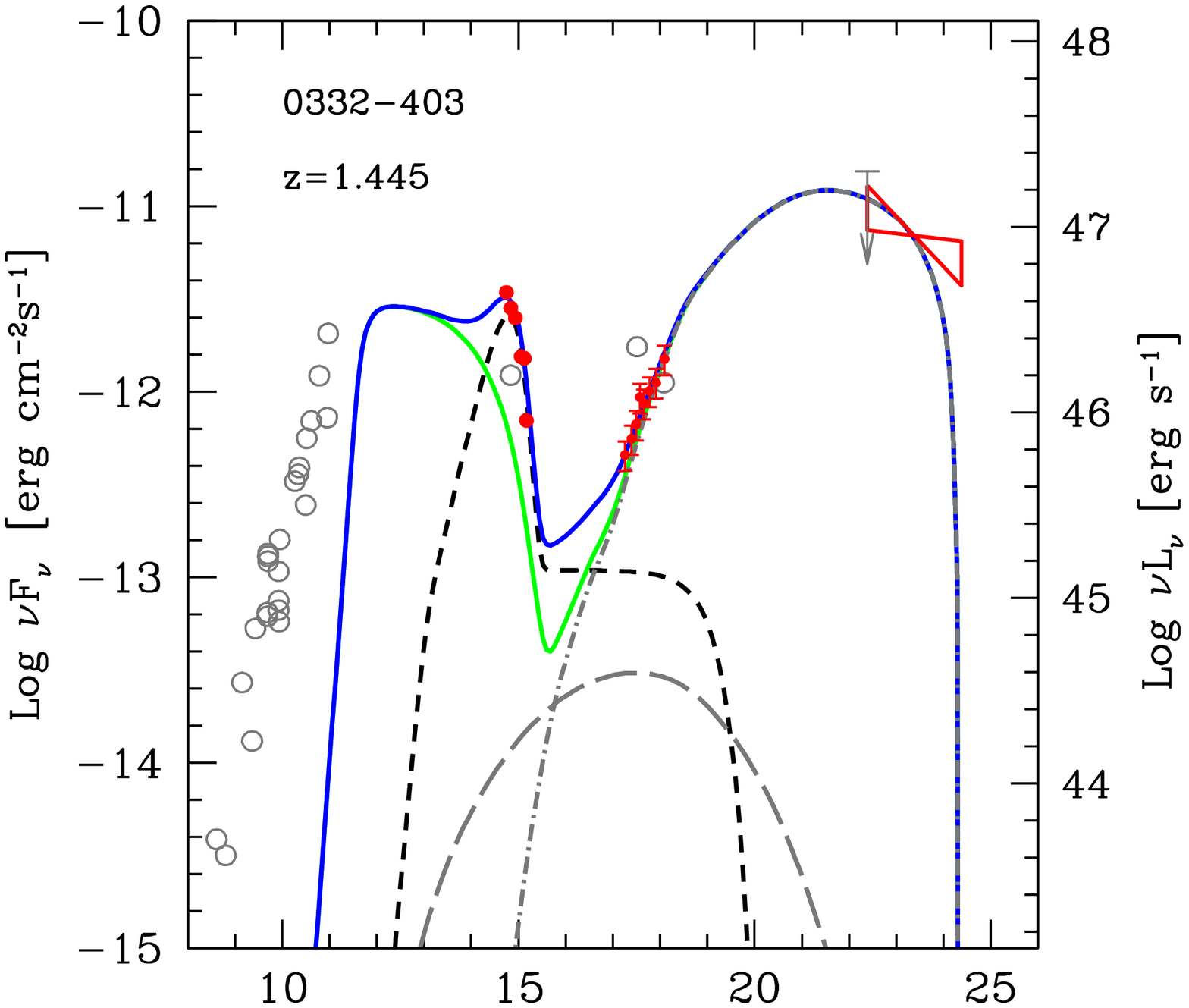,width=9cm,height=7cm}
\vskip -1.4cm
\psfig{figure=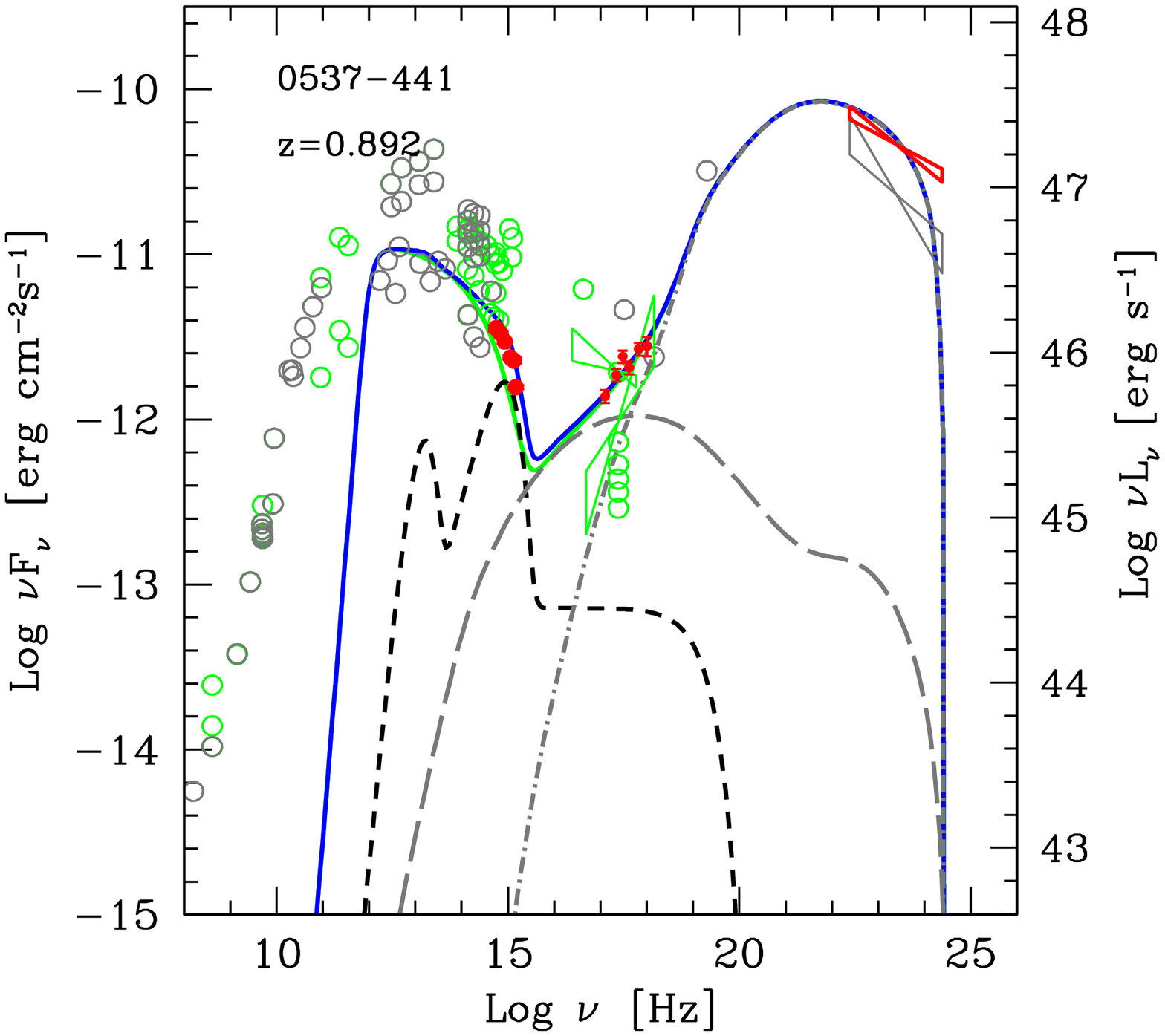,width=9cm,height=7cm}
\vskip -0.6 cm
\caption{SED of PKS 0208--512, B2 0218+35,
PKS 0332--403 and PKS 0537--441.
Symbols and lines as in Fig. \ref{f1}.
}
\label{f2}
\end{figure}

\begin{figure}
\vskip -0.2cm
\psfig{figure=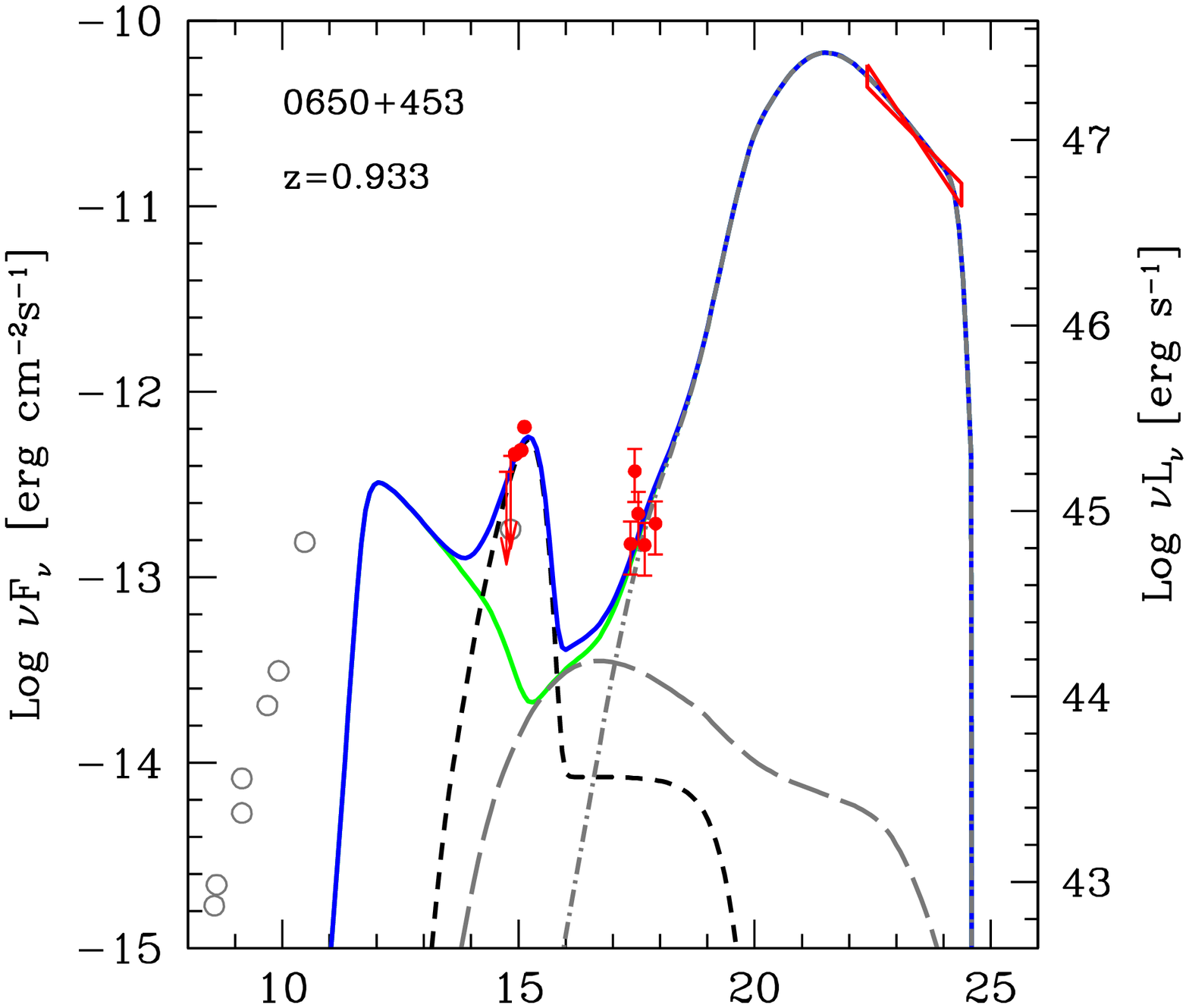,width=9cm,height=7cm}
\vskip -1.4cm
\psfig{figure=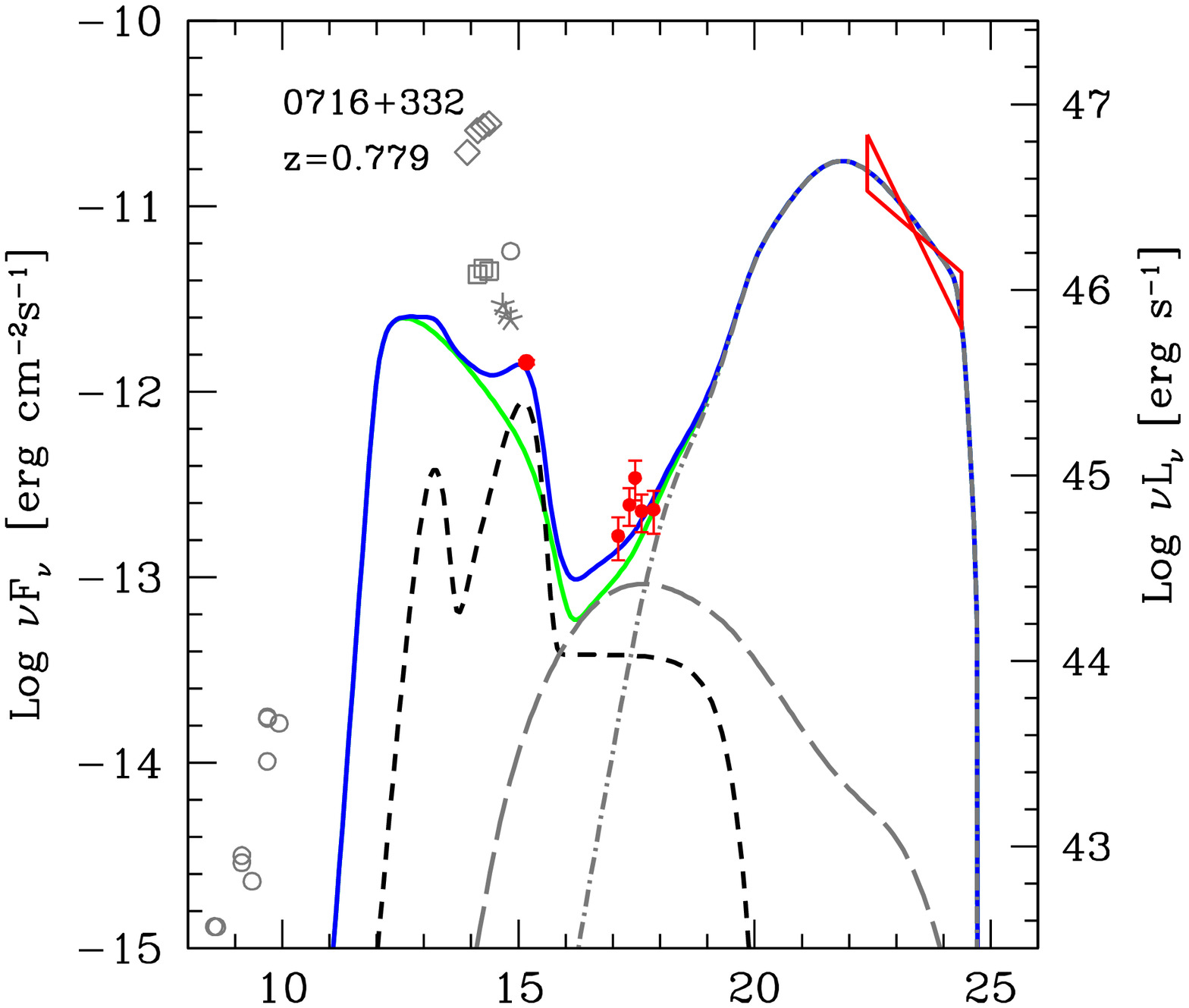,width=9cm,height=7cm}
\vskip -1.4cm
\psfig{figure=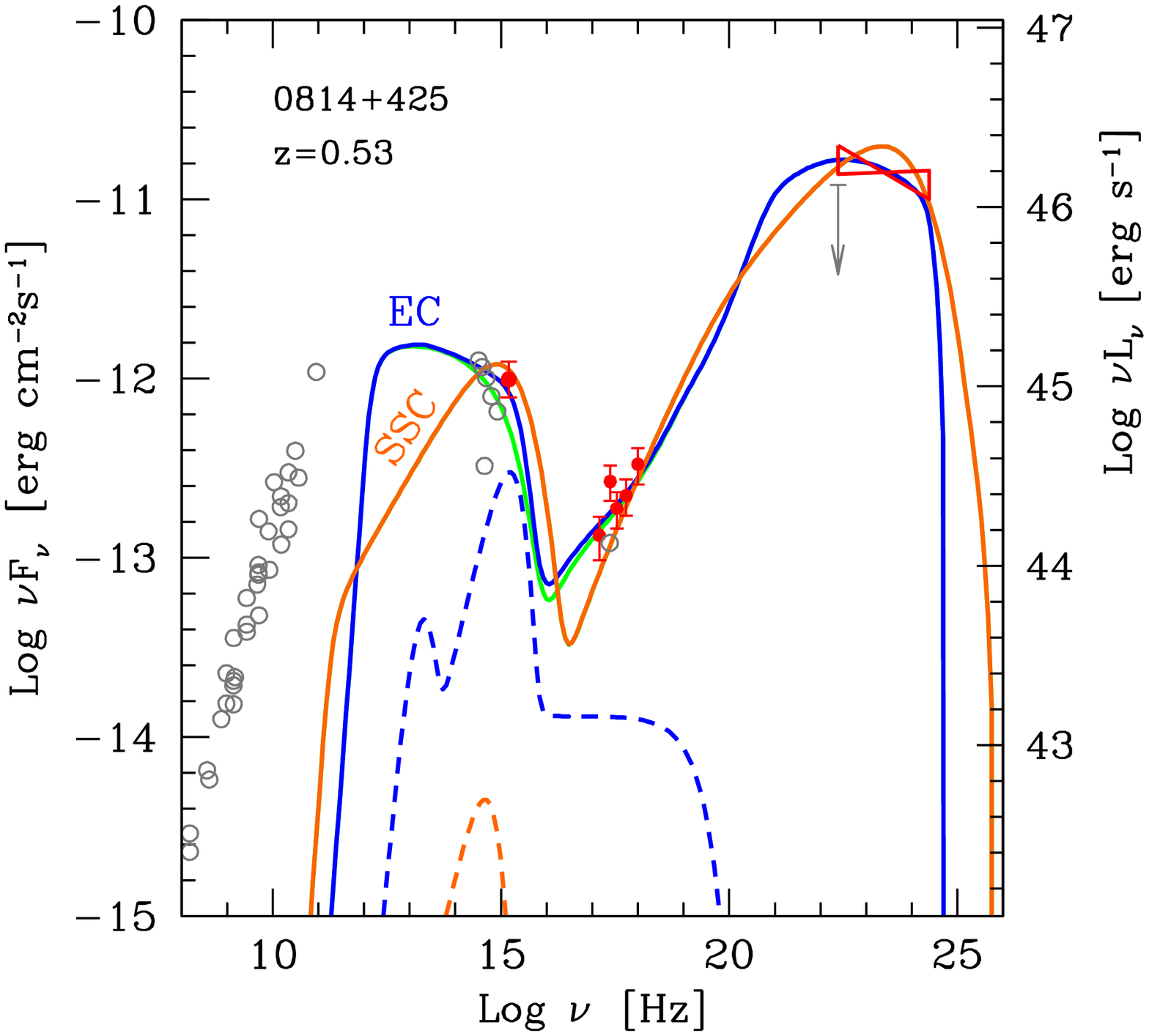,width=9cm,height=7cm}
\vskip -0.6 cm
\caption{SED of B3 0650+453, TXS 0716+332
and 0814+425 (OJ 425).
For the latter source we show a pure SSC model
and an EC one, as indicated.
Symbols and lines as in Fig. \ref{f1}.
}
\label{f3}
\end{figure}

\begin{figure}
\vskip -0.2cm
\psfig{figure=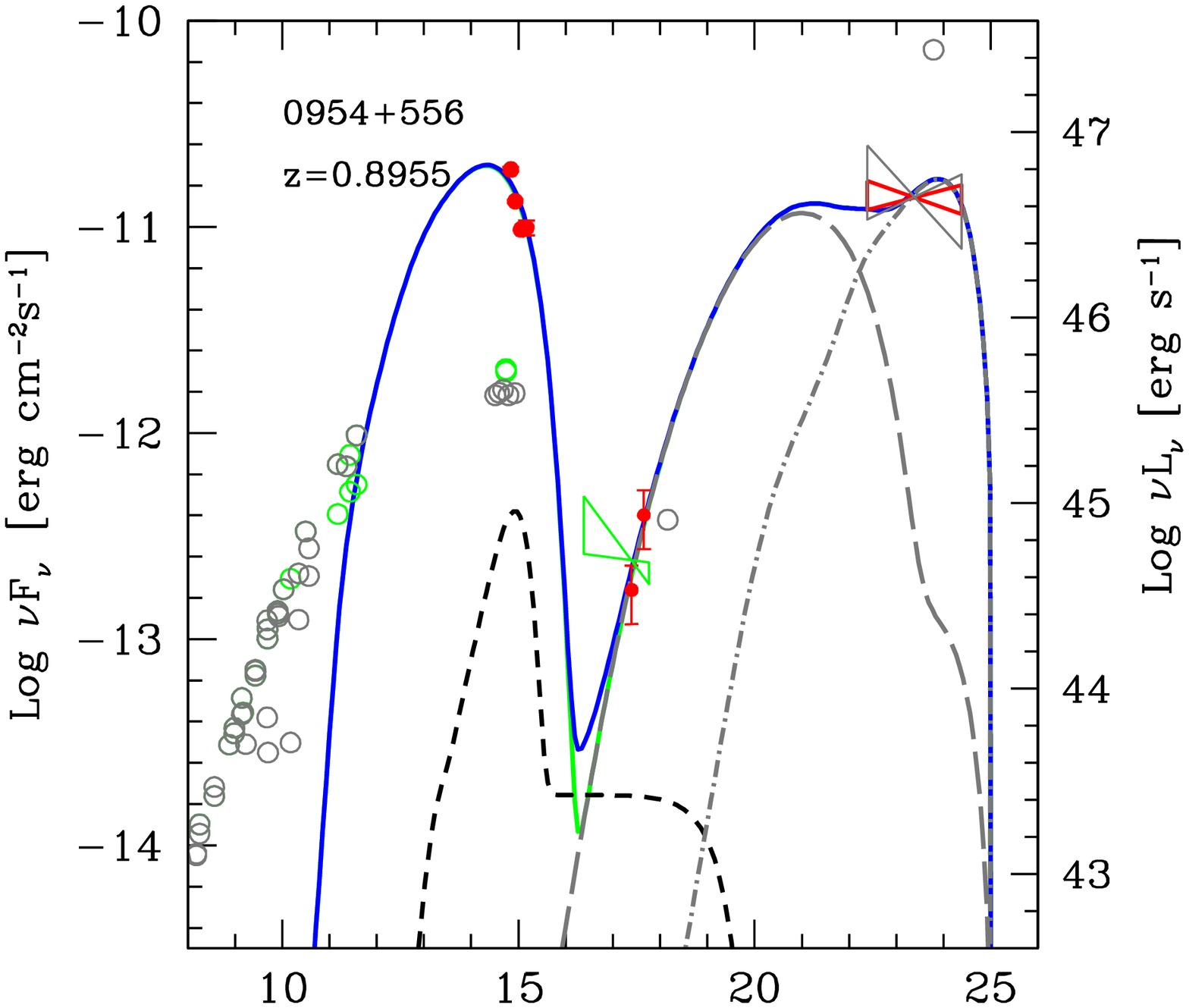,width=9cm,height=7cm}
\vskip -1.4cm
\psfig{figure=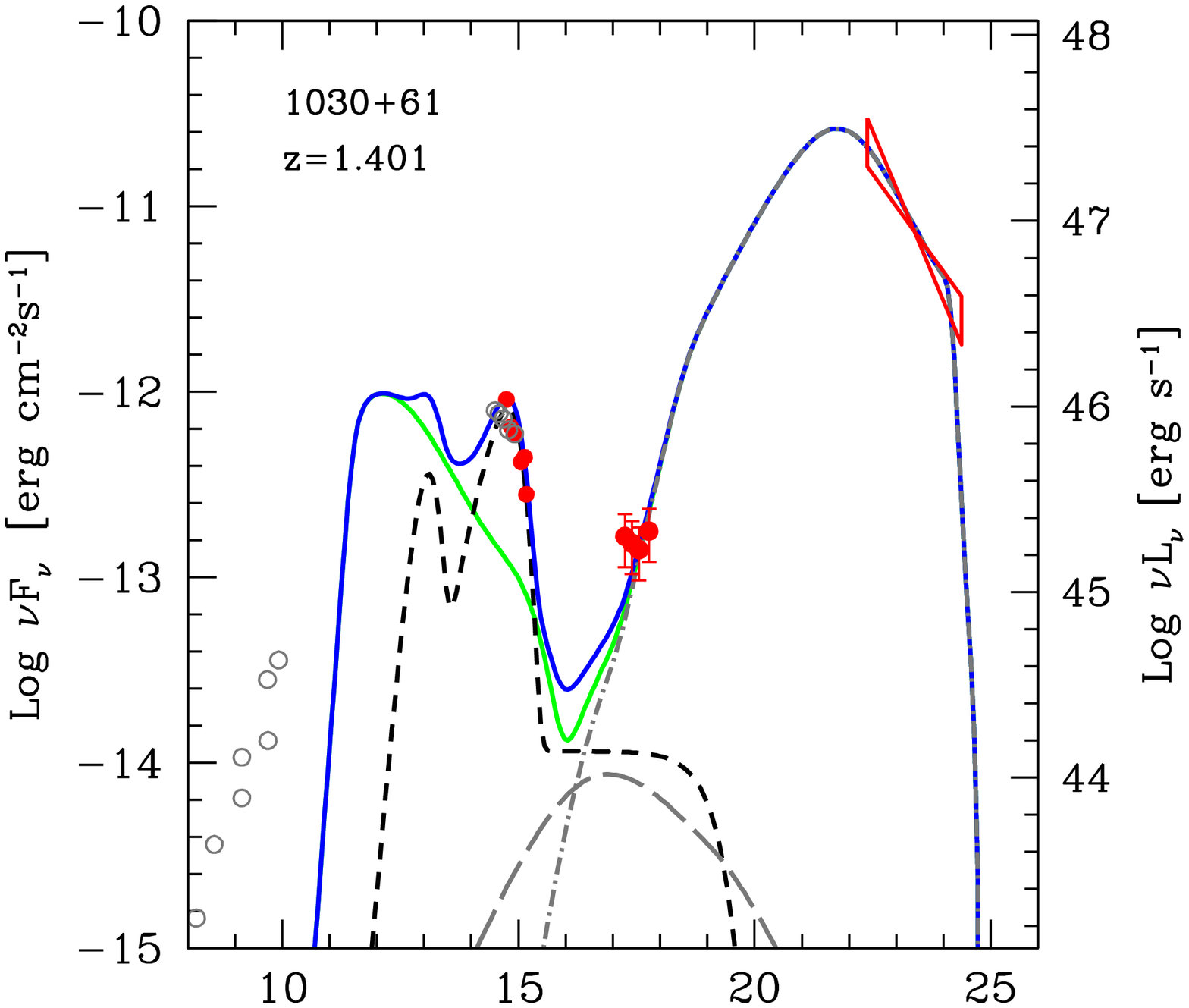,width=9cm,height=7cm}
\vskip -1.4 cm
\psfig{figure=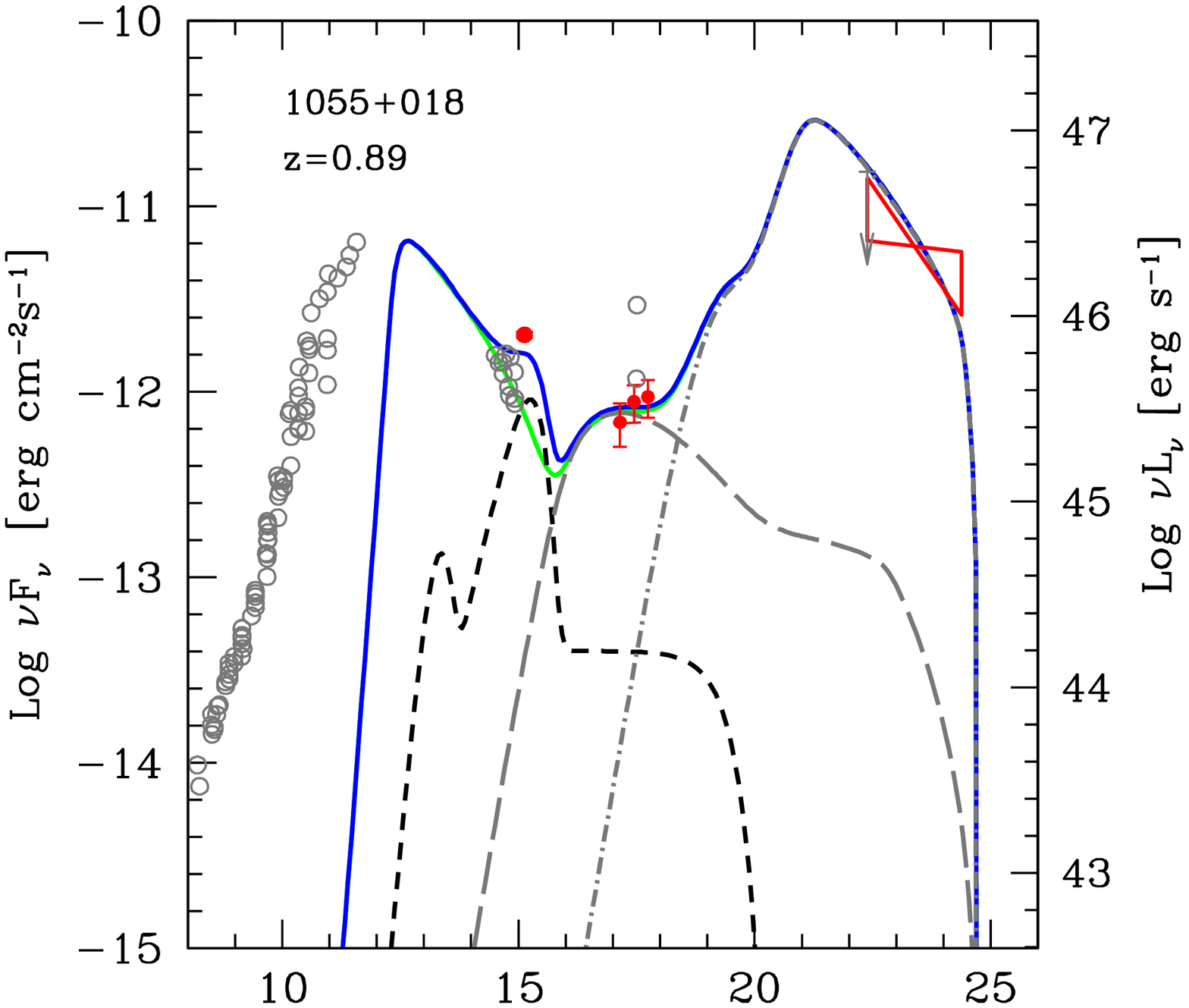,width=9cm,height=7cm}
\vskip -1.4 cm
\psfig{figure=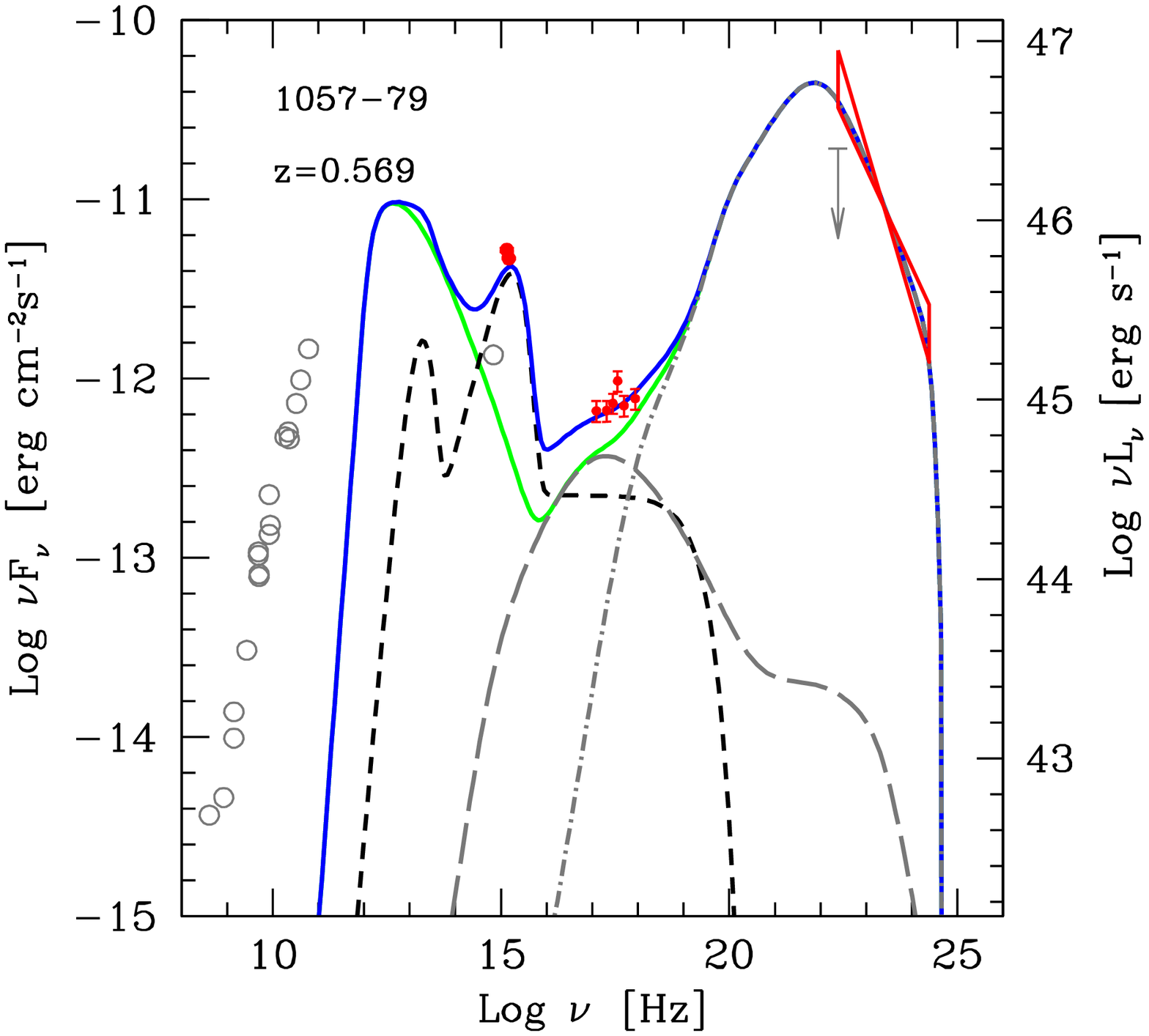,width=9cm,height=7cm}
\vskip -0.6 cm
\caption{SED of 0954+556 (4C 55.17),
S4 1030+61, PKS 1055+018 and PKS 1057--79.
Symbols and lines as in Fig. \ref{f1}.
}
\label{f4}
\end{figure}

\begin{figure}
\vskip -0.2cm
\psfig{figure=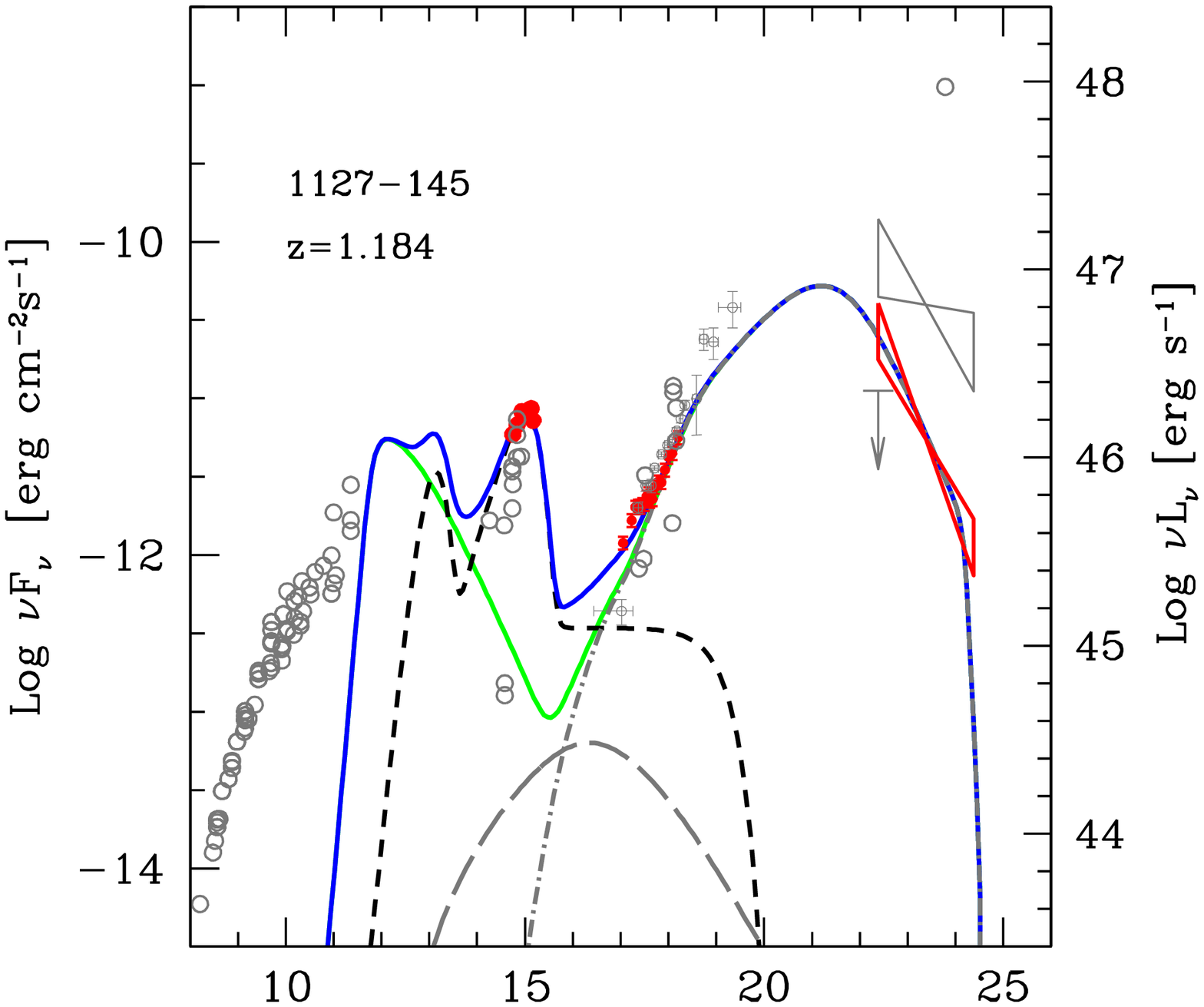,width=9cm,height=7cm}
\vskip -1.4 cm
\psfig{figure=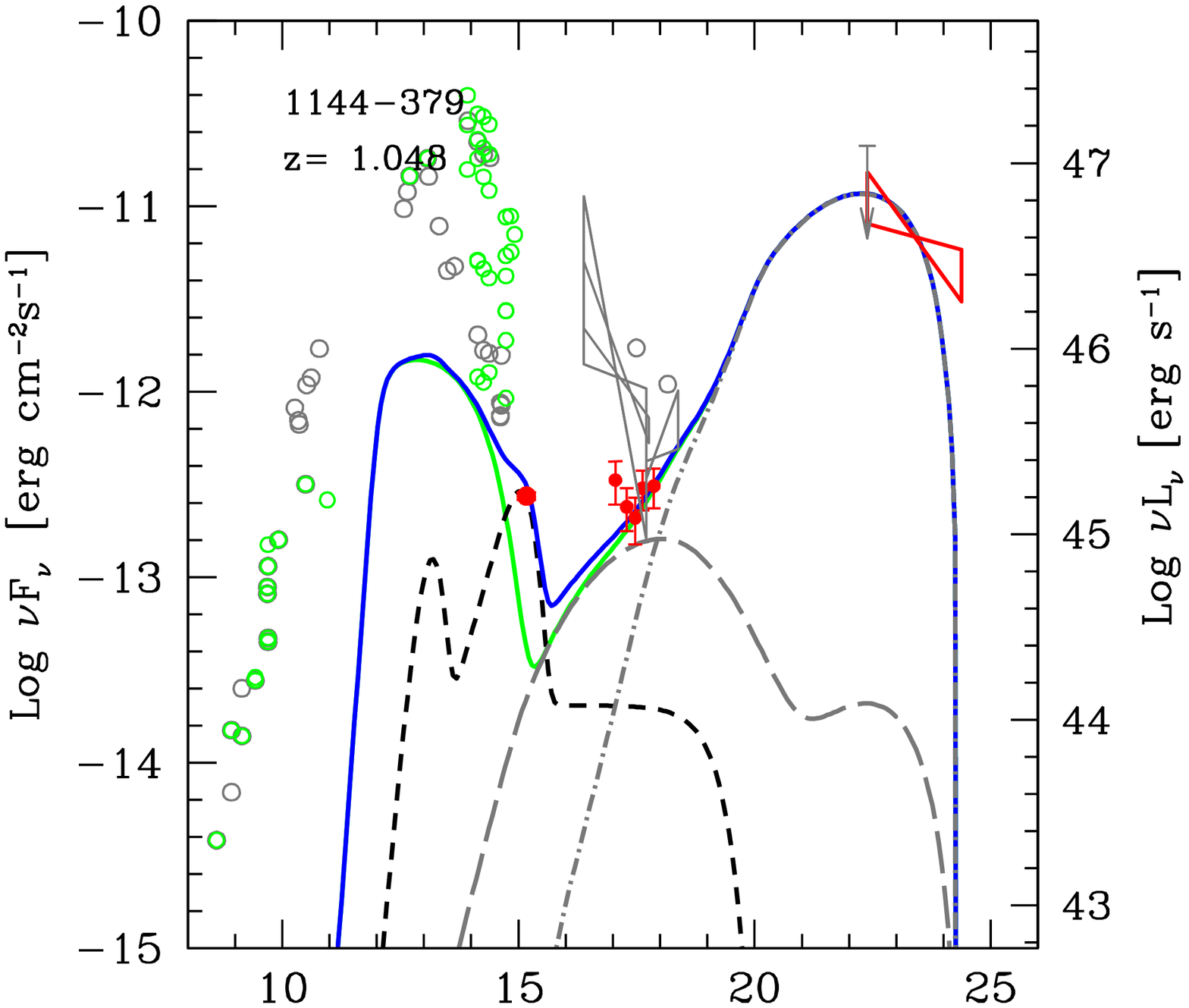,width=9cm,height=7cm}
\vskip -1.4 cm
\psfig{figure=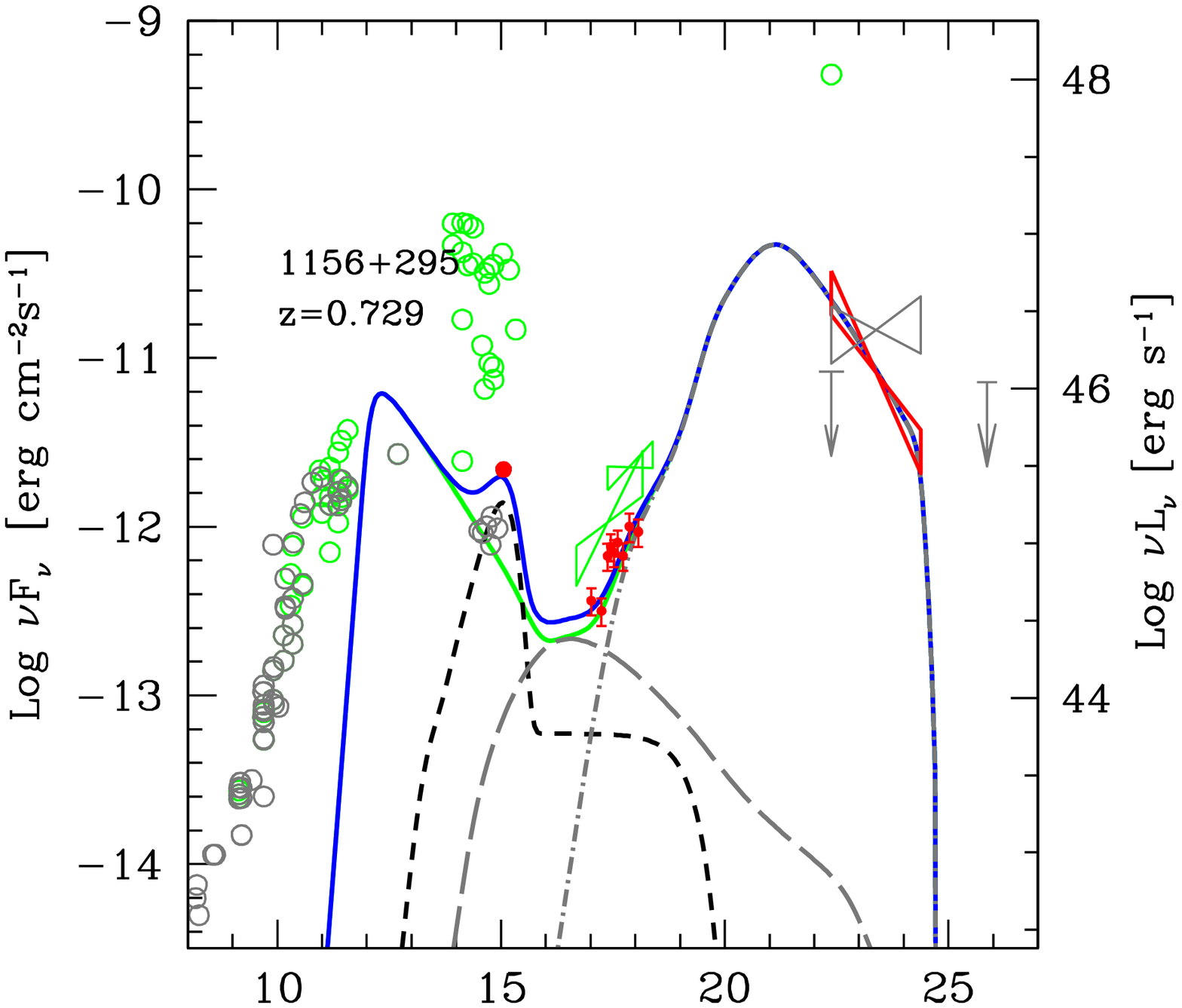,width=9cm,height=7cm}
\vskip -1.4 cm
\psfig{figure=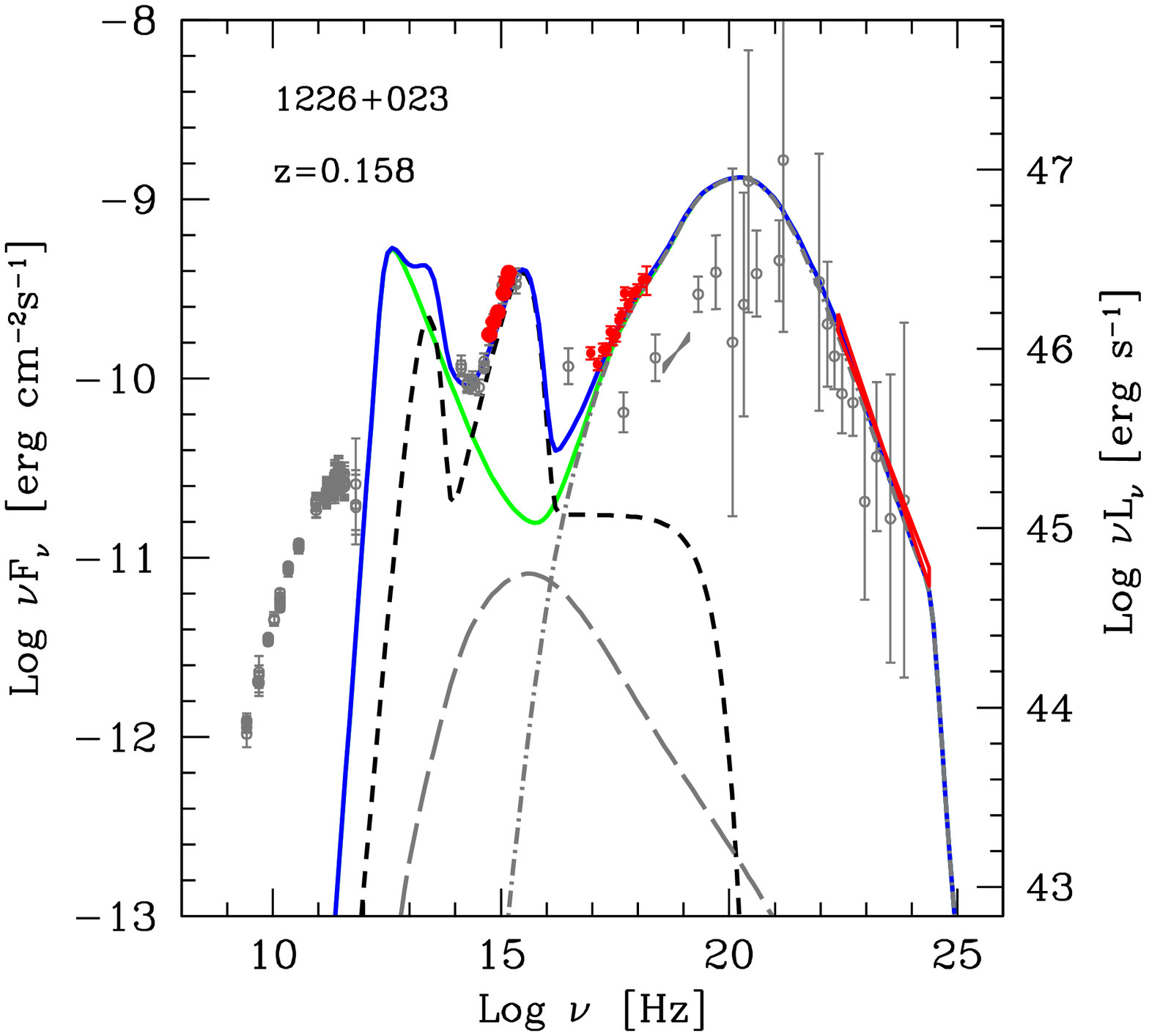,width=9cm,height=7cm}
\vskip -0.5 cm
\caption{SED of PKS 1127--145, PKS 1144--379, 1156+295 (4C 29.45)
and 1226+023 (3C 273).
Symbols and lines as in Fig. \ref{f1}.
}
\label{f5}  
\end{figure}

\begin{figure}
\vskip -0.2cm
\psfig{figure=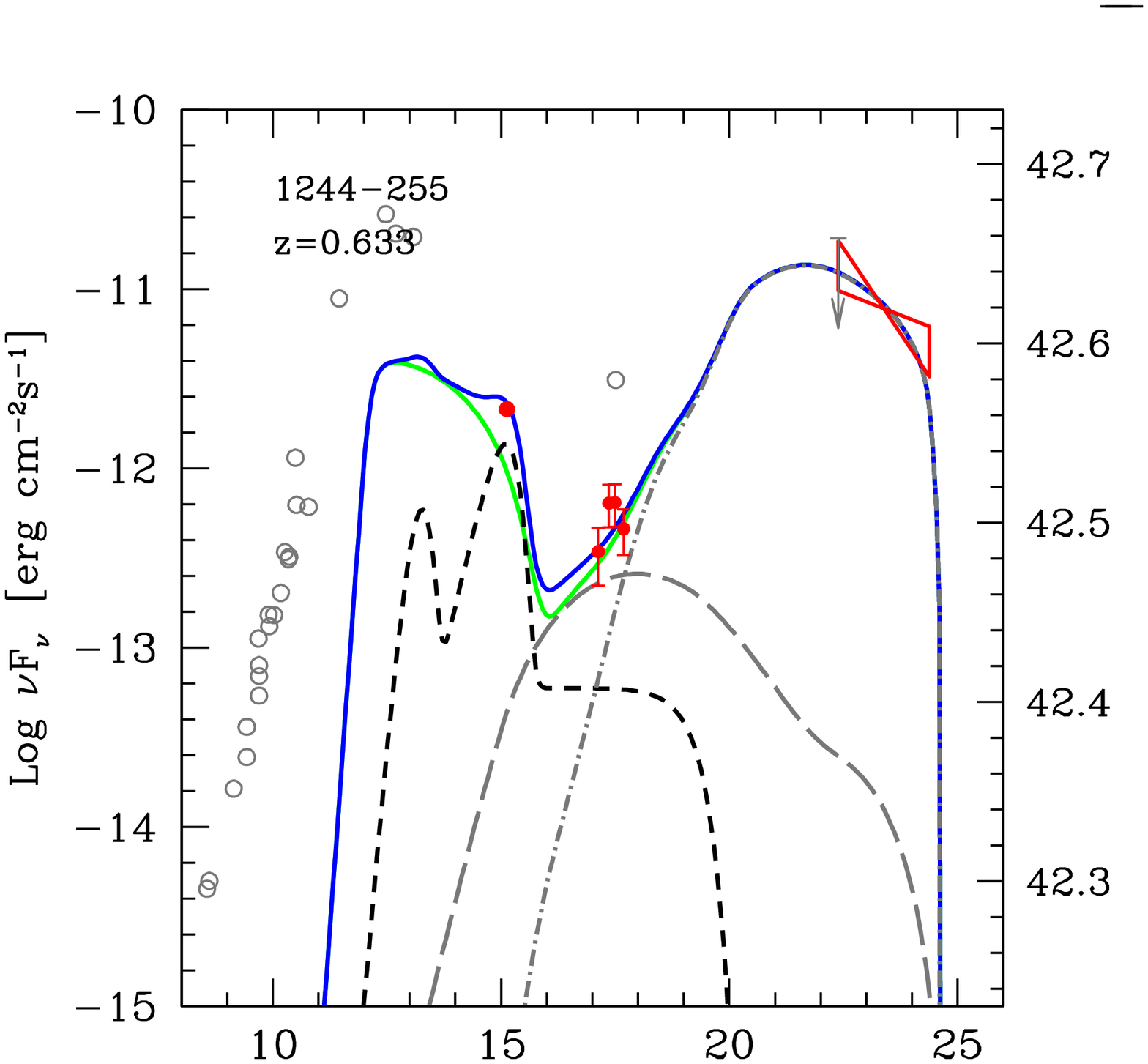,width=9cm,height=7cm}
\vskip -1.4 cm
\psfig{figure=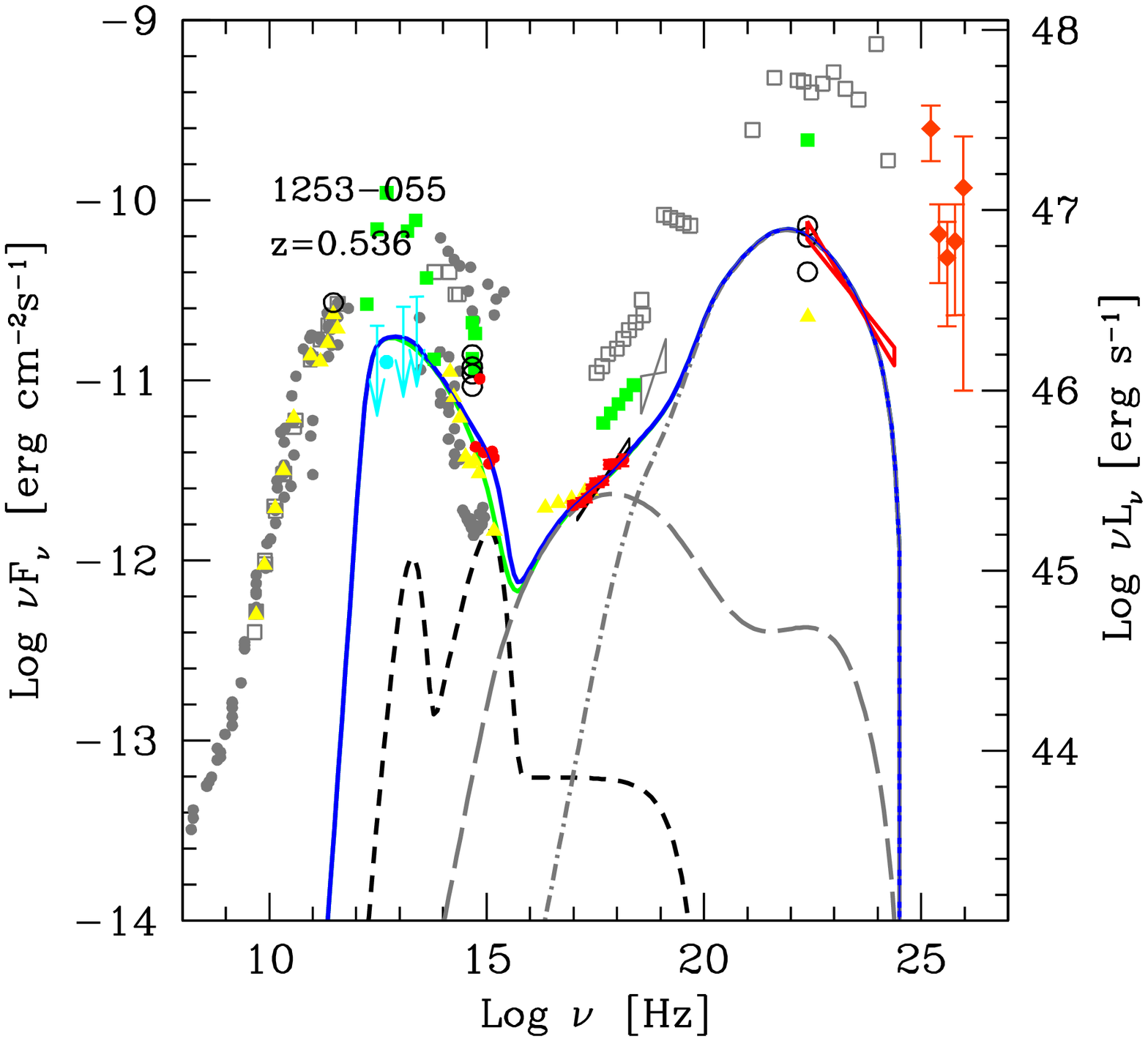,width=9cm,height=7cm}
\vskip -1.4 cm
\psfig{figure=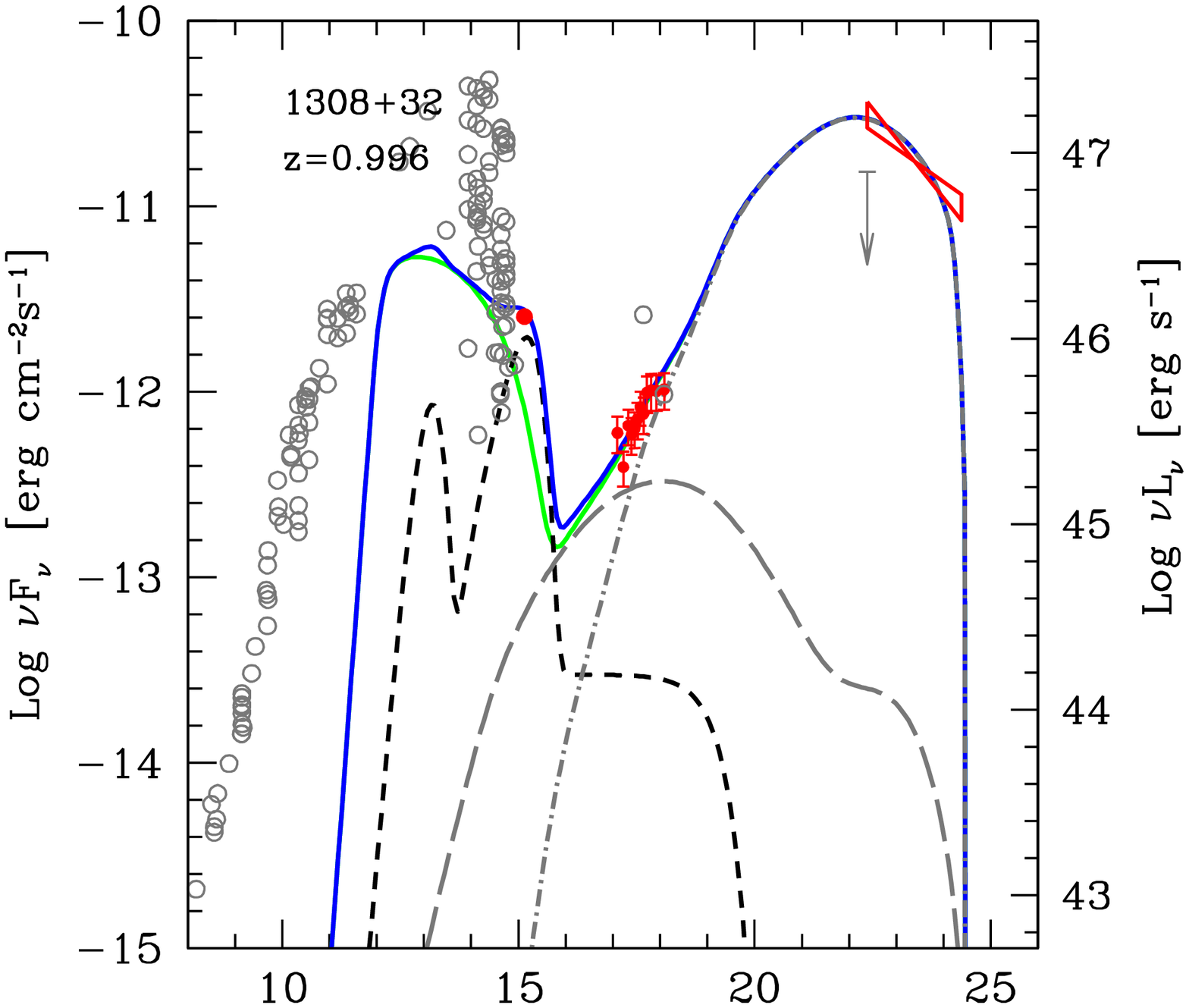,width=9cm,height=7cm}
\vskip -1.4 cm
\psfig{figure=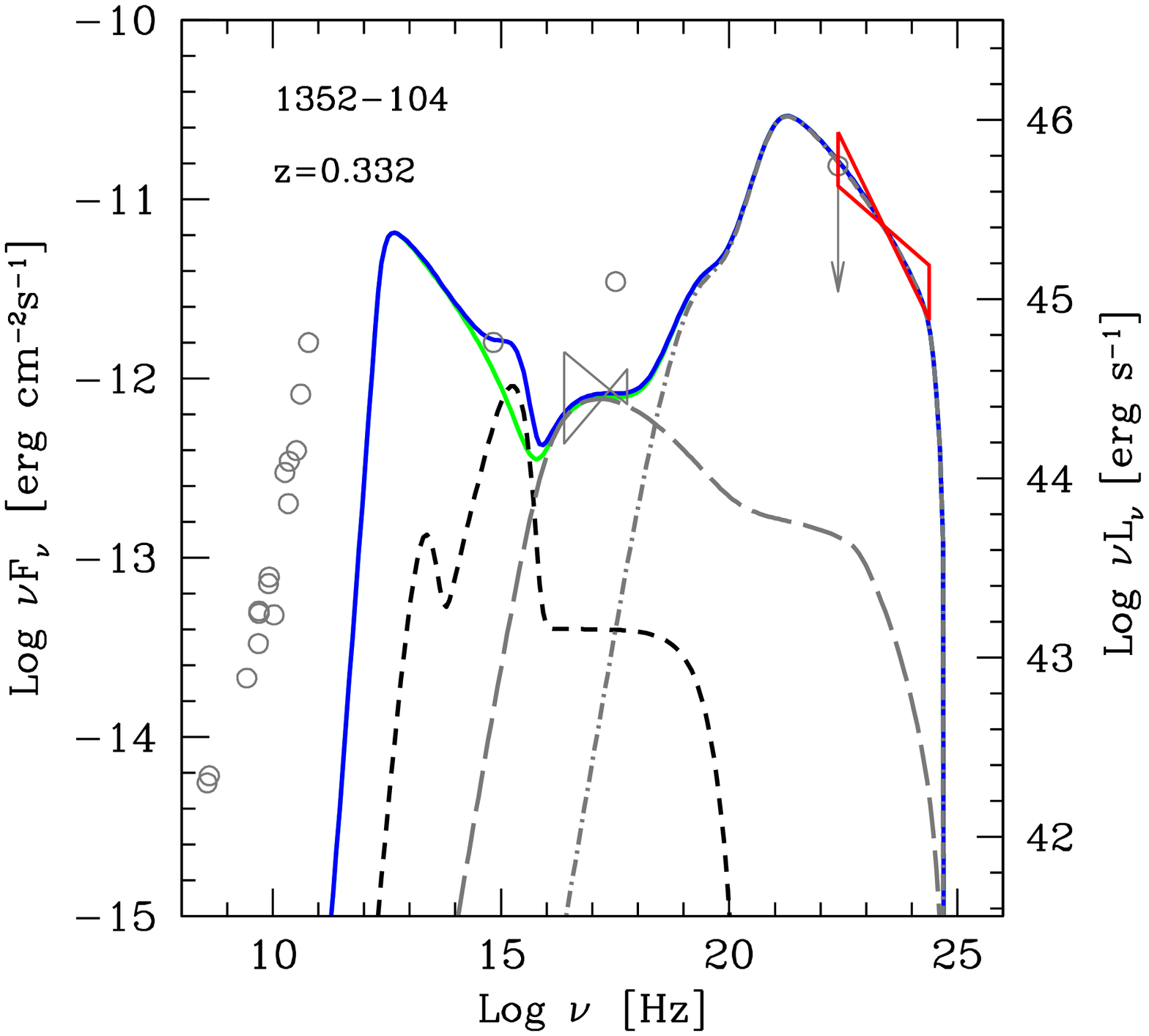,width=9cm,height=7cm}
\vskip -0.5 cm
\caption{SED of PKS 1244--255, 1253--055 (3C 279),
B2 1308+32 and PKS 1352--104.
Symbols and lines as in Fig. \ref{f1}.
}
\label{f6}  
\end{figure}

\begin{figure}
\vskip -0.2cm
\psfig{figure=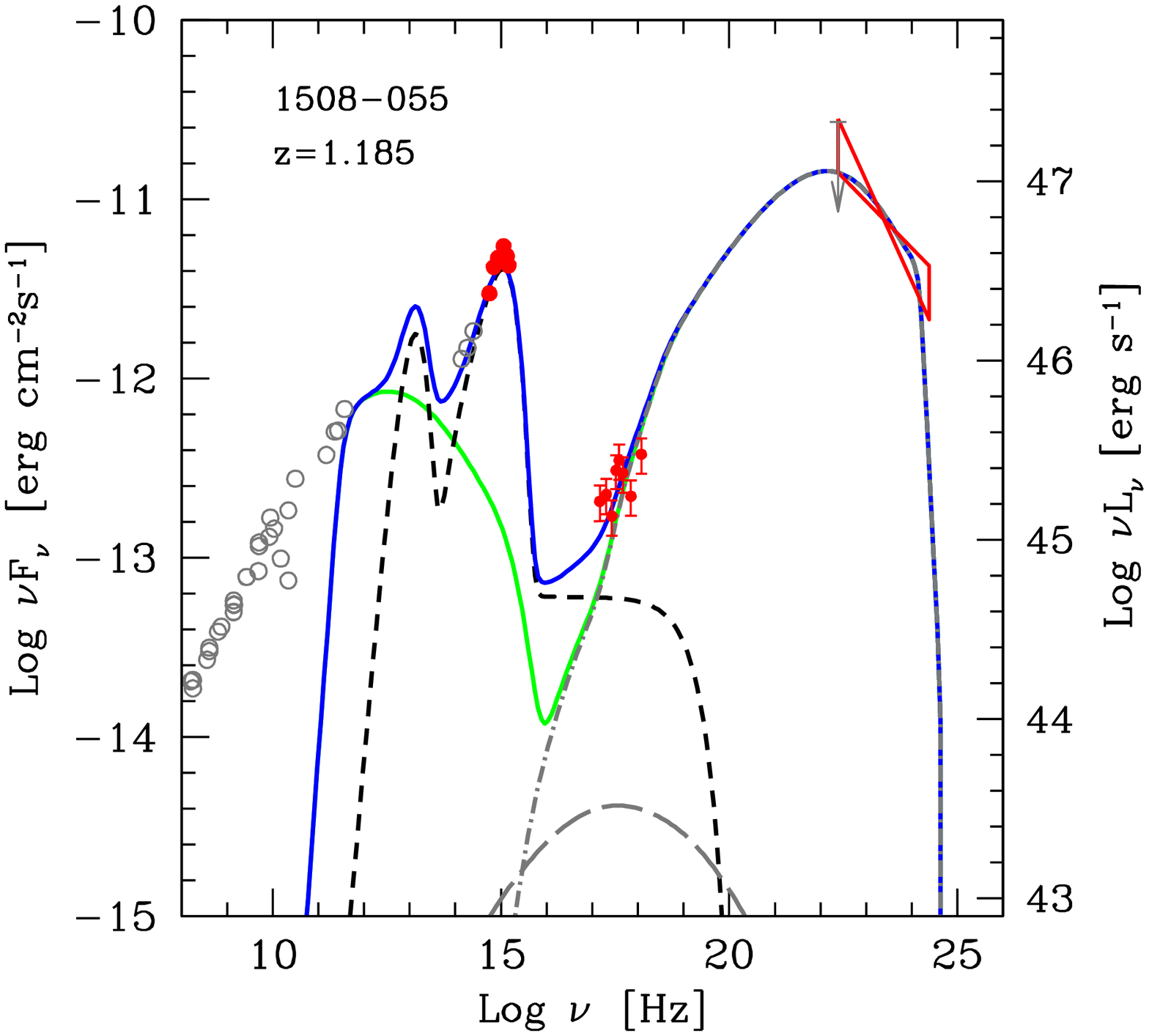,width=9cm,height=7cm}
\vskip -1.4 cm
\psfig{figure=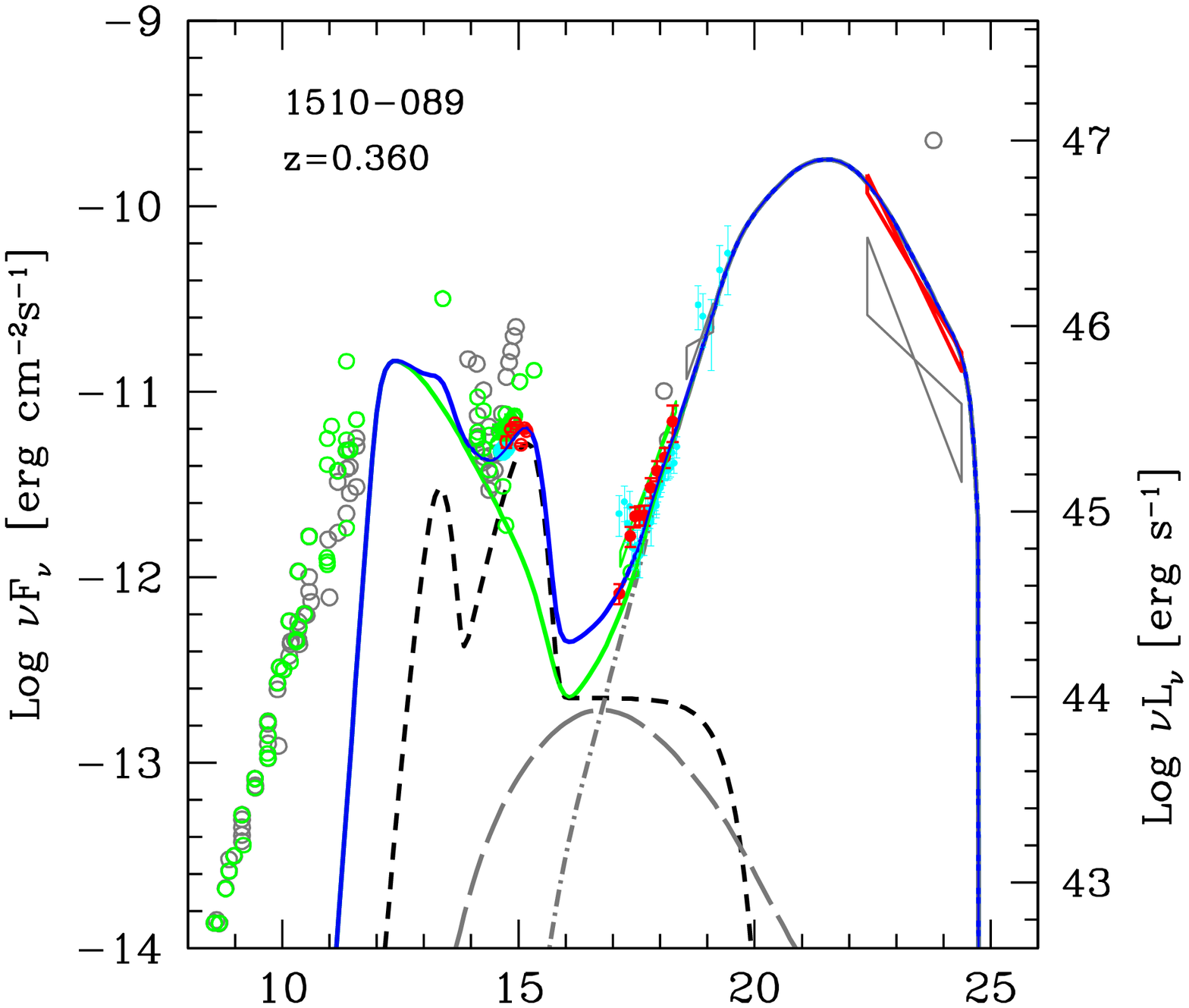,width=9cm,height=7cm}
\vskip -1.4 cm
\psfig{figure=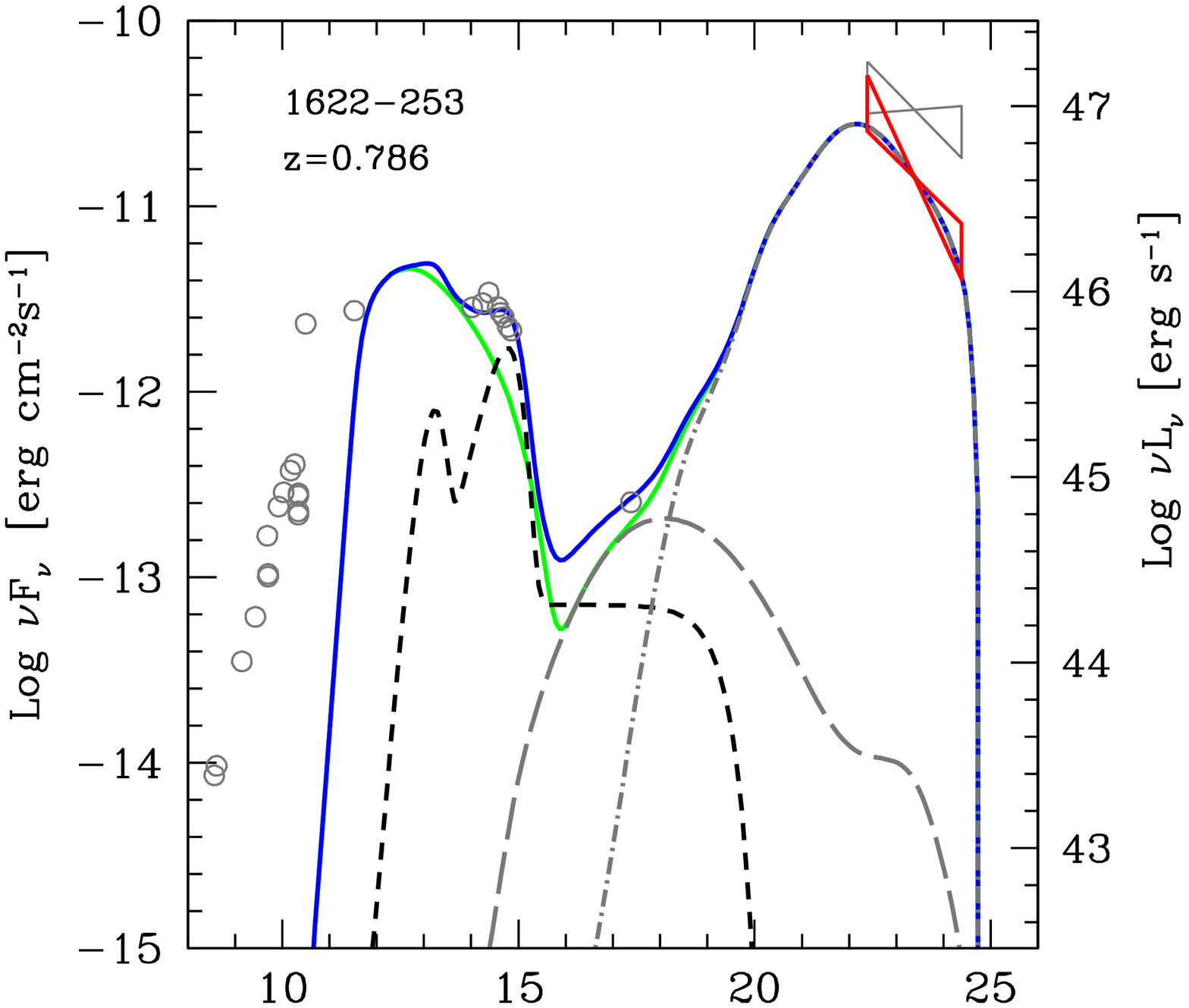,width=9cm,height=7cm}
\vskip -1.4 cm
\psfig{figure=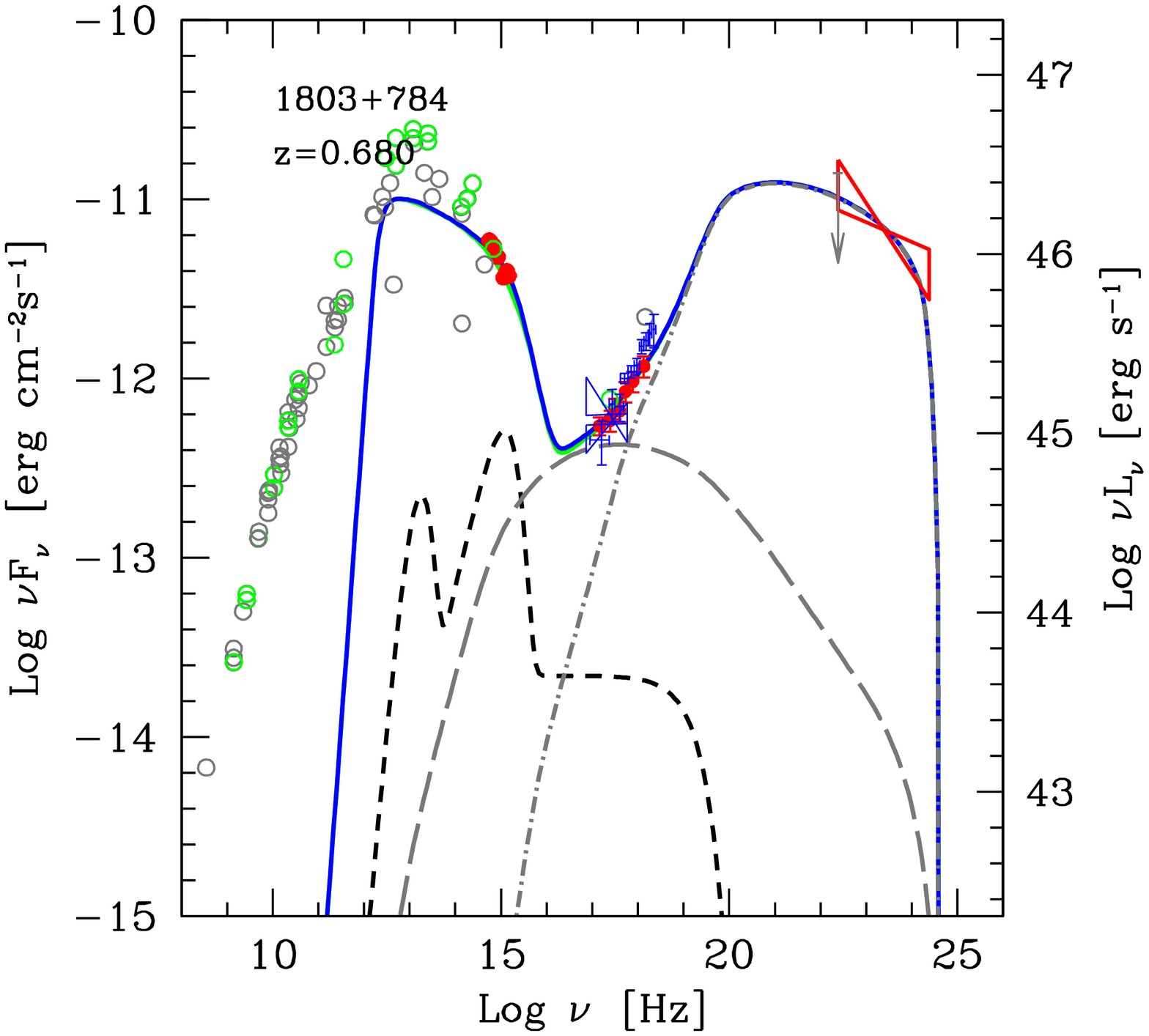,width=9cm,height=7cm}
\vskip -0.5 cm
\caption{SED of PKS 1508--055, PKS 1510--089,
PKS 1622--253 and S5 1803+784.
Symbols and lines as in Fig. \ref{f1}.
}
\label{f7}  
\end{figure}

\begin{figure}
\vskip -0.2cm
\psfig{figure=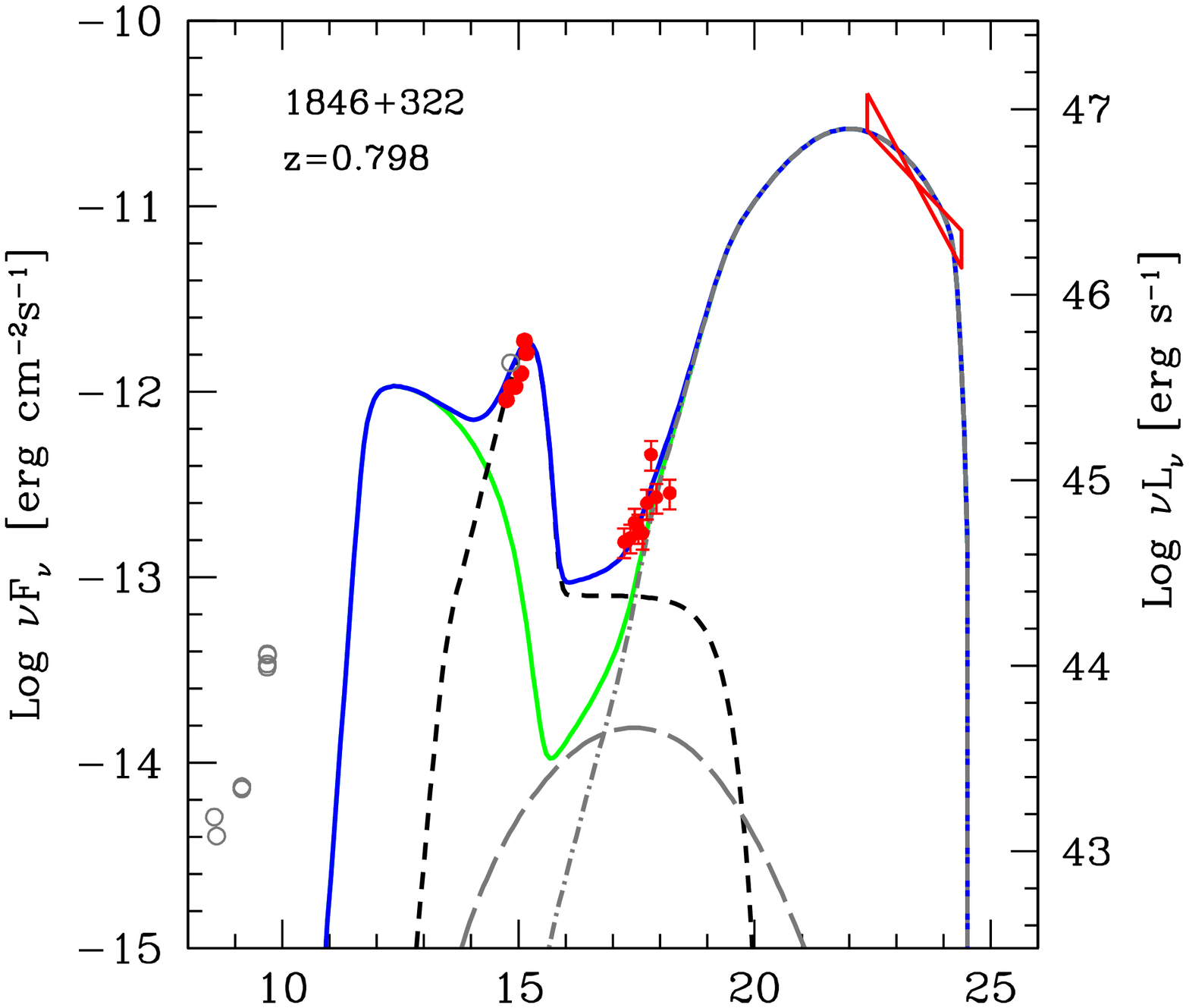,width=9cm,height=7cm}
\vskip -1.4 cm
\psfig{figure=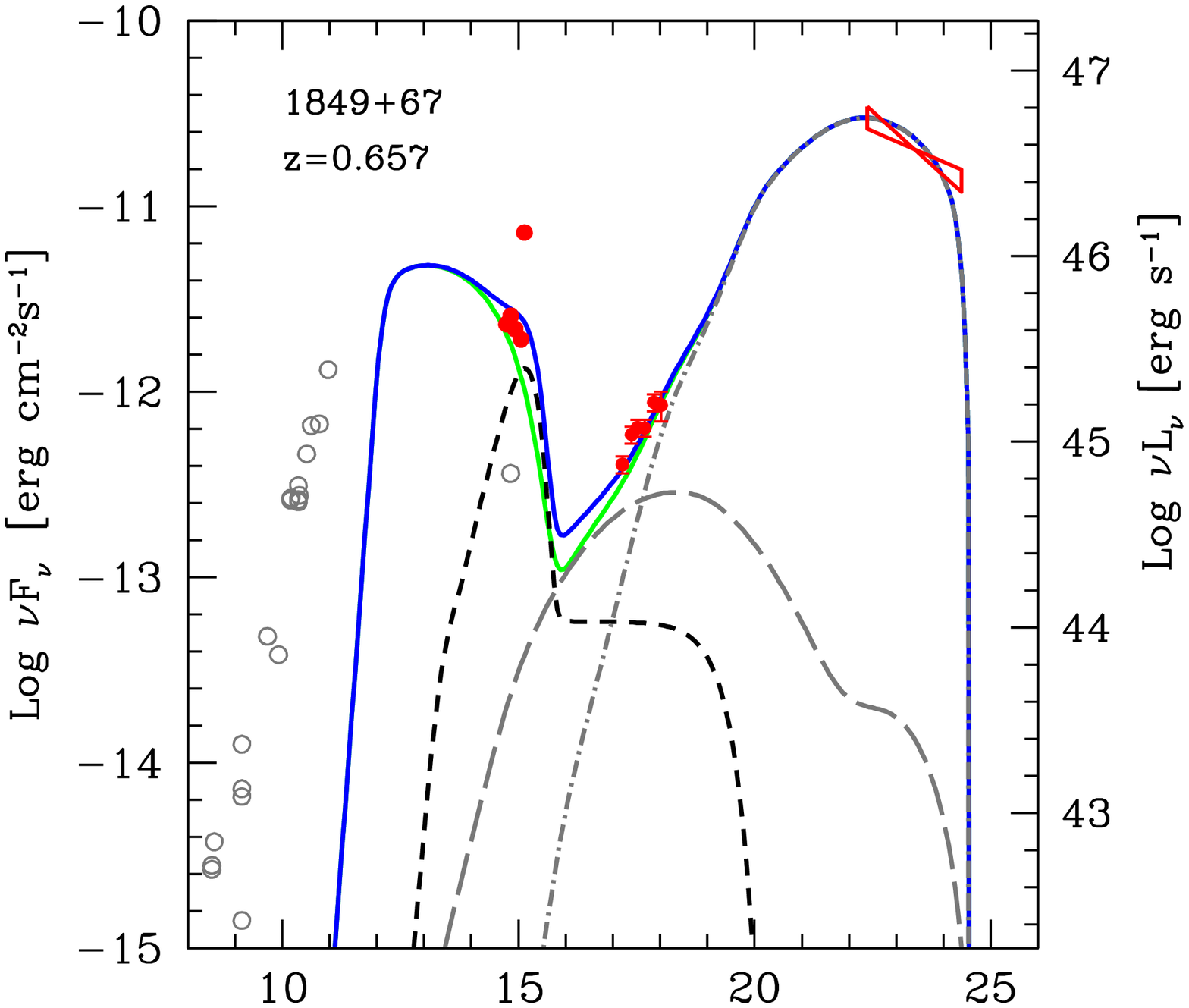,width=9cm,height=7cm}
\vskip -1.4 cm
\psfig{figure=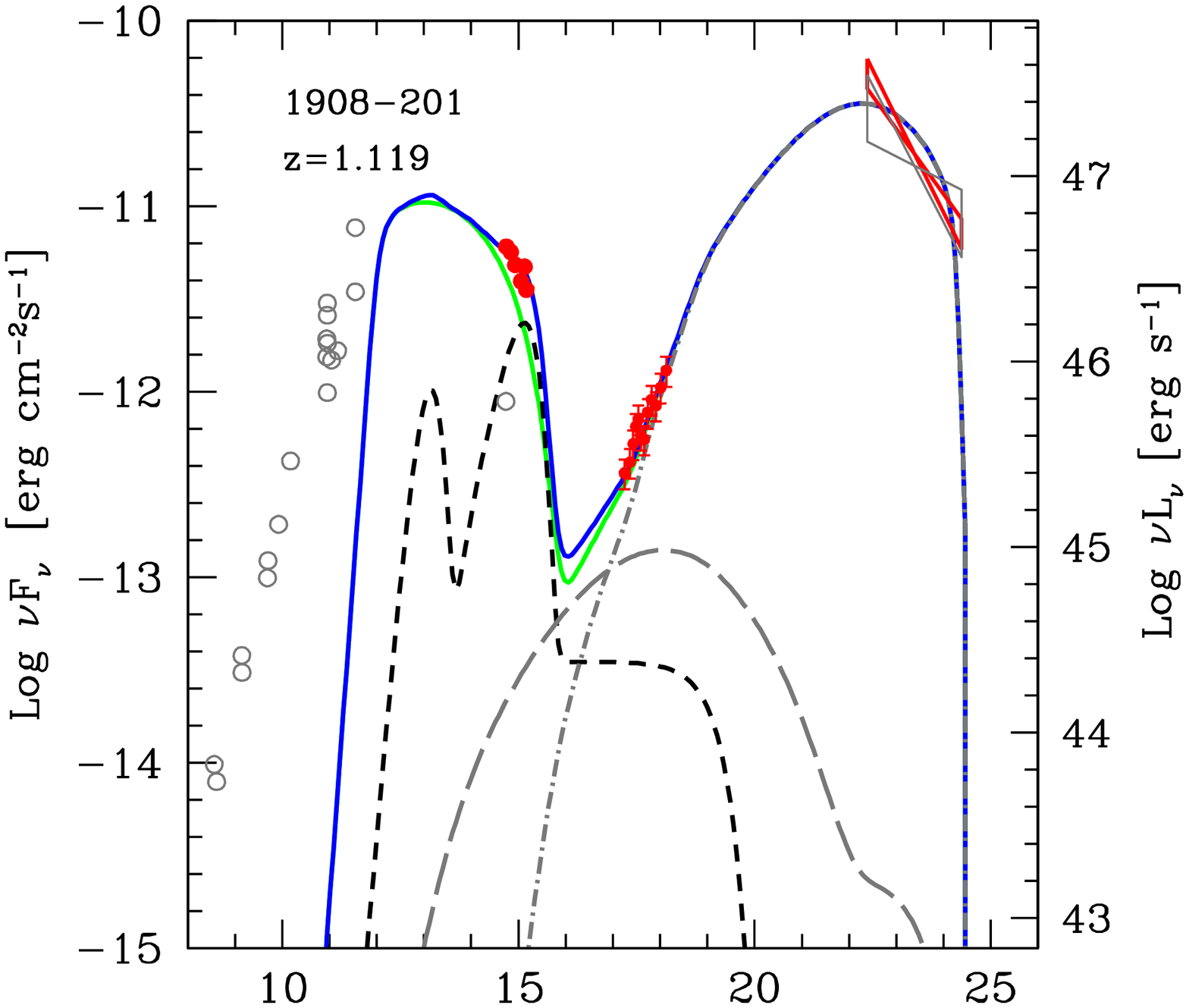,width=9cm,height=7cm}
\vskip -1.4 cm
\psfig{figure=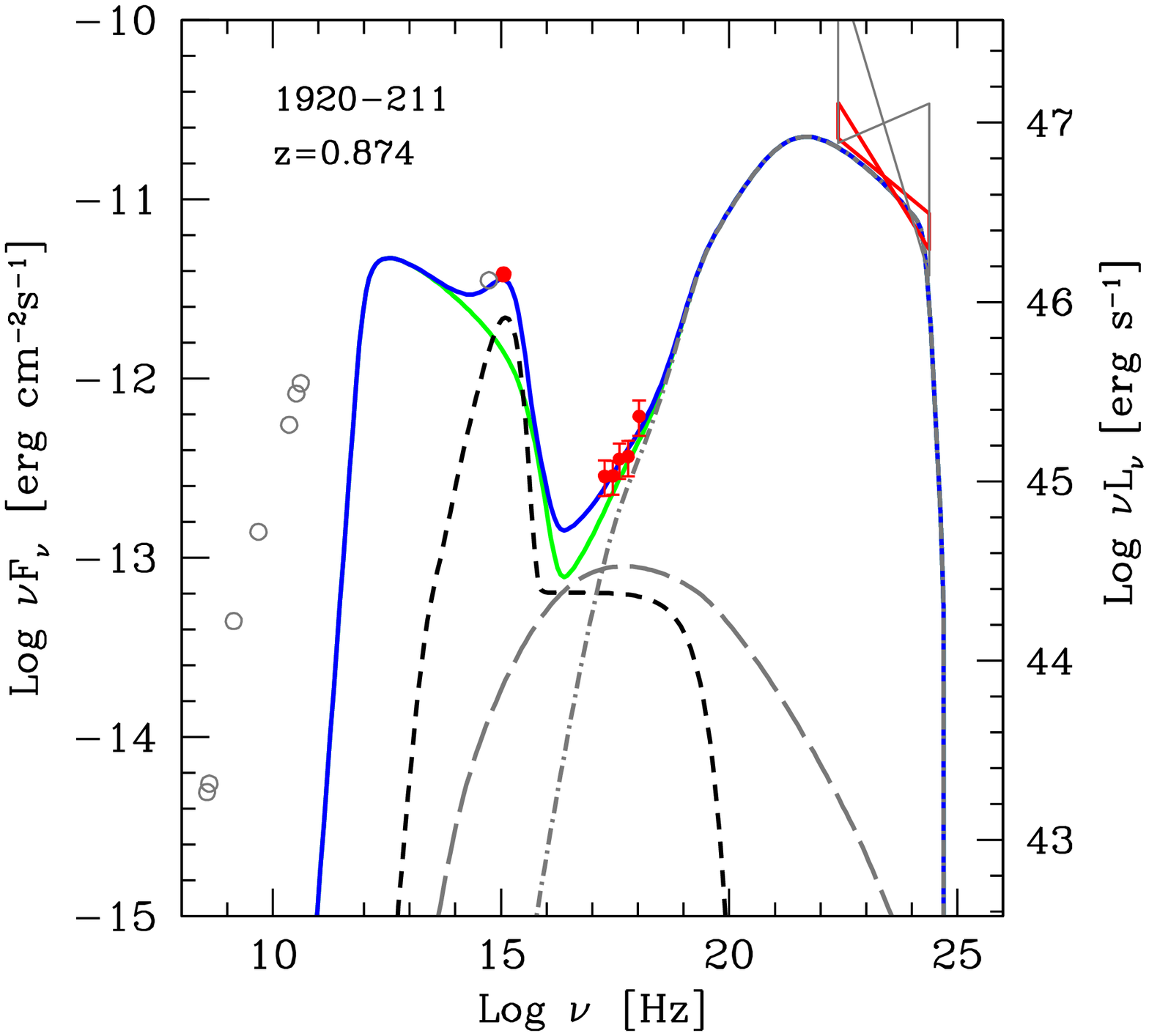,width=9cm,height=7cm}
\vskip -0.5 cm
\caption{SED of TXS 1846+322, S4 1849+67,
PKS 1908--201 and TXS 1920--211.
Symbols and lines as in Fig. \ref{f1}.
}
\label{f8}  
\end{figure}

\begin{figure}
\vskip -0.2cm
\psfig{figure=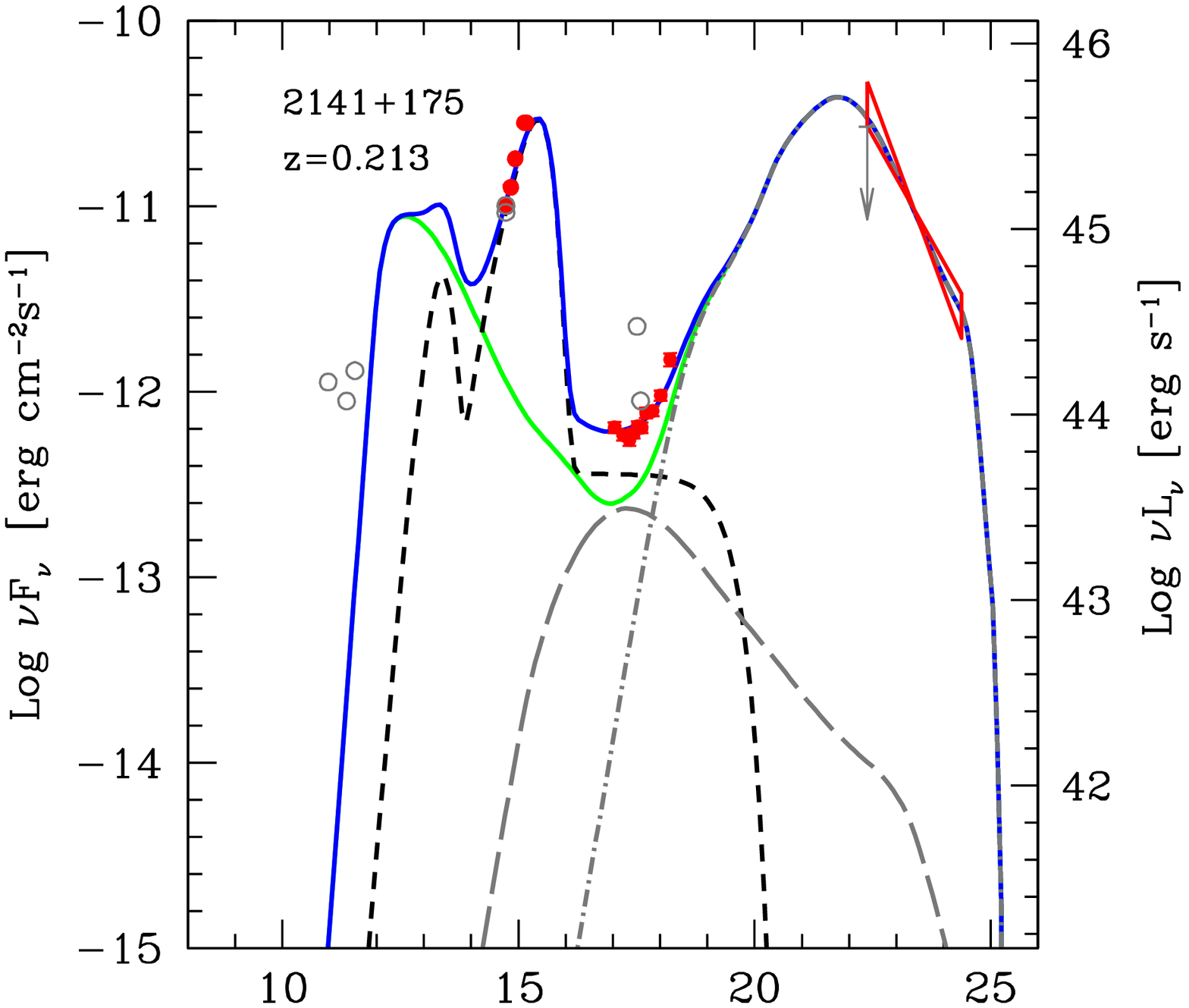,width=9cm,height=7cm}
\vskip -1.4 cm
\psfig{figure=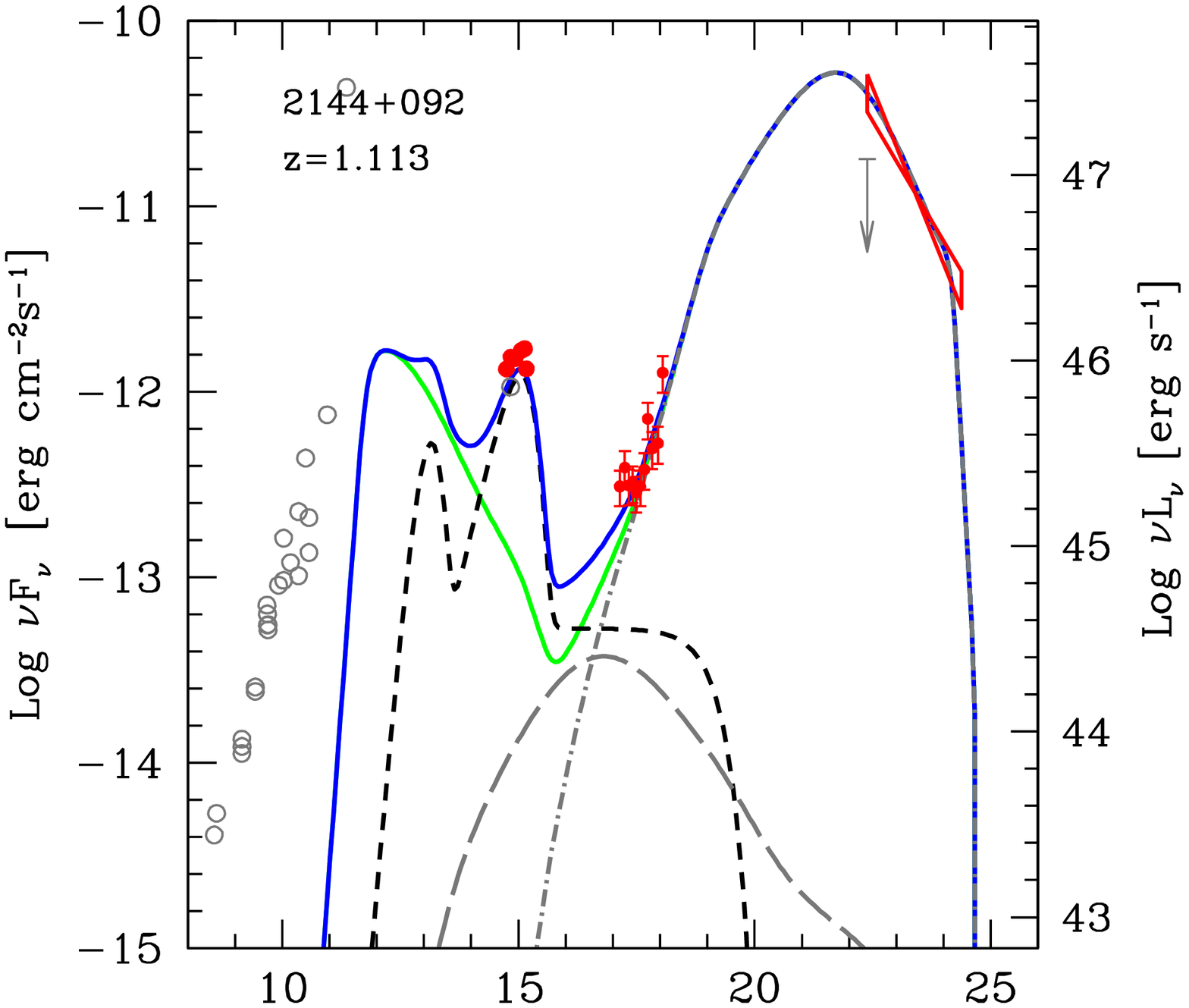,width=9cm,height=7cm}
\vskip -1.4 cm
\psfig{figure=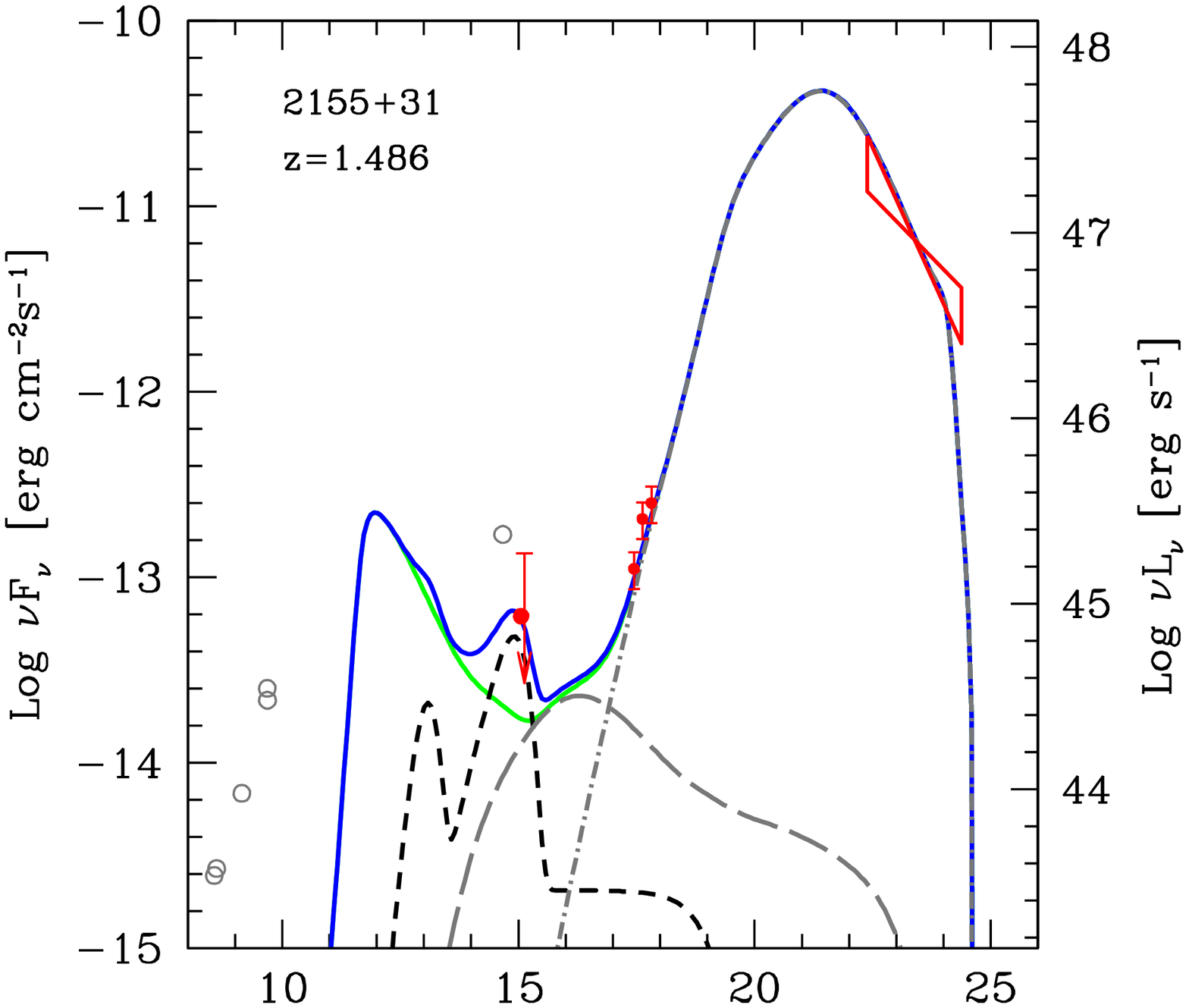,width=9cm,height=7cm}
\vskip -1.4 cm
\psfig{figure=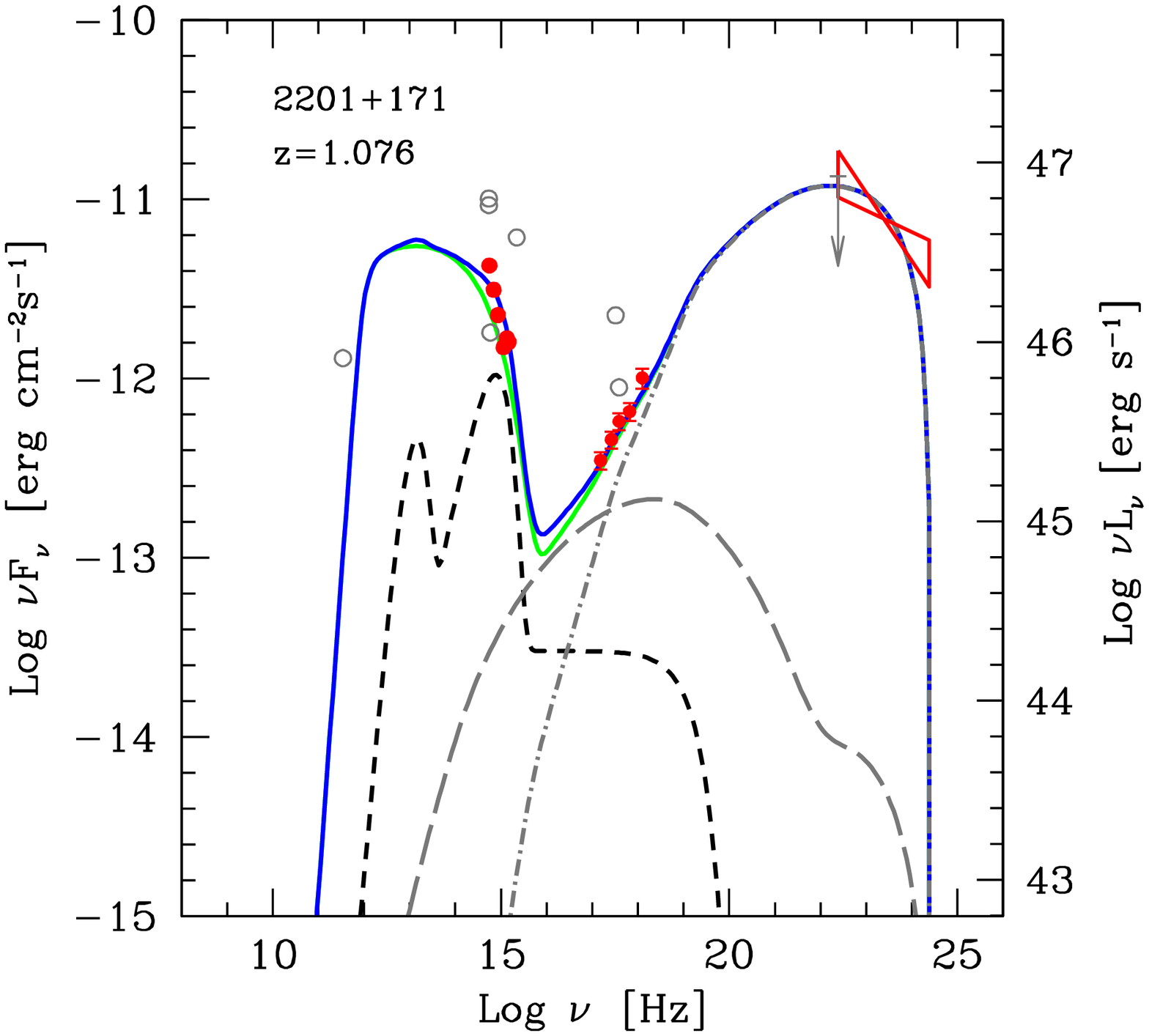,width=9cm,height=7cm}
\vskip -0.5 cm
\caption{SED of 2141+175 (OX 169), PKS 2144+092,
B2 2155+31 and PKS 2201+171.
Symbols and lines as in Fig. \ref{f1}.
}
\label{f9}  
\end{figure}

\begin{figure}
\vskip -0.2cm
\psfig{figure=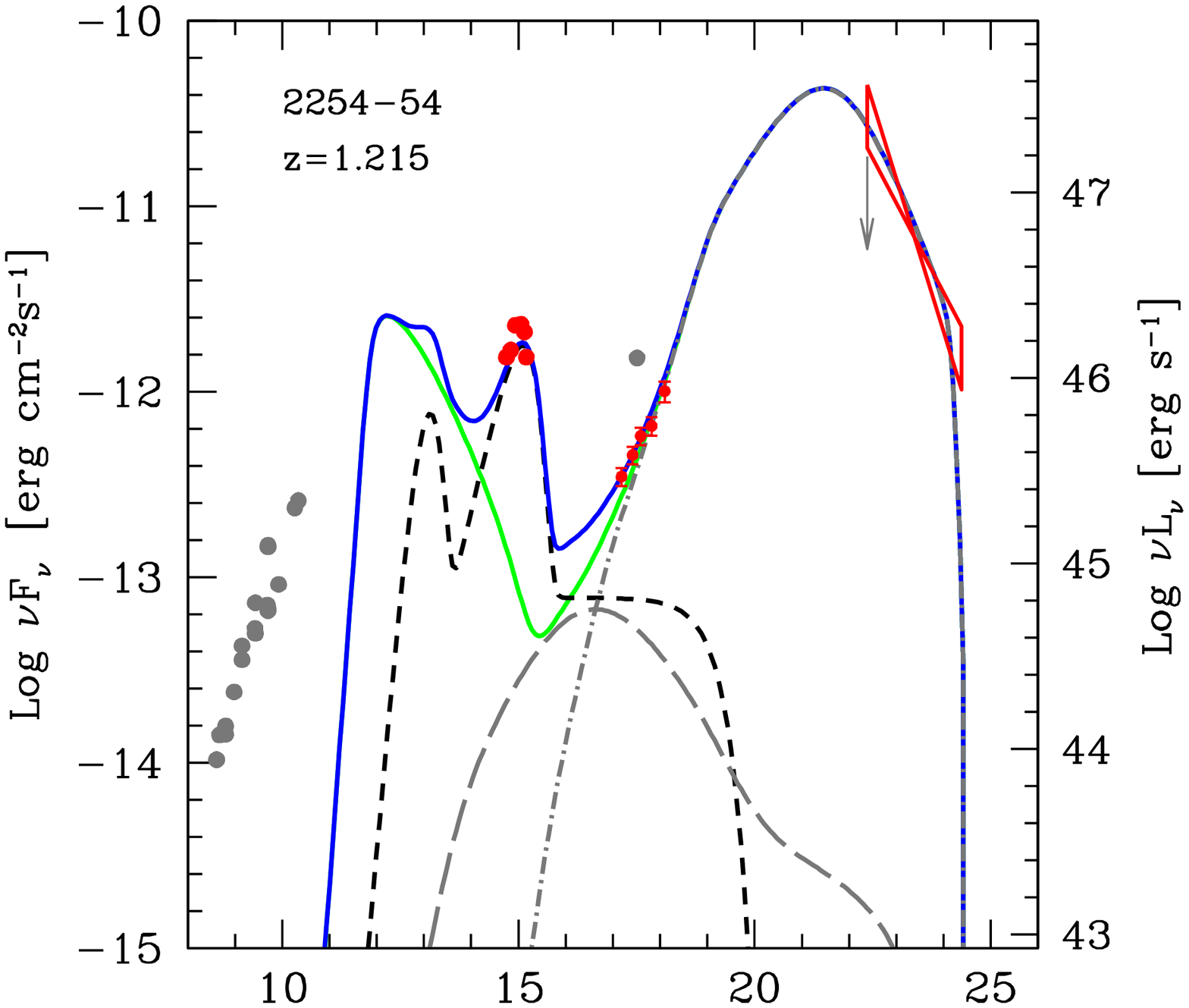,width=9cm,height=7cm}
\vskip -1.4 cm
\psfig{figure=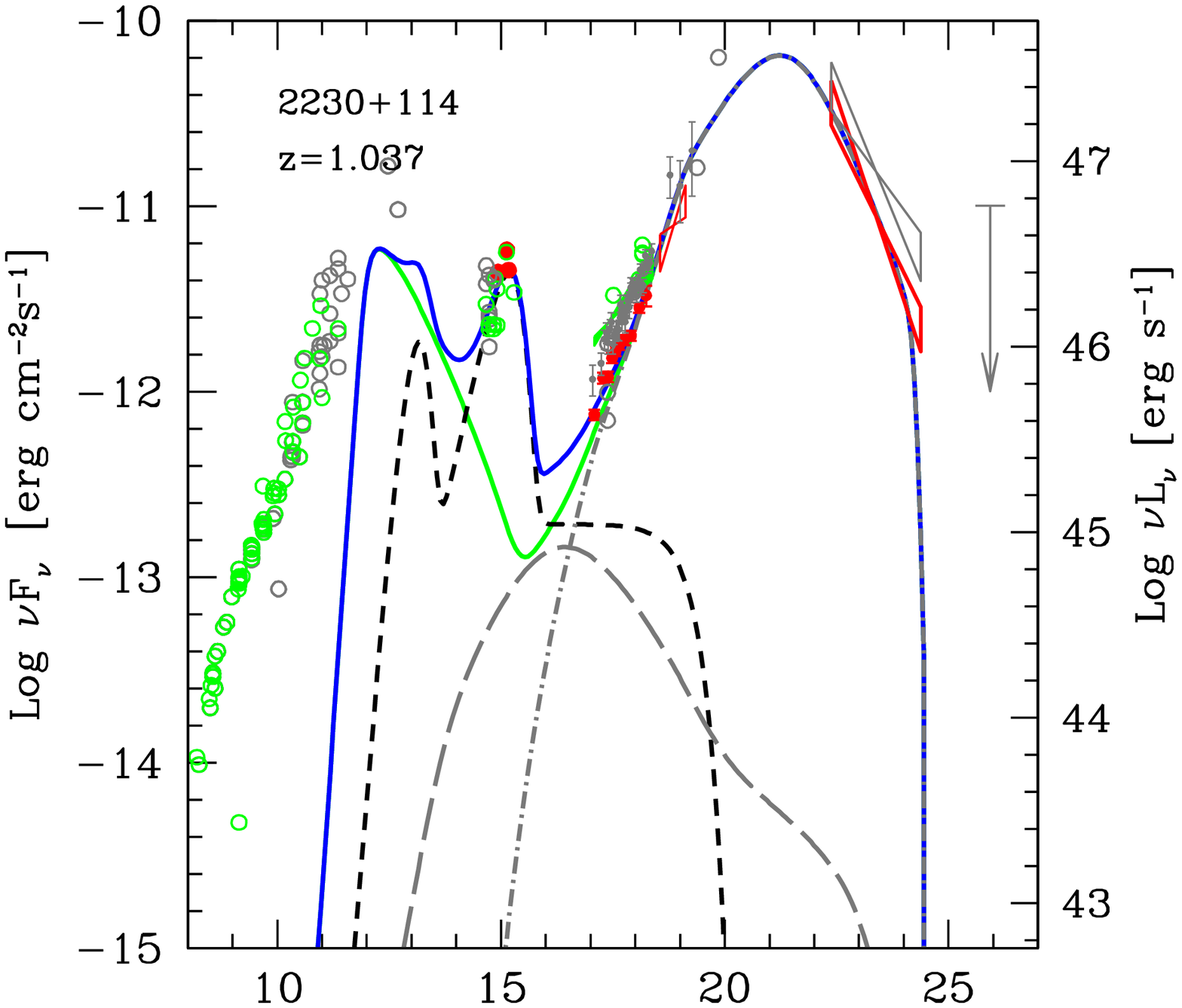,width=9cm,height=7cm}
\vskip -1.4 cm
\psfig{figure=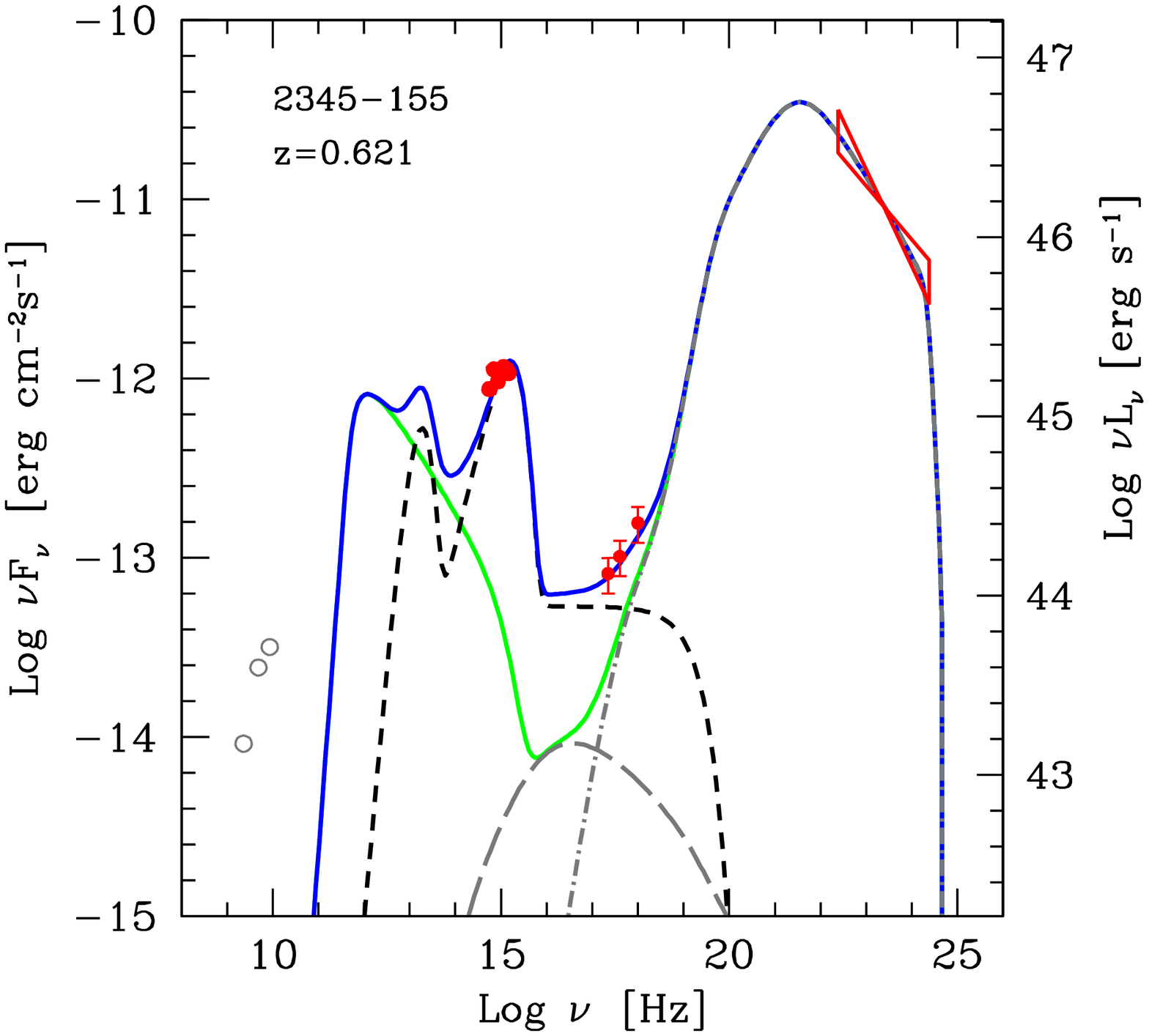,width=9cm,height=7cm}
\vskip -0.5 cm
\caption{SED of PKS 2204--54, 2230+114 (CTA102) and
PMN 2345--1555.
Symbols and lines as in Fig. \ref{f1}.
}
\label{f10}  
\end{figure}

\end{document}